%% file: main.tex
\RequirePackage{fix-cm}

\documentclass{article}
\usepackage{preprint}
\usepackage{hyperref}
\usepackage{url}
\usepackage{graphicx}
\usepackage{booktabs}
\usepackage{adjustbox}

\usepackage{amsmath,amssymb,amsthm}
\usepackage{microtype}
\hypersetup{colorlinks=true,linkcolor=preprintlink,citecolor=preprintlink,urlcolor=preprintlink,
  bookmarksnumbered=true,bookmarksopen=true,
  pdftitle={The Predictability Bias of Joint-Embedding Predictive Learning in Medical Imaging},
  pdfauthor={Fabian Drexel, Bernhard Kainz, Jakob N. Kather, Daniel Rueckert, Lisa C. Adams, Keno K. Bressem, Cosmin I. Bercea},
  pdfkeywords={self-supervised learning, representation learning, joint-embedding predictive architectures, medical imaging, predictability bias}}
\usepackage{multirow}
\renewcommand{\topfraction}{0.7}
\renewcommand{\textfraction}{0.2}
\newcommand{\method}{\textsc{TAP-JEPA}}
\newcommand{\context}{\mathcal{C}}
\newcommand{\target}{\mathcal{T}}
\newcommand{\best}[1]{\textbf{#1}}
\newcommand{\second}[1]{\underline{#1}}
\newcommand{\metric}[2]{\begin{tabular}[c]{@{}c@{}}#1\\[-0.25ex]{\scriptsize #2}\end{tabular}}
\newcommand{\inlineMetric}[2]{#1\,{\scriptsize #2}}
\theoremstyle{plain}
\newtheorem{proposition}{Proposition}
\newtheorem{lemma}{Lemma}
\theoremstyle{definition}
\newtheorem{definition}{Definition}
\usepackage{array,tikz}
\definecolor{milteal}{HTML}{087E83}
\definecolor{milcoral}{HTML}{C3654D}
\definecolor{milgray}{HTML}{707982}
\definecolor{milaxis}{HTML}{C2C7CC}
\definecolor{effectteal}{HTML}{087E83}
\definecolor{effectgray}{HTML}{707982}
\definecolor{axisgray}{HTML}{C2C7CC}
\newcommand{\cellmetric}[3]{\textcolor{#3}{#1}\,{\fontsize{6.5}{7.5}\selectfont #2}}
\newcommand{\effectplot}[5]{%
  \begin{tikzpicture}[x=1.85pt,y=1pt,baseline=-2pt]
    \path[use as bounding box] (-5,-5.5) rectangle (35,5.5);
    \draw[axisgray,densely dashed,line width=.35pt] (0,-5.5)--(0,5.5);
    \draw[#4,line width=.8pt] (#2,0)--(#3,0);
    \draw[#4,line width=.6pt] (#2,-2)--(#2,2) (#3,-2)--(#3,2);
    \ifnum#5=1
      \filldraw[draw=#4,fill=#4] (#1,0) circle[radius=2.2pt];
    \else
      \filldraw[draw=#4,fill=white,line width=.8pt] (#1,0) circle[radius=2.2pt];
    \fi
  \end{tikzpicture}%
}
\newcommand{\plotaxis}{%
  \begin{tikzpicture}[x=1.85pt,y=1pt,baseline=0pt]
    \path[use as bounding box] (-5,-13) rectangle (35,12);
    \draw[axisgray,line width=.4pt] (-5,0)--(35,0);
    \foreach \x in {0,10,20,30}{
      \draw[effectgray,line width=.4pt] (\x,0)--(\x,-2);
      \node[font=\fontsize{6.5}{7}\selectfont,anchor=north,inner sep=1pt] at (\x,-2) {$\x$};
    }
    \node[font=\fontsize{6.2}{7}\selectfont,anchor=south,inner sep=1pt,text=effectteal] at (15,3) {Control cost higher $\rightarrow$};
  \end{tikzpicture}%
}

\newcommand{\milAttention}[3]{%
\begin{tikzpicture}[x=9.3pt,y=1pt,baseline=-2pt]
\path[use as bounding box] (-1.15,-5.5) rectangle (4.15,5.5);
\draw[milaxis!55,line width=.3pt] (-1,0)--(4,0);
\draw[milaxis,densely dashed,line width=.4pt] (0,-5.5)--(0,5.5);
\pgfmathsetmacro{\est}{ln(#1)/ln(2)}
\pgfmathsetmacro{\low}{ln(#2)/ln(2)}
\pgfmathsetmacro{\high}{ln(#3)/ln(2)}
\draw[milcoral,line width=.8pt] (\low,0)--(\high,0);
\draw[milcoral,line width=.55pt] (\low,-1.6)--(\low,1.6) (\high,-1.6)--(\high,1.6);
\fill[milcoral] (\est,0) circle[radius=1.6pt];
\end{tikzpicture}}

\newcommand{\milAttentionAxis}{%
\begin{tikzpicture}[x=9.3pt,y=1pt,baseline=0pt]
\path[use as bounding box] (-1.15,-15) rectangle (4.15,6);
\draw[milaxis,line width=.4pt] (-1,0)--(4,0);
\foreach \x/\lab in {-1/{.5},0/{1},2/{4},4/{16}}{
\draw[milgray,line width=.4pt] (\x,0)--(\x,-2);
\node[font=\scriptsize,anchor=north,inner sep=.5pt] at (\x,-2) {$\lab$};}
\node[font=\scriptsize,anchor=north west,inner sep=0pt,text=milgray] at (-1,-10) {Less};
\node[font=\scriptsize,anchor=north east,inner sep=0pt,text=milcoral] at (4,-10) {More};
\end{tikzpicture}}

\newcommand{\milEffect}[3]{%
\begin{tikzpicture}[x=.80pt,y=1pt,baseline=-2pt]
\path[use as bounding box] (-4,-5.5) rectangle (104,5.5);
\draw[milaxis!55,line width=.35pt] (0,0)--(100,0);
\draw[milaxis,densely dashed,line width=.35pt] (0,-5.5)--(0,5.5) (100,-5.5)--(100,5.5);
\draw[milteal,line width=.7pt] (#2,0)--(#3,0);
\draw[milteal,line width=.5pt] (#2,-1.7)--(#2,1.7) (#3,-1.7)--(#3,1.7);
\fill[milteal] (#1,0) circle[radius=1.7pt];
\end{tikzpicture}}

\newcommand{\milAxis}{%
\begin{tikzpicture}[x=.80pt,y=1pt,baseline=0pt]
\path[use as bounding box] (-4,-15) rectangle (104,6);
\draw[milaxis,line width=.4pt] (0,0)--(100,0);
\foreach \x in {0,50,100}{
\draw[milgray,line width=.4pt] (\x,0)--(\x,-2);
\node[font=\scriptsize,anchor=north,inner sep=.5pt] at (\x,-2) {$\x$};}
\node[font=\scriptsize,anchor=north west,inner sep=0pt,text=milgray] at (0,-10) {No loss};
\node[font=\scriptsize,anchor=north east,inner sep=0pt,text=milgray] at (100,-10) {Full loss};
\end{tikzpicture}}

\newcommand{\milAttentionValue}[1]{\makebox[3.2em][r]{$#1\times$}}
\newcommand{\milEffectValue}[1]{\makebox[2.8em][r]{$#1$}}
\newcommand{\milAttentionCell}[4]{%
\begin{tabular}{@{}c@{\hspace{3pt}}c@{}}\milAttention{#1}{#2}{#3} & \milAttentionValue{#4}\end{tabular}}
\newcommand{\milEffectCell}[4]{%
\begin{tabular}{@{}c@{\hspace{3pt}}c@{}}\milEffect{#1}{#2}{#3} & \milEffectValue{#4}\end{tabular}}
\newcommand{\milAttentionHeaderAxis}{%
\begin{tabular}{@{}c@{\hspace{3pt}}c@{}}\milAttentionAxis & \makebox[3.2em][r]{}\end{tabular}}
\newcommand{\milEffectHeaderAxis}{%
\begin{tabular}{@{}c@{\hspace{3pt}}c@{}}\milAxis & \makebox[2.8em][r]{}\end{tabular}}

\title{The Predictability Bias of Joint-Embedding Predictive Learning in Medical Imaging}
\shorttitle{The Predictability Bias of Joint-Embedding Predictive Learning in Medical Imaging}

\author{%
  Fabian~Drexel\textsuperscript{1,2,$\ast$},\enspace\allowbreak
  Bernhard~Kainz\textsuperscript{3,4},\enspace\allowbreak
  Jakob~N.~Kather\textsuperscript{5,6},\enspace\allowbreak
  Daniel~Rueckert\textsuperscript{1,4,7},\enspace\allowbreak
  Lisa~C.~Adams\textsuperscript{2},\enspace\allowbreak
  Keno~K.~Bressem\textsuperscript{2,8,9,10,$\ddagger$},\enspace\allowbreak
  Cosmin~I.~Bercea\textsuperscript{2,$\ddagger$}%
}
\affiliations{%
  \textsuperscript{1}Chair for AI in Healthcare and Medicine, Technical University of Munich (TUM) and TUM University Hospital, Munich, Germany;
  \textsuperscript{2}Technical University of Munich, School of Medicine and Health, TUM University Hospital, Klinikum rechts der Isar, Department of Diagnostic and Interventional Radiology, Munich, Germany;
  \textsuperscript{3}Dept. AIBE, Friedrich-Alexander University Erlangen-N\"urnberg, Erlangen, Germany;
  \textsuperscript{4}Department of Computing, Imperial College London, London, UK;
  \textsuperscript{5}Institute for Medical AI (IMAI), Heidelberg University Hospital and Medical Faculty of Heidelberg University, Heidelberg, Germany;
  \textsuperscript{6}Else Kroener Fresenius Center for Digital Health (EKFZ), Faculty of Medicine Carl Gustav Carus, TUD Dresden University of Technology, Dresden, Germany;
  \textsuperscript{7}Munich Center for Machine Learning (MCML), Munich, Germany;
  \textsuperscript{8}Institute for Artificial Intelligence in Medicine (IKIM), University Hospital Essen (A\"oR), Essen, Germany;
  \textsuperscript{9}Institute of Interventional and Diagnostic Radiology and Neuroradiology, University Hospital Essen, Essen, Germany;
  \textsuperscript{10}National Center for Tumor Diseases West, Essen, Germany.%
}
\titlenotes{\textsuperscript{$\ast$}Corresponding author: \href{mailto:fabian.drexel@tum.de}{\texttt{fabian.drexel@tum.de}}\qquad
  \textsuperscript{$\ddagger$}These authors share last authorship.}

\begin{document}
\maketitle

\begin{abstract}
Joint-embedding predictive architectures (JEPAs) learn by predicting teacher representations of masked regions from visible context. Point prediction summarizes the conditional target distribution and cannot reproduce variation that the context does not resolve. We call this constraint \emph{predictability bias}. In computed tomography (CT), recurrent anatomy is predictable from context, whereas the sporadic, localized pathology that determines clinical interpretation often is not. We introduce \method, a ViT-B/3D trained on 113{,}611 CT volumes that is competitive with state-of-the-art encoders (COLIPRI, Merlin, VoxelFM) in scan-level classification. We then audit the learned predictive pathway by replaying the exact masking, teacher, and predictor computations used during training.
Prediction lowers the mean Skill (prevalence-normalized average precision) from 0.388 to 0.271 for pathology, but only from 0.921 to 0.873 for anatomy. Along the teacher direction, separating pathology from matched control regions in the same organ, prediction moves pathology 34.7\% toward the controls and controls 13.6\% toward pathology, with larger shifts farther from visible context. Replacing visible matched-control representations increases surrounding-target prediction loss more than fully replacing pathology representations across four pathology categories. Downstream, removing all final-layer pathology tokens reduces the decision margin by at most 0.45\% in four of seven scan-level probes. Together, these results show that contextual predictability is not a sufficient proxy for clinical relevance in medical representation learning. Training objectives should therefore retain diagnostic features regardless of their predictability from surrounding anatomy.
\end{abstract}

\section{Introduction}
\label{sec:introduction}
Joint-embedding predictive architectures (JEPAs) learn visual representations by predicting teacher features at hidden locations from a visible context~\citep{assran2023ijepa,bardes2024vjepa}.
Their learning signal, therefore, depends on structure that can be inferred across partial observations. In natural images, spatial continuity makes much of a hidden region predictable from its surroundings~\citep{dinh2026paradox}, providing a strong signal for learning transferable representations.

Computed tomography (CT) provides an abundant contextual signal because anatomy recurs across patients at a predictable shape and location. Clinical interpretation, however, depends on sparse and heterogeneous pathological findings that can be absent from anatomically compatible locations. A lesion can therefore be evident in its own voxels while remaining unpredictable when those voxels are hidden. This distinction is particularly important as joint-embedding predictive learning is being adopted rapidly across medical imaging, with recent models spanning radiographs, ultrasound, neuroimaging, and volumetric CT~\citep{yue2025chexworld,khan2026radjepa,radhachandran2026usjepa,kondepudi2026neurovfm,huang2026neurojepa,trinh2026radjepa3d}.

\begin{figure}[tb]
\centering
\includegraphics[width=\linewidth]{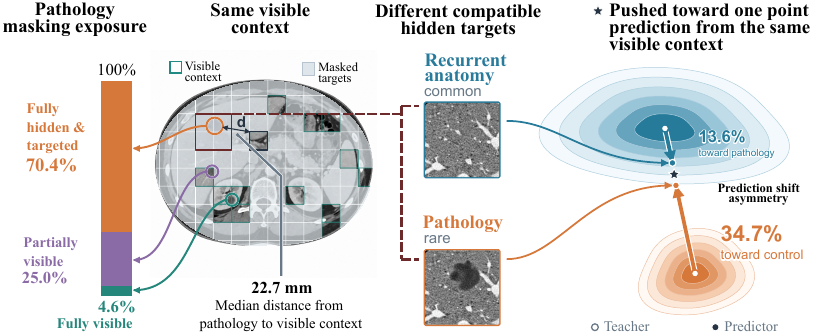}
\caption{\textbf{Predictability bias in medical imaging.}
Joint-embedding prediction preferentially preserves information that is constrained by visible context.
Left, common JEPA masking schemes can leave most localized pathology outside the visible context.
Right, hidden-target prediction in \method{} shifts pathology representations toward matched normal controls (values, see Appendix~\ref{app:summary-values}).}
\label{fig:teaser}
\end{figure}

Current transfer benchmarks establish what a representation can support downstream, but they do not reveal which local information the predictive objective preserves. We study this information preference directly and refer to its dependence on what visible context can resolve as \emph{predictability bias}. Our central comparison is how much teacher-accessible information survives hidden-target prediction for recurrent anatomy versus sporadic pathology in a strong trained encoder.
We trained \method{}, a ViT-B/3D model, for 240 epochs on 113{,}611 thoraco-abdomino-pelvic CT volumes from four public cohorts using 32 NVIDIA H200 GPUs. Its frozen representations are competitive with state-of-the-art CT encoders on thoracic and abdominal classification benchmarks. We then replay its exact mask sampler, teacher, predictor, and loss to trace annotated pathology through the training objective. The audit spans masking exposure, teacher-to-prediction retention, pathology--control displacement, interventions on visible context, and downstream dependence on pathology tokens. \autoref{fig:teaser} summarizes the central asymmetry, with recurrent anatomy preserved more strongly than pathology and predicted pathology shifted toward matched normal controls.

\section{Related Work}
\label{sec:related}

\textbf{Latent predictive learning.}
Data2vec predicts contextualized teacher representations from masked inputs
\citep{baevski2022data2vec}. I-JEPA applies latent prediction to spatial
blocks in images \citep{assran2023ijepa}, while V-JEPA and V-JEPA~2 extend
the approach to video \citep{bardes2024vjepa,assran2025vjepa2}.
Predicting features rather than pixels allows these models to learn from
structure shared between the visible and hidden regions without reproducing
every detail of the observation.

\textbf{Medical imaging.}
CheXWorld and RadJEPA apply latent prediction to chest radiographs \citep{yue2025chexworld,khan2026radjepa}; US-JEPA to ultrasound~\citep{radhachandran2026usjepa}; NeuroVFM and Neuro-JEPA to volumetric neuroimaging~\citep{kondepudi2026neurovfm,huang2026neurojepa}, and Rad-JEPA 3D~\citep{trinh2026radjepa3d} to volumetric CT scans.
Other CT encoders use self-distillation, as in VoxelFM
\citep{morenoaguado2026voxelfm} or supervision from radiology reports as in Merlin \citep{blankemeier2026merlin} and COLIPRI \citep{wald2026colipri}.
Two studies evaluated what the representations contain. \citet{ivezic2026pretext} find in ultrasound and histopathology that different self-supervised objectives favor either global anatomy or localized diagnostic features, and \citet{chevli2026big} find that small, low-contrast findings remain hard to detect for all objectives. Neither traces a finding through the training objective, so whether the objective itself removes local pathology remains open.

\textbf{Predictability and feature selection.}
\citet{sobal2022slow} show that joint-embedding architectures favor features that vary slowly across related observations. In a deep linear setting, \citet{littwin2024noisy} characterize a preference for features that are more strongly predictable from the available representation, while \citet{vanassel2025joint} derive conditions under which view construction suppresses task-irrelevant variation. \citet{pendharkar2026predictive} shows in synthetic control tasks that information can remain task-relevant even when it is unpredictable from context, while \citet{dinh2026paradox} analyze the related case in language, where the same context can admit multiple valid targets that deterministic prediction must summarize. In medical imaging, \citet{ivezic2026pretext} find that self-supervised objectives differ in how well their representations support global anatomical and localized diagnostic information, while \citet{chevli2026big} show that small and low-contrast findings remain difficult across CT foundation models. We connect these lines of work by tracing localized pathology through the predictive computation of a large-scale 3D CT JEPA.

\section{Predictability Bias}
\label{sec:theory}
At each JEPA update, the student predicts teacher features at hidden targets from the visible context. Stop-gradient fixes the teacher targets during that update, while gradients from the prediction loss update the student and predictor. The teacher subsequently follows the student by an exponential moving average. We analyze this per-update prediction problem because it is the computation that generates the encoder's training signal.
Let \(X\sim P_X\) denote the 3D volume, with
\(X\in\mathbb{R}^{H\times W\times D}\). Patchification maps \(X\) to
\(N\) input patches \((x_i)_{i\in\mathcal G}\), indexed by the token set
\(\mathcal G=\{1,\ldots,N\}\). Conditional on \(X\), the mask sampler draws
\(M=(\context,\target)\sim P_M(\cdot\mid X)\), where
\(\context,\target\subseteq\mathcal G\) are disjoint context and target
index sets. We write \(X_\context=(x_i)_{i\in\context}\) for the visible
input.

The student encoder \(E_\theta\) maps the visible input to context features
\(H_\context=E_\theta(X_\context)\). The teacher encoder
\(E_{\bar\theta}\) processes the complete training view and produces
\(Z=(Z_i)_{i\in\mathcal G}=E_{\bar\theta}(X)\), where
\(Z_i\in\mathbb R^d\). The teacher targets are \(Z_i\) for \(i\in\target\). The predictor \(P_\phi\) maps the context features
and sampled mask to predicted target features
\(\widehat Z_\target=P_\phi(H_\context,M)\), where
\(\widehat Z_i\in\mathbb R^d\) estimates \(Z_i\) for each
\(i\in\target\).

For each target index \(i\in\target\), the prediction error is measured in
teacher-feature space. The deployed objective uses the feature-wise \(L_1\)
loss
\begin{equation}
\ell_i
=
\ell(\widehat Z_i-Z_i)
=
\frac{1}{d}
\sum_{k=1}^{d}
\left|
\widehat Z_{ik}-Z_{ik}
\right|,
\label{eq:token-loss}
\end{equation}
where \(k\) indexes the \(d\) feature coordinates.

Let \(\alpha_i=\alpha_i(X,M)\geq 0\) denote the normalized target weight, with \(\sum_{i\in\target}\alpha_i=1\). In the deployed objective, \(\alpha_i\) depends only on body occupancy; clinical annotations do not enter the weighting (details in Appendices~\ref{app:importance} and \ref{app:details}). The loss for one volume and mask is \(L(X,M)=\sum_{i\in\target}\alpha_i\ell_i\).
Let \(\operatorname{sg}\) denote the stop-gradient operator. The expected training loss is
\begin{equation}
\mathcal L(\theta,\phi;\bar\theta)
=
\mathbb E_{X,M}\!\left[
\sum_{i\in\target}\alpha_i
\frac{1}{d}
\sum_{k=1}^{d}
\left|
\widehat Z_{ik}
-
\operatorname{sg}(Z_{ik})
\right|
\right].
\label{eq:objective}
\end{equation}
Gradients from Eq.~\ref{eq:objective} update the student parameters
\(\theta\) and predictor parameters \(\phi\) through \(\widehat Z_i\), while
the teacher parameters \(\bar\theta\) are held fixed. After each student
update, the teacher parameters follow the student by exponential moving
average,
\(\bar\theta\leftarrow\tau\bar\theta+(1-\tau)\theta\), where
\(\tau\in[0,1)\) is the teacher momentum. The predictive computation, therefore, provides the training signal for the student representation.

For target-level analysis, we absorb the normalized target weights into a
joint law \(\mathbb P\). We draw \(X\sim P_X\), then
\(M\sim P_M(\cdot\mid X)\), and finally a target index \(I\) with
\(
\mathbb P(I=i\mid X,M)=\alpha_i(X,M),
\qquad i\in\target.
\)
Define \(Z=Z_I\) and \(\widehat Z=\widehat Z_I\). The objective in
Eq.~\ref{eq:objective} can then be written as
\(
\mathcal L(\theta,\phi;\bar\theta)
=
\mathbb E_{X,M,I}\!\left[
\ell\!\left(\widehat Z-\operatorname{sg}(Z)\right)
\right].
\)
All unqualified probabilities and expectations below are taken under
\(\mathbb P\).

Fixing one training update, we define
\(\mathcal U=\sigma(X_\context,M,I)\) as the information available to the
predictor for the selected target. The prediction \(\widehat Z\) is
\(\mathcal U\)-measurable. The best point prediction available from the visible context is therefore
determined by the conditional law \(Z\mid\mathcal U\). Finite encoder and predictor capacity can introduce additional error (Appendix~\ref{app:locations}).

Under this target law, the Bayes-optimal point predictions follow from the corresponding loss minimizers~\citep{gneiting2011point,dinh2026paradox}.

\begin{proposition}[Optimal point predictions]
\label{prop:location}
Let \(\mathcal A\) denote an information set and let
\(Z\in\mathbb R^d\) be the teacher target under the weighted target law
defined above. If \(\mathbb E\|Z\|_2^2<\infty\), the unique
\(\mathcal A\)-measurable minimizer of
\(\mathbb E\|a-Z\|_2^2\) is the conditional mean. If
\(\mathbb E\|Z\|_1<\infty\), an \(\mathcal A\)-measurable prediction
minimizes \(\mathbb E\|a-Z\|_1\) if and only if each coordinate is a
conditional median:
\begin{equation}
\begin{aligned}
a_2^\star
&=
\mathbb E[Z\mid\mathcal A],
\\
\mathbb P(Z_k\leq a_{1,k}^\star\mid\mathcal A)
&\geq\tfrac12,
\qquad
\mathbb P(Z_k\geq a_{1,k}^\star\mid\mathcal A)
\geq\tfrac12,
\quad k=1,\ldots,d.
\end{aligned}
\label{eq:conditional-location}
\end{equation}
\end{proposition}

Prediction can therefore preserve subject-specific anatomy or pathology to
the extent that it is resolved by the visible context. Variation that remains
unresolved conditional on \(\mathcal U\) is represented only through the
corresponding conditional summary. Importantly, a finding can be explicit in
the hidden target while remaining unresolved from the information available
to the predictor. Appendix~\ref{app:locations} applies both characterizations
to this setting.

\begin{lemma}[Conditional training signal]
\label{lem:gradient}
At the same training update, let \(\omega=(\theta,\phi)\) and let
\(S\in\partial_{\widehat Z}\ell(\widehat Z-Z)\) be a measurable choice of
the loss derivative or subgradient with respect to the prediction. Since
\(\widehat Z\) is \(\mathcal U\)-measurable,
\(J_\omega=\partial\widehat Z/\partial\omega\) is also
\(\mathcal U\)-measurable. Define the corresponding parameter-space training
signal as \(G_\omega=J_\omega^\top S\). If the required expectations are
finite, then
\begin{equation}
\mathbb E[G_\omega\mid\mathcal U]
=
J_\omega^\top\mathbb E[S\mid\mathcal U],
\qquad
\mathbb E\!\left[
J_\omega^\top
\bigl(S-\mathbb E[S\mid\mathcal U]\bigr)
\mid\mathcal U
\right]
=0.
\label{eq:conditional-gradient}
\end{equation}
\end{lemma}

The target-dependent component of the loss score around its conditional expectation can affect individual updates, but contributes no conditional mean training signal through the visible path. Under \(L_2\), this component is proportional to \(Z-\mathbb E[Z\mid\mathcal U]\); under feature-wise \(L_1\), it is the corresponding centered sign/subgradient score. Appendix~\ref{app:gradient} proves the lemma and gives both forms.

\begin{definition}[Predictability bias]
\label{def:bias}
\emph{Predictability bias} is the dependence of point-prediction learning on what the visible context can resolve about the target. Optimal predictions summarize the conditional target distribution (Eq.~\ref{eq:conditional-location}), while target-specific variation unresolved by the visible context contributes no conditional mean training signal through the visible path (Eq.~\ref{eq:conditional-gradient}).
\end{definition}

\textbf{Empirical implications.} Applied to medical CT, this characterization yields several measurable signatures. If recurrent anatomy is more strongly constrained by visible context than sporadic pathology, hidden-target prediction should preserve more teacher-accessible anatomical than pathological discrimination. Pathology--control differences should contract when the corresponding regions are predicted from context, with stronger attenuation as the visible context becomes less informative about the target. The same information preference can be examined from the context side by intervening on visible pathology representations and measuring their contribution to hidden-target prediction.
\section{Experimental Setup}
\label{sec:methods}

\textbf{Model and training.}
\method{} consists of a ViT-B/3D student \citep{dosovitskiy2021vit}, an exponential-moving-average
teacher, and a six-layer predictor. We trained the model on 113{,}611 thoraco-abdomino-pelvic CT volumes from NLST,
CT-RATE, Merlin, and AbdomenAtlas (ATLAS, version 3.0-mini)
\citep{nlst2011,hamamci2024ctclip,blankemeier2026merlin,bassi2025radgpt}.
Training ran for 240 epochs on 32 NVIDIA H200 GPUs, using
1{,}622 GPU-hours. Each input token covers a 24\,mm isotropic patch, following prior volumetric JEPA designs; the same tokenization is used across findings of varying spatial extent.
The mask sampler combines multi-block targets, cropped visible context,
and multi-scale blocks, drawing on volumetric JEPA masking in NeuroVFM
and Neuro-JEPA \citep{kondepudi2026neurovfm,huang2026neurojepa}.
The realized masks expose approximately 16\% of in-body tokens as
context and assign 84\% as hidden targets. Training uses the
occupancy-weighted feature-wise \(L_1\) objective in
Eq.~\ref{eq:objective}. Architecture, sampling probabilities, and
training settings are given in Appendix~\ref{app:details}.

\textbf{Evaluation design.}
We examine prediction by replaying the training tokenization,
mask sampler, teacher, and predictor at a fixed checkpoint, without parameter updates.
Downstream evaluations use frozen encoder features from fully
visible inputs. Encoder benchmarks use each model's native preprocessing and the readout protocols detailed in Appendix~\ref{app:benchmark-protocol}. The regional analyses draw on 1{,}863 abdominal ATLAS scans and 950 thoracic ReXGroundingCT (ReX) scans, all unseen during pretraining
\citep{baharoon2025rexgroundingct}.
Hidden-target discrimination uses annotation-negative control tokens from the
corresponding organ. For paired displacement and
visible-context analyses, controls are constructed by translating
each pathology annotation within its parent organ,
preserving its shape and voxel count. We refer to these as matched controls, constructed within the same anatomical region and excluding annotated pathology. Each analysis specifies its eligible cohort,
mask replays, and overlap criteria
(Appendices~\ref{app:e4}--\ref{app:e6}).

\textbf{Metrics and uncertainty.}
We report macro AUROC and AUPRC for encoder benchmarks; AUPRC denotes stepwise average precision (AP).
For the regional analyses, we use prevalence-normalized AP,
$\mathrm{Skill}=(\mathrm{AP}-\pi)/(1-\pi)$, where $\pi$ is class prevalence (further details in Appendix \ref{app:skill_aggregation}).
Skill is zero when AP equals prevalence and one for perfect discrimination.
Appendix~\ref{app:estimands} and evaluation-specific sections define the estimates; uncertainty and multiplicity corrections accompany each analysis.

\section{Results}
\label{sec:results}

\subsection{Competitive transfer across thoracic and abdominal CT}
\label{sec:results-benchmark}
We trained scan-level classification probes on frozen encoder features and selected them on source-domain validation data. Thoracic probes fitted on CT-RATE were applied unchanged to the external RAD-ChestCT cohort \citep{draelos2021radchestct}; abdominal probes were tested on the SSL-held-out official Merlin test split. Table~\ref{tab:benchmarks} reports these held-out transfer results; thoracic source-domain performance is provided in Appendix~\ref{app:benchmark-results}.
Frozen \method{} features are competitive on both benchmarks
(Table~\ref{tab:benchmarks}). On external RAD-ChestCT, \method{} gives the highest AUPRC,
0.5268, and an AUROC of 0.7590. Patient-level paired differences, VoxelFM
minus \method{}, are $+0.0013$ (95\% CI $[-0.0027,0.0058]$) for AUROC and
$-0.0075$ ($[-0.0151,-0.0006]$) for AUPRC.
On Merlin, \method{} gives the highest AUROC and AUPRC, 0.8714 and 0.6965,
ahead of the locally evaluated report-supervised Merlin encoder and VoxelFM.

\begin{table}[tb]
\centering
\small
\setlength{\tabcolsep}{1.5pt}
\caption{Frozen scan-level classification on held-out thoracic and abdominal benchmarks: macro AUROC/AUPRC with 95\% bootstrap intervals; bold marks the highest point estimate.
$^\dagger$ Published official COLIPRI results; other rows are local evaluations. Comparator sources: \citet{wald2026colipri},
\citet{morenoaguado2026voxelfm} and \citet{blankemeier2026merlin}.
SSL: self-supervised learning; VL: vision--language training; EHR: electronic health record supervision.}
\label{tab:benchmarks}
\begin{tabular*}{\textwidth}{@{\extracolsep{\fill}}llccc@{}}
\toprule
Domain & Encoder & Pretraining & AUROC $\uparrow$ & AUPRC $\uparrow$\\
\midrule
\multirow{3}{*}{\shortstack{Thorax\\{\scriptsize (RAD-ChestCT)}}}
& \method{} (ours) & Image SSL & \inlineMetric{0.7590}{[0.7533,0.7649]} &
  \inlineMetric{\best{0.5268}}{[0.5178,0.5368]}\\
& COLIPRI$^\dagger$ & VL &
  \inlineMetric{\best{0.7657}}{[0.7608,0.7718]} &
  \inlineMetric{0.5255}{[0.5172,0.5365]}\\
& VoxelFM & Image SSL &
  \inlineMetric{0.7603}{[0.7553,0.7658]} &
  \inlineMetric{0.5192}{[0.5117,0.5288]}\\
\addlinespace
\multirow{3}{*}{\shortstack{Abdomen\\{\scriptsize (Merlin)}}}
& \method{} (ours) & Image SSL & \inlineMetric{\best{0.8714}}{[0.8622,0.8796]} &
  \inlineMetric{\best{0.6965}}{[0.6818,0.7158]}\\
& Merlin & VL + EHR &
  \inlineMetric{0.8668}{[0.8562,0.8750]} &
  \inlineMetric{0.6825}{[0.6675,0.7013]}\\
& VoxelFM & Image SSL &
  \inlineMetric{0.8379}{[0.8269,0.8472]} &
  \inlineMetric{0.6422}{[0.6276,0.6634]}\\
\bottomrule
\end{tabular*}
\end{table}

\subsection{Training masks frequently hide local pathology evidence}
\label{sec:results-exposure}

We replayed 1{,}600 production masks per annotated scan and classified each
crop-retained pathology--mask event as fully visible, partially visible, or
fully hidden with its representation targeted for prediction. Across seven
pathology categories, 61.5--81.4\% of events are fully hidden and targeted, with
median annotation-boundary distances to visible context of 20.25--25.11\,mm.
Pathology-containing tokens are rare, accounting for only 0.41--7.04 per 1,000 target tokens in these annotated cohorts. Aggregate exposure across visibility states is shown in Figure~\ref{fig:teaser}, with additional category-level statistics in Appendix~\ref{app:e3-exposure}.
Thus, \textit{many pathology targets must be predicted from surrounding evidence}. This establishes
the exposure condition under which predictability bias is relevant.
\subsection{Target prediction reduces pathology discrimination more than anatomy}
\label{sec:results-hidden}
\begin{figure}[!t]
\centering
\includegraphics[width=\linewidth]{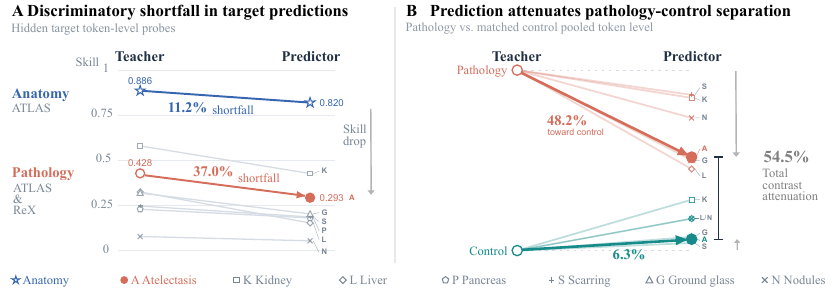}
\caption{Prediction reduces pathology discrimination and pathology--control separation. \textbf{A:} Teacher and predictor Skill; percentages show normalized shortfall (Eq.~\ref{eq:recovery}). Geometry baselines are omitted from the plot.
Anatomy uses class-macro Skill; pathology uses positive-class Skill on distinct token populations.
\textbf{B:} Projections onto cross-fitted teacher pathology--control axes. Teacher control/pathology positions are 0/1; predictor positions are $D_-$ and $1-D_+$ (Eq.~\ref{eq:retained-contrast}).
Atelectasis/consolidation is highlighted in both panels. Pancreas lacks sufficient matched pairs for B.
Further details and uncertainties
are provided in Appendices~\ref{app:e4}--\ref{app:e5}.}
\label{fig:e4-ladder}
\label{fig:e4-anatomy-finding}
\label{fig:e5-shrinkage}
\end{figure}

We compared how much anatomical and pathological discrimination is preserved from complete-view teacher targets to their corresponding hidden-target predictions.
Separate linear probes read three states at the same hidden target tokens:
a geometry-only baseline $\mathcal B$ (mask mode, token budgets, grid
position, body occupancy, and nearest-context distance), the deployed
prediction $\widehat Z$, and the normalized teacher target $Z$.
Rows, folds, and masks were identical across states within each task.
We define the
\emph{normalized Skill shortfall} as
\begin{equation}
Q_{\mathrm{short}}=
\frac{\mathrm{Skill}(Z)-\mathrm{Skill}(\widehat Z)}
     {\mathrm{Skill}(Z)-\mathrm{Skill}(\mathcal B)}.
\label{eq:recovery}
\end{equation}
This normalizes the teacher-to-prediction Skill reduction by the geometry-to-teacher Skill gap, measuring the fraction of teacher-accessible discrimination lost under prediction.
Figure~\ref{fig:e4-ladder}A shows the comparison for atelectasis/consolidation; geometry baselines, full intervals, and category-level results are reported in Appendix~\ref{app:e4}.

Relative to the teacher-accessible signal, pathology shows a larger normalized Skill shortfall
than anatomical identity in both datasets: 36.7\% versus 11.2\% in ATLAS,
and 31.8\% versus 6.2\% in ReX. This pattern is consistent with greater contextual constraint on anatomy than pathology.

The asymmetry adds to an already unequal teacher separability. In ATLAS,
Skill falls from 0.434 in the teacher to 0.298 in the prediction for pathology and from
0.886 to 0.820 for anatomy, and in ReX, from 0.341 to
0.243 and from 0.956 to 0.925. Predictions remain well above geometry
baselines across tasks, yet widen the anatomy--pathology Skill gap already
present in the teacher.

Hidden-target prediction preserves substantially more teacher-accessible anatomical than pathological discrimination. The paired
pathology-minus-anatomy shortfall difference is 25.5 and 25.6 percentage
points in ATLAS and ReX (95\% CIs $[20.2,32.5]$ and $[21.4,29.9]$) and
persists under a richer geometry baseline (Appendix~\ref{app:e4}).

\subsection{Hidden-target prediction attenuates pathology--control separation}
\label{sec:results-feature}

We measured how prediction changes pathology--control separation by hiding a pathology region
and its matched annotation-negative control in the same organ,
and projected their teacher and predicted features onto a cross-fitted teacher
pathology--control direction (Figure~\ref{fig:e5-shrinkage}B).

For pathology-control pair $p$, let $x_p$ and $y_p$ be the projected teacher and predicted
pathology-minus-control separations. We define the \emph{fractional separation
attenuation} as
\begin{equation}
A_{\mathrm{ctr}}
=\frac{\sum_p x_p(x_p-y_p)}{\sum_p x_p^2}
=1-\frac{\sum_p x_p y_p}{\sum_p x_p^2}
=D_++D_-.
\label{eq:retained-contrast}
\end{equation}
This is the $x_p^2$-weighted mean signed fractional reduction in teacher separation.
$D_+$ measures pathology displacement toward the teacher control; $D_-$ measures control displacement toward the teacher pathology, with the same units and weights.
We report both as percentages relative to the teacher separation.
When visible context does not resolve pathology, optimal point prediction summarizes the conditional target distribution. Greater pathology-side than control-side displacement would be consistent with this summary favoring pathology-free representations.
Separation is attenuated in all six pathology categories, from 17.9\% for
scarring/fibrosis to 72.4\% for liver lesions. Pathology-side displacement
exceeds control-side displacement in five categories, matching the expected
signature. Atelectasis/consolidation shows 54.5\% attenuation, with 48.2\%
pathology-side and 6.3\% control-side displacement
(Figure~\ref{fig:e5-shrinkage}B); liver lesions and ground-glass opacity show
similarly dominant pathology-side contributions. Kidney lesions are the
exception, as their 43.6\% attenuation has a larger control-side contribution
(28.2\% versus 15.4\%). Attenuation also grows with distance to visible
context. In all six categories, it is larger in the farthest than in the nearest
distance bin; for liver lesions, it rises from 67.4\% below 24\,mm to 83.4\%
at 96\,mm or more (Appendix~\ref{app:e5}).

To distinguish pathology-directed attenuation from generic contraction, we compared the same pairs with 4{,}096 sets of random axes. Attenuation is greater along the liver, ground-glass, and atelectasis directions, smaller along the kidney and scarring directions, and similar for nodules. Random-axis calibration therefore shows attenuation beyond generic prediction-space contraction for several pathology directions, with the magnitude varying across categories (Appendix~\ref{app:e5}).

\begin{table}[!t]
\centering
\caption{Replacing visible control representations disrupts prediction of shared hidden regions more than replacing pathology representations in all six pathology categories, with four differences significant after multiplicity correction. Positive values indicate higher control replacement cost. Costs are adjusted for equal-count unrelated-region replacement and normalized by unadjusted control replacement cost or unperturbed target loss. Whiskers show pointwise 95\% CIs; filled markers and $^{*}$ denote joint-adjusted $p<0.05$. $^{\dagger}$ Below 100 pairs, descriptive. Full protocol in Appendix~\ref{app:e6}.}
\label{tab:context-main}
\small
\begin{adjustbox}{max width=\linewidth}
\begin{tabular}{@{}llrccc@{}}
\toprule
& & & \multicolumn{3}{c}{Replacement cost: control $-$ pathology}\\[2pt]
\cmidrule(lr){4-6}
Dataset & Pathology & Pairs & \shortstack{Effect\\\plotaxis} & \shortstack{\% of control\\cost} & \shortstack{\% of baseline\\loss}\\
\midrule
\multirow{2}{*}{ATLAS}
& Kidney lesion & $60^{\dagger}$
& \effectplot{11.6}{-2.0}{24.0}{effectgray}{0}
& \cellmetric{$+11.6$}{$[-2.0,\,24.0]$}{effectgray}
& \cellmetric{$+0.423$}{$[-0.068,\,0.913]$}{effectgray}\\
& Liver lesion & 402
& \effectplot{12.8}{8.1}{17.3}{effectteal}{1}
& \cellmetric{$+12.8^{*}$}{$[8.1,\,17.3]$}{effectteal}
& \cellmetric{$+0.599$}{$[0.362,\,0.860]$}{effectgray}\\
\addlinespace[2pt]
\multirow{4}{*}{ReX}
& Scarring/fibrosis & 242
& \effectplot{27.3}{22.4}{32.0}{effectteal}{1}
& \cellmetric{$+27.3^{*}$}{$[22.4,\,32.0]$}{effectteal}
& \cellmetric{$+2.433$}{$[1.914,\,2.989]$}{effectgray}\\
& Atelectasis/consol. & 280
& \effectplot{17.6}{12.8}{22.0}{effectteal}{1}
& \cellmetric{$+17.6^{*}$}{$[12.8,\,22.0]$}{effectteal}
& \cellmetric{$+1.616$}{$[1.136,\,2.103]$}{effectgray}\\
& Ground-glass opacity & 444
& \effectplot{12.6}{9.0}{15.9}{effectteal}{1}
& \cellmetric{$+12.6^{*}$}{$[9.0,\,15.9]$}{effectteal}
& \cellmetric{$+1.152$}{$[0.812,\,1.475]$}{effectgray}\\
& Nodules/masses & 288
& \effectplot{6.0}{0.3}{11.4}{effectgray}{0}
& \cellmetric{$+6.0$}{$[0.3,\,11.4]$}{effectgray}
& \cellmetric{$+0.297$}{$[0.013,\,0.589]$}{effectgray}\\
\bottomrule
\end{tabular}
\end{adjustbox}
\end{table}

\subsection{Pathology context contributes less to surrounding prediction}
\label{sec:results-context-role}

To measure how much visible pathology supports the prediction of surrounding
hidden targets, we kept a pathology region and its matched control visible
and separately replaced each region's final student activations with the mean
of other annotation-negative same-organ context tokens. We then
re-evaluated the trained predictor on shared clean hidden targets whose centers
lie 24--96\,mm from both region-token centroids, subtracting the cost of an
equal-count unrelated-region replacement as a reference adjustment. We use the same six categories as in the displacement analysis.

Replacing control features increases shared-target loss more than replacing
pathology features in all six categories, with four differences significant after joint correction (Table~\ref{tab:context-main}). For
liver, scarring, atelectasis, and ground-glass, the control-minus-pathology
difference is $+12.6\%$ to $+27.3\%$ of control replacement cost ($+0.60\%$ to
$+2.43\%$ of baseline loss) ($p<0.001$ after joint correction). The
nodule difference ($+6.0\%$) does not remain significant after joint correction ($p=0.0656$), and kidney
is descriptive at $n=60$. Under full replacement on shared targets, this provides a context-side counterpart to the target-side asymmetry in the four supported categories: replacing visible pathology disrupts prediction of surrounding anatomy less than replacing matched control regions.

Pathology--control separation is detectable in the student features before intervention for liver and all ReX categories. Sampled-donor replacement and equal-norm
perturbations preserve the higher control replacement cost in the four
supported categories.
The ordering is strongest for full replacement on shared surrounding targets; own-neighbourhood targets and partial replacement doses show weaker or reversed effects (Appendix~\ref{app:e6}).

\begin{figure}[tb]
\centering
\includegraphics[width=\textwidth]{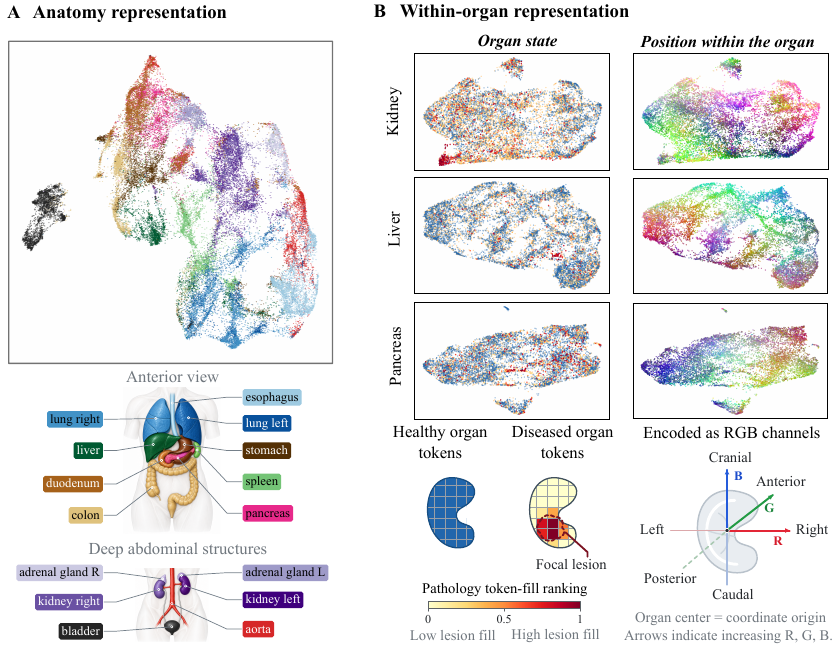}
\caption{ATLAS token embeddings show anatomical and positional organization.
\textbf{A:} Complete-view encoder tokens colored by anatomical structure.
\textbf{B:} Each organ-specific UMAP is recolored by organ state and position.
Disease-state coloring ranks pathology token fill within diseased organs, including zero-fill tokens; ``healthy'' denotes annotation-negative organs.
Position uses organ-normalized coordinates mapped to RGB, with the right kidney mirrored to the left-side frame.
Plotting and token-fill protocols: Appendix~\ref{app:umap-protocol}.}
\label{fig:umap}
\end{figure}

\subsection{Strong anatomical organization and uneven local pathology dependence}
\label{sec:results-phenotype}
\label{sec:results-route}

We visualized complete-view encoder tokens, as used downstream without the
predictor, with UMAP \citep{mcinnes2018umap}. ATLAS token embeddings group by anatomical structure and
show smooth within-organ position gradients (Figure~\ref{fig:umap}; thoracic
counterpart in Appendix~\ref{app:e2-rex}). Coloring organ populations by
disease status shows greater intermixing, with separation varying with
pathology token fill.
The UMAPs show clear anatomical and within-organ positional organization, while disease-state separation is less apparent; quantitative organ-state probes are reported in
Appendix~\ref{app:atlas-organ-state}.

To relate these representations to scan-level decisions, we fitted attentive
multiple-instance learning (MIL) probes that pool frozen tokens into
scan-level pathology scores \citep{ilse2018attention, kondepudi2026neurovfm}. Heads were fitted
and selected on separate SSL-included annotated scans and tested on
SSL-held-out ATLAS/ReX scans (Appendices~\ref{app:routes}
and~\ref{app:mil-localization}). \emph{Relative attention} compares the mean
attention per pathology token with that of surrounding parent-organ/lobe
tokens. For the \emph{deletion effect}, we deleted all same-category final-layer encoder pathology
tokens, renormalized attention without refitting, and expressed the median
logit drop as a percentage of the reference decision margin.

Scan-level Skill spans 0.198--0.607, while deleting final-layer pathology-overlapping tokens reduces the reference margin by at most 0.45\%  for liver, kidney, scarring, and nodules
(Table~\ref{tab:mil-main}). Kidney reaches 0.584 Skill with $0.65\times$
relative attention and a 0.18\% deletion effect. Atelectasis and ground-glass
show higher relative attention ($9.3\times$/$8.2\times$) and deletion effects
(10.47\%/2.39\%); pancreas is intermediate (1.69\%). Alternative scan-level cues are themselves predictive, with probes using only acquisition metadata or co-occurring labels reaching 0.68--1.24 times encoder Skill in five categories (Appendix~\ref{app:routes}).

We related the deletion effect to anatomical location typicality, defined by the concentration of same-category pathology at a location relative to the corresponding organ occupancy. After volume adjustment, pathology at more typical organ-relative locations has a slightly larger deletion effect:
$\rho_{\mathrm{pos}}=+0.085$ (analytic Fisher 95\% CI $[0.042,0.127]$; descriptive pooled estimate),
positive in six of seven categories and nominally significant in three (Table~\ref{tab:mil-main};
Appendix~\ref{app:positional}). The direction of this exploratory association is consistent with the predictability-bias hypothesis, linking greater anatomical regularity to greater local pathology dependence.

\begin{table}[t]
\centering
\caption{Scan-level performance and dependence on local pathology tokens.
\textcolor{milcoral}{Relative attention} is the geometric mean of per-scan attention-per-token ratios for pathology versus surrounding organ/lobe tokens (log scale; $1\times$ = equal).
\textcolor{milteal}{Deletion effect} is the median logit drop after removing all final-layer tokens overlapping same-category pathology annotations and renormalizing attention, expressed as a percentage of the positive--negative median-logit margin; 0/100\% denotes no/full reference-margin loss.
Whiskers show 95\% group-bootstrap CIs.
Exploratory $\rho_{\rm pos}$ is the volume-adjusted Spearman correlation between location typicality and deletion effect; $^{*}$ denotes nominal $p<0.05$.
Details in Appendices~\ref{app:routes}--\ref{app:mil-localization}.}
\label{tab:mil-main}

\small
\begin{adjustbox}{max width=\linewidth}
\begin{tabular}{@{}llcccc@{}}
\toprule
Dataset & Pathology & \shortstack{Scan-level\\Skill}
& \shortstack{\textcolor{milcoral}{Relative attention}\\{\scriptsize $1\times$ = surrounding tissue}\\[-2pt]\milAttentionHeaderAxis}
& \shortstack{\textcolor{milteal}{Deletion effect}\\{\scriptsize Margin loss (\%)}\\[-2pt]\milEffectHeaderAxis}
& \shortstack{Location\\$\rho_{\rm pos}$} \\
\midrule
\multirow{3}{*}{\rotatebox{90}{\textcolor{milgray}{ATLAS}}}
& Liver lesion & 0.376
& \milAttentionCell{2.1}{1.7}{2.5}{2.1}
& \milEffectCell{0.45}{0.28}{0.65}{0.45}
& $+0.178^{*}$ \\
& Kidney lesion & 0.584
& \milAttentionCell{0.65}{0.57}{0.75}{0.65}
& \milEffectCell{0.18}{0.12}{0.28}{0.18}
& $+0.129^{*}$ \\
& Pancreatic lesion & 0.386
& \milAttentionCell{0.99}{0.79}{1.24}{0.99}
& \milEffectCell{1.69}{1.14}{2.51}{1.69}
& $+0.006$ \\
\addlinespace[3pt]
\multirow{4}{*}{\rotatebox{90}{\textcolor{milgray}{ReX}}}
& Scarring/fibrosis & 0.198
& \milAttentionCell{0.84}{0.78}{0.91}{0.84}
& \milEffectCell{-0.04}{-0.22}{0.02}{-0.04}
& $+0.032$ \\
& Atelectasis/consol. & 0.558
& \milAttentionCell{9.3}{7.6}{11.7}{9.3}
& \milEffectCell{10.47}{6.02}{14.47}{10.47}
& $-0.036$ \\
& Ground-glass opacity & 0.607
& \milAttentionCell{8.2}{6.9}{9.9}{8.2}
& \milEffectCell{2.39}{1.38}{3.77}{2.39}
& $+0.100$ \\
& Nodules/masses & 0.386
& \milAttentionCell{1.2}{1.0}{1.4}{1.2}
& \milEffectCell{0.08}{0.05}{0.09}{0.08}
& $+0.124^{*}$ \\
\bottomrule
\end{tabular}
\end{adjustbox}
\end{table}

These results reveal a marked disconnect between scan-level pathology prediction and reliance on the local finding itself. In several categories, MIL attention to pathology tokens is limited and deleting all final-layer pathology-overlapping tokens leaves the decision margin almost unchanged; acquisition metadata and co-occurring labels can support much of the same scan-level prediction.

\textbf{Scope and extensions.} Our analyses characterize predictability bias in a large-scale 3D CT JEPA. The downstream interventions measure direct dependence on final-layer pathology tokens, while the context-side analysis concerns full replacement on shared surrounding targets. Analyses across representation depth and perturbation strength can further trace where local pathology information is retained, redistributed, or attenuated. Controlled changes to masking and predictive objectives can then identify which design choices preserve context-unresolved clinical variation while retaining the anatomical structure learned from context. More broadly, the same tension between contextual predictability and clinically important residual variation may arise in other self-supervised objectives based on prediction, reconstruction, or agreement across partial observations. Determining how such objectives treat this variation is an important direction beyond JEPA.

\section{Conclusion}
\label{sec:discussion}

We introduced predictability bias and showed how it shapes a large-scale 3D CT JEPA. \method{} learns a representation with strong anatomical and positional structure, while hidden-target prediction preserves substantially less of the signal carried by sporadic pathology. Pathology is shifted toward matched normal controls, contributes less than matched normal tissue to predicting surrounding anatomy, and in several downstream tasks the local finding can be removed with almost no change in the decision. Taken together, the representation is organized around what anatomy makes predictable, while clinically decisive deviations are less reliably retained.

This suggests a productive direction for medical predictive pretraining. Learning the conditional structure of anatomy is valuable for tasks that depend on localization, shape, and spatial organization. The next step is to complement this strength with objectives that explicitly preserve clinically important variation that surrounding anatomy cannot resolve. Medical representation learning must capture the regularity of anatomy without losing the sparse, subtle pathological events that determine clinical interpretation.

\section*{Acknowledgments}
\addcontentsline{toc}{section}{Acknowledgments}
The authors gratefully acknowledge the scientific support and HPC resources provided by the Erlangen National High Performance Computing Center (NHR@FAU) of the Friedrich-Alexander-Universit\"at Erlangen-N\"urnberg (FAU) under the NHR project ``AI-BAY-HEALTH -- Clinical Data Curation \& MRI-Pathology''. NHR funding is provided by federal and Bavarian state authorities. NHR@FAU hardware is partially funded by the German Research Foundation (DFG).
Additional support was received from the state
of Bavaria under the Bavarian Foundation Model Initiative.
We thank the National Cancer Institute and the NLST Research Team, as well as the creators and maintainers of CT-RATE, Merlin, AbdomenAtlas, ReXGroundingCT and RAD-ChestCT, for making their data available to the research community.

\section*{Declarations}
\addcontentsline{toc}{section}{Declarations}

\paragraph{Funding.}
This work was supported by the Wilhelm Sander-Stiftung, Munich, Germany
(grant 2025.013.1), and by the German Federal Ministry of
Research, Technology and Space (BMFTR) within the joint project DECIPHER-M
(01KD2420D, 01KD2420A). The responsibility for the content of this publication
lies with the authors. The funders had no role in study design, data collection,
data analysis and interpretation, or the writing of the manuscript.

\paragraph{Author contributions.}
All authors contributed to the conception and design of the study. FD developed the methodology, conducted the experiments and analyses, and wrote the original draft of the manuscript. BK, JNK, DR, and LCA critically reviewed and revised the manuscript. KKB secured funding, contributed to the supervision of the project, and critically revised the manuscript. CIB contributed to the study design, scientific direction, supervision, and drafting of the manuscript. All authors read and approved the final manuscript.

\paragraph{Ethics statement.}
This study uses de-identified CT data from existing research datasets (see Data availability)
under their respective licenses, access conditions and approved research
uses. No additional proprietary clinical data were used.

\paragraph{Data availability.}
All data used in this study are previously released, de-identified research datasets, available from their publishers under the access route and terms listed below.

\begin{table}[htbp]
\centering
\small
\caption*{Datasets used in this study, their role, and the access route and terms under which they are distributed by their publishers.}
\begin{tabular}{@{}>{\raggedright\arraybackslash}p{0.23\linewidth}>{\raggedright\arraybackslash}p{0.33\linewidth}>{\raggedright\arraybackslash}p{0.38\linewidth}@{}}
\toprule
Dataset & Use in this study & Access and terms \\
\midrule
NLST \citep{nlst2011,aberle2011reduced} & Pretraining & The Cancer Imaging Archive (TCIA), public collection \citep{nlst2013tciadata,clark2013tcia}; TCIA Data Usage Policy, CC BY 4.0 \\
\addlinespace
CT-RATE \citep{hamamci2024ctclip} & Pretraining; thoracic probe fitting, selection and source-domain evaluation & Hugging Face (gated); CC BY-NC-SA 4.0 \\
\addlinespace
Merlin \citep{blankemeier2026merlin} & Pretraining; abdominal benchmark (SSL-held-out official test split) & Stanford AIMI; Stanford University School of Medicine Merlin Abdominal CT Dataset Research Use Agreement (personal, non-commercial research use) \\
\addlinespace
AbdomenAtlas 3.0-mini (ATLAS) \citep{bassi2025radgpt} & Pretraining; abdominal regional analyses and MIL probes & Hugging Face; CC BY-NC-SA 4.0 \\
\addlinespace
ReXGroundingCT (ReX) \citep{baharoon2025rexgroundingct} & Thoracic regional analyses and MIL probes & Hugging Face (gated); CC BY-NC-SA 4.0 \\
\addlinespace
RAD-ChestCT \citep{draelos2021radchestct,draelos2020radchestctdata} & External thoracic evaluation only & Zenodo, restricted access approved by Duke University under CC BY-NC-ND 4.0; no further sharing of the data \\
\bottomrule
\end{tabular}
\end{table}

\paragraph{Use of artificial intelligence.}
An LLM (Claude, Anthropic) was used for drafting and editing text. The authors reviewed and verified all content and take full responsibility for it.

\bibliography{references}
\bibliographystyle{preprint}

\clearpage
\renewcommand{\topfraction}{0.95}
\renewcommand{\textfraction}{0.05}
\appendix
\appendixcontents
\clearpage
\input{appendix/supplementary.tex}

\end{document}

%% file: appendix/supplementary.tex
\section{Importance Invariance}
\label{app:importance}
Let $Q$ collect every variable consumed by pretraining: the image,
augmentations, mask, occupancy weights, teacher targets and training
randomness. Let $V$ denote an externally assigned downstream importance
annotation or utility. Equation~\ref{eq:objective} contains no explicit
$V$-dependent sampling or loss weight.
\begin{proposition}[Importance invariance]
\label{prop:importance-app}
Fix the distribution of $Q$, the model state and the loss. Any reassignment
of $V$ that preserves the law of $Q$ leaves the population risk and
stochastic gradient law unchanged.
\end{proposition}
Both loss and gradient are functions of $Q$, which proves the claim.
Clinical relevance can still influence learning through correlations in
$Q$, including sampling, masks, teacher features, occupancy and feature
scale. More independent data can improve estimation, coverage and training;
at a fixed data law and model state, it introduces no independent utility
term. This is an invariance statement about the criterion, not a claim
that additional data cannot improve pathology representations.

\section{Optimal Predictions and Gradient Identities}
\label{app:proofs}
We use the notation, weighted target law $\mathbb P$ and integrability
assumptions of Section~\ref{sec:theory}, with the teacher and current model
parameters fixed. The visible information $\mathcal U$ includes any
metadata or stochastic state consumed by the visible forward path.

\subsection{Optimal predictions and the deployed loss}
\label{app:locations}
Proposition~\ref{prop:location} applies the standard optimality of the mean
under squared error and of medians under absolute error
\citep{gneiting2011point} to the conditional target distribution given
$\mathcal A$. Both losses separate by feature coordinate, so their
vector-valued minimizers follow by applying the scalar results to each
coordinate. The squared-loss conditional-mean characterization is also
used in the JEPA analysis of \citet{dinh2026paradox}. For the deployed
feature-wise $L_1$ loss, the factor $1/d$ in Eq.~\ref{eq:token-loss} does
not change the minimizers; conditional medians need not be unique.

For $\mathcal A=\mathcal U$, these optima range over all predictions
measurable with respect to the raw visible information. The context
encoder and finite predictor class may prevent attaining this optimum,
and optimization can introduce further error. A deployed prediction
residual, therefore, does not identify irreducible uncertainty alone.

\subsection{Proof and loss-specific forms of Lemma~\ref{lem:gradient}}
\label{app:gradient}
At the fixed step, $\widehat Z$ and
$J_\omega=\partial\widehat Z/\partial\omega$ are
$\mathcal U$-measurable (where the Jacobian exists). Consequently,
\[
\mathbb E[G_\omega\mid\mathcal U]
=\mathbb E[J_\omega^\top S\mid\mathcal U]
=J_\omega^\top\mathbb E[S\mid\mathcal U].
\]
Subtracting this conditional mean gives the second identity in
Eq.~\ref{eq:conditional-gradient}.

\paragraph{Squared loss.}
For the unnormalized squared loss, $S=2(\widehat Z-Z)$, so
\begin{equation}
\mathbb E[G_\omega\mid\mathcal U]
=2J_\omega^\top(\widehat Z-\mathbb E[Z\mid\mathcal U]).
\label{eq:l2-gradient}
\end{equation}
The centered score is
$S-\mathbb E[S\mid\mathcal U]=-2(Z-\mathbb E[Z\mid\mathcal U])$,
giving the residual form stated after Lemma~\ref{lem:gradient}.

\paragraph{Deployed absolute loss.}
For the feature-wise $L_1$ loss, choosing zero at an exact tie gives
$S_k=d^{-1}\operatorname{sign}(\widehat Z_k-Z_k)$, with
$\operatorname{sign}(0)=0$. Hence
\begin{equation}
\mathbb E[S_k\mid\mathcal U]
=\frac{1}{d}\bigl\{
\mathbb P(Z_k<\widehat Z_k\mid\mathcal U)
-\mathbb P(Z_k>\widehat Z_k\mid\mathcal U)\bigr\}.
\label{eq:l1-score}
\end{equation}
The conditional score depends on sign probabilities, not merely the
conditional mean of $Z$. The centered score is
$S-\mathbb E[S\mid\mathcal U]$, rather than a centered target residual.
Lemma~\ref{lem:gradient} centers the chosen score and remains valid for
any measurable tie convention.

\paragraph{Scope of the gradient identities.}
These identities concern the loss signal for a target selected under
$\mathbb P$ at a fixed model state. Actual training sums weighted target
losses and updates shared parameters. A zero conditional mean for the
centered score does not eliminate its contribution to individual updates.

\section{Estimands and Protocols}
\label{app:estimands}
Sections~\ref{sec:results-benchmark}--\ref{sec:results-phenotype} distinguish
scan-level transfer, pathology exposure, hidden-target recovery, matched-region
displacement, visible-context contribution and complete-volume representations.
AP is stepwise average precision, including the benchmark columns labelled
AUPRC; it is not trapezoidal integration of the precision--recall curve.
Skill uses the prevalence of each task's eligible labelled rows.
The main text reports normalized Skill shortfall
$Q_{\mathrm{short}}=1-R_{\mathrm{rec}}$ and fractional contrast attenuation
$A_{\mathrm{ctr}}=1-C_{\mathrm{ret}}$.
Appendices~\ref{app:e4} and
\ref{app:e5} report the corresponding normalized Skill recovery
$R_{\mathrm{rec}}$ and retained contrast $C_{\mathrm{ret}}$, respectively.
For visible-context interventions, Table~\ref{tab:context-main} reports
control-minus-pathology differences, whereas Appendix~\ref{app:e6} retains the
signed pathology-minus-control contrast $\Delta$. For scan-level MIL, the
deletion effect and relative attention in Table~\ref{tab:mil-main} are the
token-ablation cost (own-token necessity) and pathology-token attention
density of Appendices~\ref{app:routes} and~\ref{app:mil-localization}.

\paragraph{Complete-volume readouts.}
The final student processes the complete prepared volume without the predictor.
The native CT-RATE/RAD-ChestCT and Merlin benchmarks, and annotated-cohort ATLAS/ReX
MIL use distinct readout protocols. Token probes classify anatomy or
\emph{parent-organ state}; a token in a diseased organ need not overlap pathology.
Alternative-route models measure the
sufficiency of acquisition metadata or co-occurring labels, without identifying
which information the encoder readout uses.

\paragraph{Matched-region feature displacement.}
Each prediction is located along a cross-fitted pathology-sensitive teacher
direction, comparing a pathology region with its translated, shape-matched
control. The origin-constrained slope of predicted on teacher contrast
is the retained contrast $C_{\mathrm{ret}}$, with one denoting unit average
calibration and zero
denoting no average linear alignment, and matched isotropic axes calibrate the
generic attenuation of arbitrary feature directions. Signed displacements
$D_+$ and $D_-$ quantify the pathology-side and control-side contributions to
contrast attenuation, with $C_{\mathrm{ret}}=1-D_+-D_-$
(Appendix~\ref{app:e5}).

\paragraph{Visible-context intervention.}
When a pathology region and its matched control tokens are both visible in context, each region's final
student activations are replaced with the mean activation of other visible
tokens from the same organ, and the fixed predictor loss is re-read on
same-organ targets 24--96\,mm away. Subtracting
an equal-count unrelated same-organ/lobe replacement calibrates the
loss increase against a reference perturbation. The
intervention occurs after student encoding, so information already propagated
through self-attention remains available and the estimand is the direct
predictive contribution through the region's final context tokens.

\paragraph{Summary values in the abstract and figures.}\label{app:summary-values}
Several headline values summarize dataset- or category-specific estimates
reported in this appendix. Each is an unweighted mean that gives every dataset
or pathology category equal weight, irrespective of its number of events, pairs
or scans; means are computed from unrounded estimates and are descriptive,
without separate uncertainty intervals.
(i)~\emph{Masking exposure} (Figure~1, left): the fully hidden-and-targeted,
partially visible and fully visible fractions (70.4\%, 25.0\% and 4.6\%) average
the per-category event fractions of the seven pathology categories in
Table~\ref{tab:e3-exposure}; the 22.7\,mm distance averages the seven per-category
median any-context distances and is therefore a mean of medians, not a pooled
median.
(ii)~\emph{Hidden-target recovery} (abstract): the pathology Skill values of
0.388 (teacher) and 0.271 (prediction) and the anatomy values of 0.921 and 0.873
average the pooled ATLAS and ReX tasks in Table~\ref{tab:recovery}.
(iii)~\emph{Matched-region displacement} (abstract; Figure~1, right): the
pathology-side (34.7\%) and control-side (13.6\%) displacements are the means of
$D_+$ and $D_-$ over the six categories in Table~\ref{tab:e5-primary}.
Figure~2 shows no averages: its highlighted values are ATLAS pooled anatomy and
ReX atelectasis/consolidation (Tables~\ref{tab:recovery},
\ref{tab:e4-families} and~\ref{tab:e5-primary}).

\section{Pretraining and Evaluation Details}
\label{app:details}

The production cohort contains 56{,}897 stored NLST volumes, 37{,}126 CT-RATE,
12{,}255 Merlin and 7{,}333 AbdomenAtlas3.0-mini volumes (113{,}611 total).
Dataset quotas are proportional to the square root of \emph{acquisition} counts,
with largest-remainder allocation, a two-acquisition patient cap per sampling
round and cycling through reconstructions; stored volumes are not unique patients.

\begin{table}[htbp]
\centering\small
\caption{Production configuration and retained checkpoint. Compute is logged active
training time, excluding queue/restart gaps, rather than complete scheduler billing.}
\label{tab:production-config}
\begin{tabular}{@{}lp{0.70\linewidth}@{}}
\toprule
Component & Specification \\
\midrule
Input/patch & Three HU windows; voxel spacing XYZ $(1,1,1.5)$\,mm; convolution kernel/stride ZYX $(16,24,24)$, i.e.\ 24\,mm isotropic tokens \\
Student/teacher & ViT-B: width 768, 12 layers, 12 heads; EMA teacher with saved output LayerNorm \\
Predictor/blocks & Width 384, 6 layers/heads; MLP ratio 4, GELU, LayerNorm, LayerScale \citep{touvron2021cait} 0.1, no attention dropout/drop-path \\
Mask/loss & Approx.\ 16\% context; remaining retained tokens targeted; feature-wise $L_1$, weight $w_i = 0.1+0.9\,o_i$ for body occupancy $o_i$, normalized within volume (then: $\alpha_i$) then averaged equally across volumes \\
Optimizer & AdamW \citep{loshchilov2019adamw}, betas $(0.9,0.95)$, $\epsilon=10^{-8}$, gradient clip 1; no weight decay on nonmatrix parameters or gradient accumulation \\
Batch & At most 48{,}000 tokens and 16 volumes/GPU; mean global batch 467.35 volumes on 32 GPUs; activation checkpointing \\
Schedules & Nominal LR $10^{-3}\to10^{-6}$, 10-epoch warmup then cosine decay over the 240-epoch horizon; WD $0.04\to0.4$ over 200 epochs then fixed; EMA $0.9995\to1$ \\
Endpoint & 240 epochs, 49{,}200 updates, 95{,}806 presentations/epoch; seed 20260729; BF16 training/extraction \\
Compute & $8\times4$ H200 GPUs; 50.685 logged active hours, approximately 1{,}622 GPU-hours \\
\bottomrule
\end{tabular}
\end{table}

The retained checkpoint is the epoch-240/step-49{,}200 final state at the end of the learning-rate schedule.
LR has square-root batch scaling (reference batch 467).
Complete-volume evaluations use the final student as the EMA momentum is annealed to 1, so the teacher is a lagged average of the student; masked-path analyses use the teacher (after output LayerNorm, not per-token unit-$L_2$ normalization) because it defines the training targets.

\paragraph{Input preparation and augmentation.}
Volumes are oriented RAS, stored ZYX, resampled with linear image/nearest-label
interpolation and clipped to $[-1024,3071]$ HU (invalid values below $-1024$
become $-1000$). Body extraction uses a $-500$ HU threshold, 3\,mm closing,
6\,mm table-disconnection erosion, slice hole filling and the largest 3D
component. QC requires at least 20 slices, 100\,mm z extent, matrix size 128,
z spacing at most 5\,mm, in-plane spacing at most 2\,mm and foreground fraction
0.05--0.98. The body bounding box receives a 10\,mm margin. Runtime crops
are at most 480\,mm per axis (ZYX $320\times480\times480$ voxels); smaller
volumes are not resized. Training uses random feasible starts and complete-volume
evaluation a central crop. End padding is $-1024$ HU with zero body occupancy.
Tokens with any body overlap are retained; occupancy includes padded voxels in
its denominator. Physical RoPE \citep{su2024roformer} uses the prepared body frame in 24\,mm units,
periods 48--960\,mm and position-scale jitter 0.75--1.33.

Augmentation uses in-plane rotation $\pm10^\circ$, isotropic scale 0.9--1.1 and
translation $\pm5$ voxels per axis, with trilinear images, nearest-neighbour masks
and no flips. With probability 0.25, Gaussian noise (SD 8 HU) precedes Poisson
noise with rate $\max(I+1024,0)/2$ and inverse mapping $2k-1024$. A separate
probability-0.25 event equally chooses blur ($0.75I+0.25\bar I$) or sharpening
($1.25I-0.25\bar I$), where $\bar I$ is a three-voxel average; HU values are
re-clipped. Windows $[-1350,150]$, $[-160,240]$ and $[-500,1300]$ map to $[0,1]$.

\paragraph{Mask construction.}
For $N$ retained body tokens, initial context size is
$K=\min(N-1,\max(1,\operatorname{round}(0.2N)))$; a further
$\min(K-1,\operatorname{round}(0.2K))$ context tokens are dropped uniformly.
Dropped context tokens become targets.
Thus 16\% visibility is approximate on finite grids. One family is selected
per volume: multi-block (0.45), cropped context (0.35), or multi-scale (0.20),
inspired by \citet{kondepudi2026neurovfm} and \citet{huang2026neurojepa}.
Multi-block targets are overlapping blocks centered on uniformly sampled
retained tokens, with independent ZYX side fractions $U[0.45,0.9]$ of the
occupied bounding-box extent. Multi-scale blocks uniformly choose an XY bucket
$[0,0.2]$, $[0.2,0.5]$ or $[0.5,0.7]$, share its sampled fraction across XY
and use depth fraction $U[0,1]$. Side lengths are rounded/clamped and offsets
use $\lfloor\mathrm{side}/2\rfloor$. Both target-union samplers stop at $N-K$
tokens or 256 attempts and randomly trim/fill to budget. Cropped context selects
the $K$ nearest tokens to a retained anchor by Chebyshev grid distance, adding
$U[0,0.001]$ tie noise; it is not a second image crop.
Appendix~\ref{app:controls} distinguishes the frozen replay draws.

\paragraph{Cohorts and annotation units.}
ATLAS has 1{,}863 SSL-excluded cases out of 9{,}196; ReX has 950 excluded scans
from 870 patients (909 partial and 41 exhaustive annotations). Intersections
with saved SSL sample/group identifiers are zero; this does not test
image-content duplicates or unknown cross-release identities.
Exclusion was applied at the patient level across all CT-RATE reconstructions, so no ReX evaluation patient contributes any reconstruction to pretraining.
ATLAS groups are
released cases/studies because patient IDs are unavailable. Annotated MIL fits
on separate SSL-included scans and tests on these excluded cohorts
(Appendix~\ref{app:routes}); permitted image reuse for SSL and probe fitting
also occurs in the native benchmarks, without SSL label use.
ATLAS entities are connected components; ReX entities are released annotations.
Scan-family outcomes use their union. ReX regions are assigned to the lung lobe
with greatest pathology overlap, with nearest-boundary fallback; scan-family
localization unions the assigned lobes, whereas organ state uses the whole lung.
Kidney organ-state and intervention support uses both kidneys; focal matching
and positional frames use individual kidneys. Annotation-negative controls
exclude all released pathologies and are not verified clinically normal.
Counts are task-specific: anatomy uses 1{,}861 ATLAS cases; organ state, matched
regions, recovery and interventions apply separate support filters.

\subsection{Frozen scan-level benchmark protocol}
\label{app:benchmark-protocol}
Table~\ref{tab:benchmarks} and
Appendix~\ref{app:benchmark-results} compare frozen image representations. No
encoder parameter was updated by a downstream classifier, candidate selection
used validation macro AUPRC only, and the final evaluation cohorts were
accessed only after selection. Each encoder retained its native image
preprocessing. Macro AUROC and macro AUPRC are unweighted means of the
per-pathology metrics. VoxelFM retains native orientation, clips
$[-1000,1900]$ HU, crops background with kernel 21, resamples to isotropic
$\max(\min\mathrm{native\ spacing},0.75\,\mathrm{mm})$, uses patch size 14 and
at most 25{,}000 patches, and normalizes by mean $-573.8$ / SD 461.3 HU.
Its RAD inputs use the native NPZ route at 0.8\,mm. Merlin uses RAS,
$1.5\times1.5\times3$\,mm spacing, $[-1000,1000]$ HU and central
$224\times224\times160$ input to its frozen image encoder.
\paragraph{CT-RATE.}
The 18-pathology task used 44{,}163 volumes from 19{,}000 patients for probe
fitting, 2{,}217 volumes from a disjoint 1{,}000-patient subset for model
selection, and 3{,}002 quality-controlled volumes from 1{,}304 patients in the
official CT-RATE validation split for final evaluation. The latter is the test
set in the CT-CLIP/COLIPRI convention. Publisher train and validation roles are
retained; the publisher training role was subdivided for internal validation by
patient as described above. For \method{} and VoxelFM
\citep{morenoaguado2026voxelfm}, five patch-token aggregation families (mean,
max, learned-query attention, mean-query attention, and four-query learned
attention) were crossed with learning rates $\{0.003,0.01,0.03,0.1\}$.
The 20 readouts were trained jointly but with independent parameters for
12{,}500 steps using batch size 16, unweighted binary cross-entropy, SGD with
momentum 0.95 and no weight decay, cosine decay to zero, and seed 42. The selected
readouts were mean-query attention at learning rate 0.03 for \method{} and at
learning rate 0.1 for VoxelFM. Their 95\% intervals are percentile intervals
from 100 volume-level bootstrap replicates.
\paragraph{CT-RATE to RAD-ChestCT external transfer.}
RAD-ChestCT is an independent restricted-access cohort of 3{,}630 chest CT
volumes from 3{,}072 patients with abnormality labels derived from radiology
reports \citep{draelos2021radchestct}. For each
encoder, the complete 18-output readout selected on CT-RATE, token aggregator
and affine classifier together, was applied unchanged to every RAD-ChestCT
scan. RAD-ChestCT contributed no fitting, hyperparameter selection, probability
calibration, or thresholding. Targets were projected to the corresponding
CT-RATE abnormalities by logical OR over the publisher's anatomical-location
labels (mapping, see \cite{buess2026alo}). Evaluation followed the reported 16-target external view: it excluded
the deliberately all-zero mosaic-attenuation target and one of the two
calcification outputs that share identical RAD-ChestCT ground truth. The same
16 outputs and scan order were used for \method{} and VoxelFM.
For the locally evaluated RAD-ChestCT rows, the main-table intervals use 100
scan-level bootstrap replicates, matching the published comparison convention.
A sensitivity analysis with 1{,}000 patient-level bootstrap replicates gave the
same paired conclusions: VoxelFM minus \method{} was $0.0013$ (95\% CI
$[-0.0027,0.0058]$) for macro AUROC and $-0.0075$
($[-0.0151,-0.0006]$) for macro AUPRC. Patient-level resampling is the stricter
uncertainty analysis because scans are nested within patients.
The COLIPRI (best performing CRM model variant) rows are transcribed from the published classification-probe
results \citep{wald2026colipri}; the encoder was not rerun locally. The authors
used the same five pooling families and four-learning-rate search principle,
selected by validation AUPRC, and reported 95\% intervals from 100 bootstrap
replicates. Their internal-validation identities, split seed, and bootstrap
resampling unit were not published, and their CT-RATE test set contained
3{,}039 rather than the 3{,}002 locally retained volumes. The daggers in
Tables~\ref{tab:benchmarks} and \ref{tab:benchmarks-ct-rate-app} mark
these non-identical elements. COLIPRI-CRM is therefore a close literature
comparison and no paired evaluation; paired differences are reported only for
the locally evaluated \method{} and VoxelFM predictions.
\paragraph{Merlin Abdominal CT.}
The 30-pathology task used the released Merlin assignments after intersecting
studies with labels, a valid \method{} representation, and an available source
volume: 15{,}174 studies for fitting, 5{,}016 for validation, and 5{,}082 for
testing. A released target of $-1$ was treated as unknown and masked from both
loss and metric computation. Each retained study kept its publisher-assigned
role. We rely on the publisher's patient-disjoint split construction: patient
identifiers are unavailable in the released task data, so that construction
cannot be re-audited locally. Uncertainty was therefore estimated with 1{,}000
study-level bootstrap replicates.
\method{} and VoxelFM used the same 20-candidate token-readout search and
optimization protocol as on CT-RATE, with 30 output logits and masked binary
cross-entropy. Four-query learned attention at learning rate 0.1 was selected
for \method{}; mean-query attention at learning rate 0.1 was selected for
VoxelFM. For Merlin \citep{blankemeier2026merlin}, the released frozen image
encoder's official globally pooled 2{,}048-dimensional embedding was followed
by a single affine 30-output classifier. Eight candidates crossed batch sizes
$\{8,16,32,64\}$ with global or per-label masked-loss reduction and were fit
for 10 epochs with AdamW (learning rate $10^{-5}$, weight decay 0.01, seed 42); batch
size 8 with global reduction was selected by validation macro AUPRC. No
post-hoc probability calibration was applied. All three encoders used the same
study membership, target matrices, and split roles. This controls the data and
selection protocol; it is not a strict identical-head comparison because
Merlin exposes an official global embedding whereas \method{} and VoxelFM
expose variable-length patch-token sequences. VoxelFM's published Q-Former
numbers were not used in the table.
These benchmark intervals quantify discrimination on the evaluated cohorts;
they do not establish probability calibration, a clinical operating point, or
clinical utility. Each locally evaluated encoder is a single retained model
instance, not an independent training replicate; we therefore retain bootstrap
confidence intervals and do not report run-level standard deviations.
Token-level estimates weight eligible token rows; grouping controls fitting or
uncertainty, not equal-patient metric weighting.

\subsection{Frozen anatomy, organ-state protocols}
\paragraph{Anatomy-token probe.}
The evaluation unit is one final-layer foreground token and the label is the
annotated structure containing it. The 15 structures available in both
datasets were evaluated with five-fold group-cross-fitted, fold-standardized,
class-balanced $L_2$ multinomial logistic regression ($C=1$). To limit
training-scan imbalance, at most 64 class-stratified tokens per training scan
were fitted, whereas every eligible token in the held-out fold was scored.
CT-RATE contributed 6{,}371 scans from 5{,}217 patient groups and 6{,}023{,}635
held-out token predictions; AbdomenAtlas contributed 1{,}861 released
case/study groups and 1{,}491{,}541 predictions. Macro metrics average the 15
one-versus-rest estimates. Uncertainty uses 500 scan bootstrap replicates (seed 20260803), even though
CT-RATE fitting groups by patient. Macro AUROC/Skill 95\% intervals are
$[0.9844,0.9852]$/$[0.7527,0.7610]$ for ATLAS and
$[0.9757,0.9839]$/$[0.7159,0.7334]$ for CT-RATE; these condition on fitted
probes and do not account for patient clustering of repeated CT-RATE scans.
\begin{table}[h]
\centering
\caption{Structure-specific held-out anatomy-token results. Skill is
prevalence-normalized AP (Section~\ref{sec:methods}); entries are point estimates,
with scan-bootstrap uncertainty described in the protocol.}
\label{tab:e2-anatomy-detailed}
\scriptsize
\begin{tabular}{@{}lrrrr@{}}
\toprule
Structure & \multicolumn{2}{c}{CT-RATE} & \multicolumn{2}{c}{AbdomenAtlas} \\
 & AUROC & Skill & AUROC & Skill \\
\midrule
Liver & 0.993 & 0.964 & 0.978 & 0.915 \\
Spleen & 0.995 & 0.922 & 0.987 & 0.840 \\
Left kidney & 0.998 & 0.935 & 0.994 & 0.915 \\
Right kidney & 0.994 & 0.755 & 0.992 & 0.895 \\
Pancreas & 0.996 & 0.785 & 0.989 & 0.687 \\
Stomach & 0.987 & 0.800 & 0.960 & 0.672 \\
Aorta & 0.991 & 0.846 & 0.993 & 0.815 \\
Left lung & 0.962 & 0.880 & 0.964 & 0.802 \\
Right lung & 0.972 & 0.915 & 0.981 & 0.905 \\
Left adrenal gland & 0.997 & 0.469 & 0.996 & 0.528 \\
Right adrenal gland & 0.992 & 0.209 & 0.995 & 0.527 \\
Colon & 0.993 & 0.899 & 0.968 & 0.865 \\
Duodenum & 0.995 & 0.558 & 0.983 & 0.548 \\
Esophagus & 0.996 & 0.773 & 0.996 & 0.512 \\
Bladder & 0.818 & 0.097 & 0.996 & 0.923 \\
\midrule
Macro & 0.979 & 0.720 & 0.985 & 0.757 \\
\bottomrule
\end{tabular}
\end{table}
\paragraph{ATLAS organ-state token evaluation.}
\label{app:atlas-organ-state}
All retained tokens with any parent-organ overlap inherit organ disease status,
including zero-pathology-fill tokens. Initial pools have 424/288/224
kidney/liver/pancreas positive cases and 400 sampled annotation-negative cases
per family (seed 20260907). Requiring four organ tokens and capping at 120 per
scan retains 408/396, 279/395 and 214/386 positive/negative cases; bilateral
kidney tokens inherit the union's status. Five unshuffled GroupKFold splits
use cases, training-fold standardization and balanced $L_2$ logistic regression
($C=1$, maximum 3{,}000 iterations), without tuning. Metrics concatenate held-out
token scores; 200 case bootstrap replicates keep scores fixed.
Position uses global XYZ. Acquisition uses original slice count, z extent,
scanner, manufacturer, kernel, phase, original XYZ spacing and canonical X/Z
shape. Categories use sorted ordinal codes with an explicit missing level;
missing numeric values become zero. All probe arms share rows and folds.
Table~\ref{tab:atlas-organ-state} gives the principal estimates.

\begin{table}[htbp]
\centering\scriptsize
\caption{ATLAS organ-state token probes. Brackets are 95\% case-bootstrap CIs.
$N$ is evaluated tokens, $\pi$ positive-token prevalence.}
\label{tab:atlas-organ-state}
\setlength{\tabcolsep}{4pt}
\begin{tabular*}{\textwidth}{@{\extracolsep{\fill}}lrrccc@{}}
\toprule
Organ & $N$ & $\pi$ & Encoder AUROC & Encoder Skill & Acquisition Skill \\
\midrule
Kidney & 56{,}397 & 0.557 & \metric{0.815}{[0.789, 0.839]} & \metric{0.607}{[0.538, 0.662]} & \metric{0.661}{[0.599, 0.709]} \\
Liver & 79{,}845 & 0.414 & \metric{0.697}{[0.664, 0.730]} & \metric{0.275}{[0.217, 0.343]} & \metric{0.477}{[0.400, 0.554]} \\
Pancreas & 9{,}699 & 0.352 & \metric{0.744}{[0.706, 0.773]} & \metric{0.413}{[0.337, 0.477]} & \metric{0.168}{[0.089, 0.245]} \\
\bottomrule
\end{tabular*}
\end{table}
Acquisition matches/exceeds encoder Skill for kidney and liver, but not pancreas.
Neither organ-state decoding nor a metadata-only probe localizes the signal to
pathology. For a qualitative thoracic comparison, see
Appendix~\ref{app:e2-rex}.

\paragraph{UMAP protocol.}
\label{app:umap-protocol}
Visualization applies centered PCA (up to 50 components), then UMAP \citep{mcinnes2018umap} with cosine
distance, 30 neighbours and minimum distance 0.1; no separate feature
standardization or unit-$L_2$ normalization. Seeds are 20260908 (ATLAS) and
20260909 (ReX). Anatomy samples at most 3{,}000 tokens per structure; organ panels
use a uniform cap of 12{,}000 (ATLAS pancreas: 9{,}699). Each organ-state embedding
jointly fits all states and is recolored, not refitted, for position. For physical
token centre $x$, organ centroid $\mu$ and per-axis voxel-position SD $\sigma$,
$u=(x-\mu)/\max(\sigma,10^{-6})$; right-kidney x is mirrored and RGB is
$\operatorname{clip}(u_{XYZ}/5+0.5,0,1)$. Tokens without a valid organ frame are
omitted only from position coloring.

\section{Additional Frozen-Encoder Benchmark Results}
\label{app:benchmark-results}
The source-domain CT-RATE benchmark is reported separately from the main
external-transfer comparison (Table \ref{tab:benchmarks-ct-rate-app}). It evaluates the same readouts on the
quality-controlled official CT-RATE validation cohort used as the test set in
the CT-CLIP/COLIPRI convention. These results measure held-out performance within the model training domain; external-cohort transfer is evaluated separately on RAD-ChestCT.
\begin{table}[h!]
\centering
\setlength{\tabcolsep}{4pt}
\caption{Source-domain frozen scan-level classification on CT-RATE. Values are
macro AUROC and macro AUPRC with 95\% bootstrap confidence intervals. Bold
marks the best point estimate and underline the runner-up. VL denotes
vision--language pretraining; $^\dagger$ denotes a published result obtained
with a matched but non-identical readout and evaluation workflow.}
\label{tab:benchmarks-ct-rate-app}
\small
\begin{tabular}{lllrr}
\toprule
Dataset & Encoder & Pretraining & AUROC $\uparrow$ & AUPRC $\uparrow$ \\
\midrule
CT-RATE & \method{} (ours) & Image SSL & \metric{0.8533}{[0.8489, 0.8577]} & \metric{0.6078}{[0.5952, 0.6212]} \\
& COLIPRI-CRM$^{\dagger}$ \citep{wald2026colipri} & VL & \metric{\second{0.8612}}{[0.8552, 0.8685]} & \metric{\second{0.6128}}{[0.5959, 0.6310]} \\
& VoxelFM \citep{morenoaguado2026voxelfm} & Image SSL & \metric{\best{0.8653}}{[0.8612, 0.8697]} & \metric{\best{0.6261}}{[0.6148, 0.6391]} \\
\bottomrule
\end{tabular}
\end{table}

\section{Additional analyses and sensitivity checks}
\label{app:extended}
This appendix provides additional results, protocols and robustness analyses
for Sections~\ref{sec:results-exposure}--\ref{sec:results-phenotype}.
\input{appendix/thoracic_organ_state.tex}

\subsection{Scan-level MIL and alternative routes}
\label{app:routes}
\paragraph{Annotated-cohort scan-level readout.}
The reported categories are ATLAS kidney/liver/pancreatic lesions and ReX
scarring/fibrosis, atelectasis/consolidation, ground-glass opacity and nodules.
For frozen token $h_i$ and class $c$, two affine maps produce evidence $v_{ic}$
and attention logits $a_{ic}$; the scan logit is
$s_c=\sum_i\operatorname{softmax}_i(a_{ic})v_{ic}$. All retained 768-dimensional
tokens are used, without a gated-attention MLP \citep{ilse2018attention}. Heads have four ATLAS outputs
(including colon) and 14 ReX outputs. They fit on SSL-included annotations:
ATLAS 5{,}883 fitting/1{,}450 validation cases; ReX 1{,}795/447 scans from
1{,}628/407 patients. Internal validation folds are 0 and 2, respectively;
ReX selects for class support and prevalence balance within the fitting pool.
Test sets are disjoint: 1{,}863 ATLAS cases and 950 ReX scans/870 patients.

AdamW uses weight decay $10^{-4}$, batch 16, at most 50 epochs, patience 8
and seed 20260803. BCE positive weights are the training negative/positive
count ratios. Checkpoints minimize validation loss (minimum improvement
$10^{-4}$); validation macro AP selects LR $10^{-3}$ over $3\times10^{-3}$.
Test positives are 424/288/224 for ATLAS kidney/liver/pancreas and
237/306/363/421 for ReX; remaining test scans are annotation-negative.
Classification includes scans without eligible localized support. This is distinct from the native CT-RATE/Merlin benchmark readouts and masked-path audit cohorts.

\paragraph{Alternative routes and direct readout ablation.}
Acquisition uses the fields in Appendix~\ref{app:atlas-organ-state}, ordinal
categories and numeric NaNs in histogram gradient-boosted trees (learning rate
0.08, 150 iterations, seed 0). Co-occurrence uses other released labels plus
missingness indicators in logistic regression ($C=1$, 2{,}000 iterations, no
scaling/class weights). Both use five grouped cross-fitting folds on held-out
MIL rows (ATLAS cases, ReX patients), without tuning. They share evaluation
rows with the fixed MIL head, but not an exact fitting history, so they are used to measure predictive information available through alternative routes Route/encoder Skill
can exceed one or be negative. ReX co-occurrence measures co-mention.

Ablation deletes the union $L$ of all same-family pathology tokens and
renormalizes attention: $s_{-L}=\sum_{i\notin L}\alpha_i v_i/
\sum_{i\notin L}\alpha_i$. Cost is $s-s_{-L}$ in logits; the image is not
re-encoded and information spread to other tokens remains. The percentage is
100 times median cost over evaluable positives divided by the difference of
median original logits over all positive and all negative test scans.
Table~\ref{tab:mil-main} reports this percentage as the \emph{deletion effect}
and the pathology-token attention density
(Appendix~\ref{app:mil-localization}) as \emph{relative attention}; token
ablation and own-token necessity in this appendix denote the same deletion
effect. All
denominators are positive. Attention excludes empty supports/invalid ratios;
evaluable positives number 407/276/216 in ATLAS and 237/306/363/419 in ReX.
Ablation confidence intervals use 1{,}000 percentile group-bootstrap
replicates (seed 0). We resample study groups
(\texttt{study\_group\_id}) for ATLAS and patients
(\texttt{patient\_id}) for ReX with replacement, keeping observations
within each sampled group together. Each replicate recomputes the
median deletion effect over resampled evaluable positives and expresses
it as a percentage of the class's decision margin. This denominator
is the difference between the median original logits among positive
and negative scans, computed from the full-sample cached MIL
predictions and held fixed across replicates. The 2.5th and 97.5th
percentiles of the bootstrap percentages define the 95\% interval.
These intervals condition on the fitted MIL head and the estimated
full-sample margin; they do not propagate uncertainty in that margin.

\input{appendix/scan_level_localization_table.tex}

\paragraph{Results on the annotated cohorts.}
Acquisition alone reaches 0.68--1.08 times encoder Skill across the three
ATLAS classes. For kidney, encoder Skill is 0.584 and acquisition Skill is
0.632; attention density is $0.65\times$ and own-token necessity 0.18\%
$[0.12,0.28]$. ReX acquisition ratios are small ($-0.02$ to 0.09), but
co-mention reaches 1.08 times encoder Skill for scarring and 1.24 for nodules.
Their own-token necessities are $-0.04\%$ $[-0.22,0.02]$ and 0.08\%
$[0.05,0.09]$. Atelectasis/consolidation and ground-glass differ: attention
densities are $9.3\times$ and $8.2\times$, with necessities 10.47\% and
2.39\%. Pancreatic necessity is intermediate at 1.69\%, compared with the four classes at 0.45\% or below. A negative finite-sample
route Skill is retained.

\subsection{Positional recurrence}
\label{app:positional}
Centroids use the organ frame in Appendix~\ref{app:umap-protocol}, with
individual kidneys (right mirrored) and the whole lung for ReX. Crop-truncated
organs are excluded. Maps use a $32^3$ grid over $[-3,3]$ SD units, floor/clipped
indices and Gaussian smoothing of one cell (two for display). Organ null $q$
is normalized per-scan occupancy with support threshold 0.05; pathology
density $p$ weights each centroid inversely by the number in its group.
Five shuffled group folds (seed 20260902) estimate held-out typicality
$\log[(0.99p+0.01q)/q]$, without matched-null subtraction. The separate
cross-family recurrence index averages 24 subsamples of 500 pathologies and
subtracts a matched null drawn from $q$; its percentile spread describes
subsamples, not a patient bootstrap. Categories below 500 lack that index.

\paragraph{Association with local readout dependence.}
Each scan--family contributes one row: largest-pathology typicality and volume
are paired with ablation of \emph{all} same-family pathology tokens, not that
individual lesion alone. Partial rank correlation residualizes ranked typicality
and outcome against ranked volume. Table~\ref{tab:recurrence-inference} uses
99{,}999 null-imposed wild-cluster bootstrap-$t$ draws \citep{cameron2008bootstrap}: independent Rademacher
signs by case/patient multiply residuals of ranked outcome regressed on intercept
and ranked volume, followed by refitting with ranked typicality. Ranks are fixed;
CR1 variance uses $G/(G-1)(n-1)/(n-3)$ and two-sided $p$ values use the plus-one
correction. There is no multiplicity correction or nested map uncertainty.

Pooled Fisher-$z$ estimates use weights $n_f-3$ and analytic intervals
$z\pm1.96/\sqrt{\sum_f(n_f-3)}$, transformed back. Necessity gives
$\rho_{\mathrm{pos}}=0.085$ $[0.042,0.127]$ and attention $0.025$
$[-0.018,0.067]$. The analytic pooling formula does not model dependence between categories
that share scans or patients, and the intervals do not propagate
uncertainty in the estimated location maps.

\input{appendix/recurrence_inference_table.tex}

\subsection{Attentive-MIL localization protocol}
\label{app:mil-localization}
The unchanged MIL head supplies class-specific attention. Let $L$ overlap any
same-family pathology and $O$ be majority-vote parent-organ tokens (bilateral
kidneys in ATLAS; union of assigned pathology-bearing lobes in ReX).
The domain is $L\cup O$ and surrounding tissue $C=O\setminus L$. Per-scan
pathology-token density is $(\sum_L\alpha/|L|)/(\sum_C\alpha/|C|)$; the reported
$2^{\operatorname{mean}(\log_2\mathrm{density})}$ is the geometric mean of
scan ratios. Its CI exponentiates the 1{,}000 case/patient-bootstrap interval
for mean log ratio. Point estimates weight scans equally, not patients.

Volume quartiles split positives by union volume of all same-family annotations,
reusing the complete negative pool. Global attention excess is pathology attention mass minus its
fraction of all scan tokens. Additional metrics use 1{,}000 case/patient
resamples of fixed predictions.

\begin{table}[h]
\centering
\caption{Extended localization metrics. Organ-region attention is total attention
on pathology plus its parent organ/lobes; density is the mean log$_2$
pathology-to-surrounding-tissue attention ratio. Score $\sim$ vol.\ and excess
$\sim$ vol.\ correlate positive-scan scores and global pathology attention excess
with union pathology volume. Skill quartiles share negatives; prevalence differs.}

\label{tab:mil-localization}
\scriptsize
\setlength{\tabcolsep}{4pt}
\begin{tabular}{@{}lccccc@{}}
\toprule
Class & Organ-region & log$_2$ & Skill by volume & Score $\sim$ vol. & Excess $\sim$ vol. \\
 & attention & density & quartile Q1--Q4 & $\rho$ & $\rho$ \\
\midrule
Liver lesion & 0.245 & $+1.046$ & 0.091 / 0.070 / 0.184 / 0.432 & $+0.302$ & $+0.575$ \\
Kidney lesion & 0.077 & $-0.630$ & 0.124 / 0.353 / 0.506 / 0.466 & $+0.313$ & $-0.075$ \\
Pancreatic lesion & 0.170 & $-0.010$ & 0.081 / 0.271 / 0.263 / 0.237 & $+0.190$ & $+0.016$ \\
Scarring/fibrosis & 0.245 & $-0.250$ & 0.044 / 0.074 / 0.116 / 0.122 & $+0.050$ & $+0.167$ \\
Atelectasis/consol. & 0.634 & $+3.223$ & 0.136 / 0.256 / 0.428 / 0.596 & $+0.490$ & $+0.381$ \\
Ground-glass opacity & 0.667 & $+3.043$ & 0.231 / 0.415 / 0.454 / 0.527 & $+0.277$ & $+0.297$ \\
Nodules/masses & 0.325 & $+0.243$ & 0.281 / 0.228 / 0.194 / 0.127 & $-0.248$ & $+0.110$ \\
\bottomrule
\end{tabular}
\end{table}

\subsection{Production-mask exposure}
\label{app:e3-exposure}
\begin{figure}[htbp]
\centering
\includegraphics[width=\textwidth]{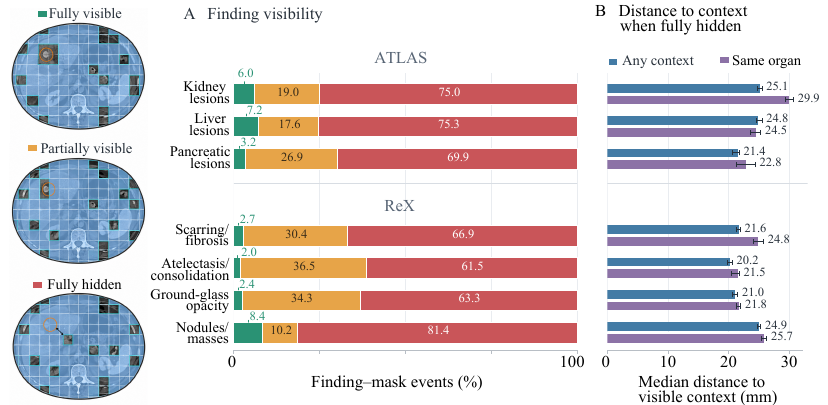}
\caption{Production masks fully conceal most annotated pathologies in SSL-held-out
cohorts. \textbf{A:} Fractions of crop-retained pathology--mask events that are
fully visible, partially visible, or fully hidden and targeted; the three
states sum to 100\% before rounding. \textbf{B:} Median
annotation-boundary-to-visible-token-box distance over fully hidden-targeted
events. Whiskers are pointwise 95\% confidence intervals from 2{,}000
case/study-group (ATLAS) or patient (ReX) bootstrap resamples, conditional on
the frozen spatial draws. ReX uses the instance's assigned dominant lung lobe
for the same-organ distance. Full numbers, IQRs, crop retention, size
associations and positive-target rates are in
Appendix~\ref{app:e3-exposure}.}
\label{fig:e3-exposure}
\end{figure}
Table~\ref{tab:e3-exposure} gives the complete numerical results underlying
Figure~\ref{fig:e3-exposure}. The bracketed distance values here are IQRs of
hidden-event distances, whereas the figure's whiskers are confidence intervals
for their medians.
\begin{table*}[t]
\centering
\caption{Production-mask exposure. State probabilities condition on crop
retention and sum to 100\%. Distances are medians over fully hidden-targeted
events with IQRs below, measured from the annotation boundary to the token box
in millimetres
after deployed preprocessing. Same-organ distances additionally condition on
visible same-organ context being present; for ReX this is the assigned dominant
lung lobe, and kidney support is the union of both kidneys. $\rho_H$ and $\rho_d$ are Spearman
correlations of instance volume with concealment probability and with median
context distance; every group-bootstrap interval excludes zero. These
associations use all available annotation protocols, including the small ReX
exhaustive stratum, as in the original analysis. ``Pos./1k'' is
the family's cohort-level positive-target-token contribution and is an
annotated-cohort rate, not a pretraining-mixture prevalence. ReX rows use the
909-scan partial representative-annotation protocol with CT-RATE lobe anatomy.
Instances retained after deployed preprocessing and cropping, out of all
eligible: kidney 1{,}255/1{,}309, liver 971/1{,}025, pancreas 281/282,
scarring/fibrosis 525/527,
atelectasis/consolidation 816/816, ground-glass 1{,}030/1{,}030, nodules/masses
947/948; crop exclusion was 4.13\%, 5.27\%, and 0.35\% for kidney, liver,
and pancreas, respectively, and 0--0.38\% across ReX.}
\label{tab:e3-exposure}
\scriptsize
\setlength{\tabcolsep}{5pt}
\begin{tabular}{@{}llccccc@{}}
\toprule
& & Visible / partial / & \multicolumn{2}{c}{Median distance to visible context (mm)} & & \\
\cmidrule(lr){4-5}
Dataset & Family & hidden-targeted (\%) & Any context & Same organ/lobe & $\rho_H$ / $\rho_d$ & Pos./1k \\
\midrule
ATLAS & Kidney lesion & 6.01 / 19.03 / 74.95 & \shortstack{25.11\\{\scriptsize [11.94, 46.50]}} & \shortstack{29.89\\{\scriptsize [12.75, 78.54]}} & $-0.815$ / $-0.589$ & 2.12 \\
ATLAS & Liver lesion & 7.17 / 17.58 / 75.25 & \shortstack{24.83\\{\scriptsize [11.50, 49.50]}} & \shortstack{24.50\\{\scriptsize [10.85, 47.25]}} & $-0.771$ / $-0.352$ & 1.19 \\
ATLAS & Pancreatic lesion & 3.22 / 26.90 / 69.88 & \shortstack{21.45\\{\scriptsize [9.75, 39.32]}} & \shortstack{22.77\\{\scriptsize [8.37, 46.21]}} & $-0.805$ / $-0.572$ & 0.41 \\
ReX & Scarring/fibrosis & 2.74 / 30.38 / 66.88 & \shortstack{21.62\\{\scriptsize [9.75, 43.50]}} & \shortstack{24.76\\{\scriptsize [10.50, 50.09]}} & $-0.808$ / $-0.435$ & 2.50 \\
ReX & Atelectasis/consolidation & 2.02 / 36.46 / 61.52 & \shortstack{20.25\\{\scriptsize [8.71, 39.45]}} & \shortstack{21.46\\{\scriptsize [8.75, 41.38]}} & $-0.895$ / $-0.602$ & 5.89 \\
ReX & Ground-glass opacity & 2.40 / 34.31 / 63.29 & \shortstack{21.02\\{\scriptsize [9.62, 40.63]}} & \shortstack{21.75\\{\scriptsize [9.50, 41.84]}} & $-0.909$ / $-0.614$ & 7.04 \\
ReX & Nodules/masses & 8.44 / 10.19 / 81.37 & \shortstack{24.91\\{\scriptsize [12.50, 45.57]}} & \shortstack{25.74\\{\scriptsize [11.72, 47.08]}} & $-0.513$ / $-0.333$ & 1.32 \\
\bottomrule
\end{tabular}
\end{table*}
\paragraph{Exposure units and distances.}
For a crop-retained instance, $M>0$ is its represented pathology-voxel count
and $V$ its count in visible tokens: fully visible means $V=M$, partial
$0<V<M$, and fully hidden-targeted $V=0$ with nonzero targeted pathology mass.
Crop exclusions are separate. State fractions equally weight instance--mask
events over 1{,}600 masks, so multi-instance scans contribute more events.
Pos./1k is 1{,}000 times target tokens intersecting the family annotation union
divided by \emph{all} retained body target tokens across the protocol cohort
and masks, including family-negative scans; each token counts once per family/mask.

Boundary samples are centres of voxels removed by $3^3$ binary erosion,
at $(\mathrm{index}+0.5)\times\mathrm{spacing}$. For each 24\,mm token box,
the 64 boundary samples nearest its centre are queried and their exact
point-to-box distances minimized. This approximates the voxel-boundary-to-box
minimum, not an exact continuous-surface distance. Overlapping visible support
has zero distance. Hidden-event medians retain instance/mask weighting.
Size associations use 1{,}000 group bootstraps of fixed instance summaries
(minimum 40 instances, 20 groups, 900 valid draws), retaining the table's
all-protocol correlation scope. Figure median-distance CIs use 2{,}000 draws.

\paragraph{Distance conditioning and anatomical support.}
Any visible context includes tokens from the pathology's parent organ or lobe.
For each event with both distances available, restricting this candidate set
cannot reduce the minimum distance. The reported medians nevertheless use
different event sets: masks without visible same-organ context are omitted
from the same-organ summary and receive no zero distance. For liver,
the any-context median of 24.83\,mm uses 1{,}169{,}151 hidden-targeted events,
whereas the same-organ median of 24.50\,mm uses 1{,}033{,}029 events (88.4\%).
Their ordering therefore does not contradict the event-wise inequality.
Same-organ context is available in 83.0\% of hidden kidney events and 55.3\% of
hidden pancreatic events; ReX same-lobe availability is 83.8\%, 82.4\%, 84.7\%,
and 87.9\% for scarring/fibrosis, atelectasis/consolidation, ground-glass opacity,
and nodules/masses, respectively.
For ReX, each instance is assigned the CT-RATE lung lobe with greatest annotated
voxel overlap; if no lobe overlaps, the nearest lobe boundary to the pathology
centroid is used. Same-organ context is then any visible token with nonzero
voxel overlap with that one lobe, including for pathologies spanning multiple
lobes. It is not the union of both lungs or of all involved lobes. ATLAS uses
liver, pancreas, or the union of left and right kidneys as the category's parent
support. These support masks undergo the same deployed spatial transform and
crop as the pathology masks.
\paragraph{Confidence intervals for the distance figure.}
Each pooled-event median is computed from the saved float32 distance arrays.
For each family, 2{,}000 bootstrap
resamples draw the original case/study groups (ATLAS) or patient IDs (ReX) with
replacement, keeping every instance and all its mask events together. Each
resample recomputes the pooled-event median; its 2.5th and 97.5th percentiles
form a pointwise 95\% interval. Both distance types use the same group draws
while retaining their original event-availability conditions. These intervals
condition on the one saved spatial draw per scan and do not estimate additional
within-scan crop variability.
\subsection{Teacher-sensitivity gate}
Before any prediction error is interpreted, the normalized EMA-teacher target
must contain the annotated distinction, or a residual cannot be read as a failure
to predict it. Table~\ref{tab:e3-teacher} reports cross-fitted held-out AUROC for
the original gate fitted without each held-out outer group. ATLAS uses binary
pathology/control discrimination; ReX uses a positive--unlabelled gate. A family passes only with at least 100 pairs, at least 10
of each role per outer fold, AUROC above 0.60 and a 2{,}000-group-bootstrap lower
bound above 0.55. All six pass.
\begin{table}[h]
\centering
\caption{Original cross-fitted normalized EMA-teacher gate. ReX uses
positive--unlabelled evaluation. Controls are unannotated, not verified
healthy. Brackets are 95\% intervals from 2{,}000 group bootstraps.}

\label{tab:e3-teacher}
\scriptsize
\begin{tabular}{@{}llrrc@{}}
\toprule
Dataset & Family & Pairs & Groups & AUROC \\
\midrule
ATLAS & Kidney & 671 & 280 & \shortstack{0.692\\{\scriptsize [0.665,0.720]}} \\
ATLAS & Liver & 808 & 261 & \shortstack{0.713\\{\scriptsize [0.684,0.741]}} \\
ReX & Scarring/fibrosis & 400 & 185 & \shortstack{0.939\\{\scriptsize [0.923,0.955]}} \\
ReX & Atelectasis/consolidation & 457 & 189 & \shortstack{0.914\\{\scriptsize [0.890,0.938]}} \\
ReX & Ground-glass opacity & 723 & 264 & \shortstack{0.883\\{\scriptsize [0.859,0.904]}} \\
ReX & Nodules/masses & 942 & 379 & \shortstack{0.644\\{\scriptsize [0.622,0.665]}} \\
\bottomrule
\end{tabular}
\end{table}
\input{appendix/hidden_target_recoverability.tex}
\input{appendix/matched_region_displacement.tex}
\input{appendix/visible_context_interventions.tex}

\section{Implementation checks and analysis limitations}
\label{app:controls}
\paragraph{Deployed replay and provenance.}
Audits use the retained production state and mask implementation. Exposure
fixes one spatial-only draw per scan; feature audits fix one spatial/intensity
draw. Stable dataset/sample-derived seed namespaces differ, so these are not
the same transformed draw across all four experiments. Recovery uses eight
masks; displacement/interventions 64; exposure 1{,}600. A held-out replay
reproduced image, occupancy, grid, masks, RoPE, features and loss without discrepancy against the deployed forward path.

\paragraph{Mask-exposure implementation.}
One frozen deployed spatial draw per scan is followed by 1{,}600 production-mask
draws, giving a worst-case Wilson 95\% half-width of at most 0.025 for mask-state
probabilities conditional on that crop; crop attrition is estimated separately
across the cohort and reported. Distances are physical
voxel-boundary-to-visible-token-box distances with the 64-candidate
approximation (Appendix~\ref{app:e3-exposure}), not token-centre distances. Soft
voxel-mass exposure is distinguished from occupied-token exposure so that
sub-token pathologies are not treated as whole tokens. Mask-exposure analysis uses no model forward
pass, so its geometry cannot be an artefact of the encoder.

\paragraph{Teacher-sensitivity gate.}
Directions are fitted without each held-out outer group; scaling and the $L_2$
penalty are selected only inside outer training partitions; the standardized
coefficient is divided by the training scaler's feature scale and $L_2$
normalized to give a fold-specific direction in raw normalized-teacher feature
space.

\paragraph{Hidden-target recovery controls.}
Information states share fitting and evaluation rows and study/patient folds
within each task. Anatomy and pathology retain individual token populations.
The expanded analysis uses organ/lobe-wide annotation-negative controls,
a spline geometry floor, a richer random-Fourier sensitivity and a
teacher--geometry gap reliability rule (Appendix~\ref{app:e4}).

\paragraph{Matched-region displacement controls.}
Translated controls, cross-fitted teacher directions, 4{,}096 random-axis
sets, alternative controls, context-distance strata and signed displacement
checks are specified in Appendix~\ref{app:e5}.

\paragraph{Visible-context intervention controls.}
Own-neighbourhood and shared-target estimates, reference-set subtraction,
support gates and joint reporting corrections are specified in
Appendix~\ref{app:e6}. Kidney is descriptive because it fails the support gate.

%% file: appendix/thoracic_organ_state.tex
\subsection{Qualitative thoracic token representations}
\label{app:e2-rex}

The thoracic counterpart to the qualitative ATLAS organ-state analysis in
Section~\ref{sec:results-phenotype} compares lung tokens from ReX scans
bearing a given pathology with tokens from report-normal CT-RATE lungs.
This is an organ-state task: a positive token need not overlap the pathology.
ReX annotations are report-derived and positive--unlabelled, while
report-normal means that the selected lung labels were not reported; neither
designation certifies healthy tissue.

\paragraph{Cohort and harmonized token support.}
The source cohort contains 950 ReX scans. Each family uses all lung tokens
from its annotated scans as positives. The common negative pool consists of
440 scan-disjoint, held-out CT-RATE scans negative for all ten lung-related labels:
emphysema, atelectasis, lung nodule, lung opacity, fibrotic sequela, mosaic
attenuation, peribronchial thickening, consolidation, bronchiectasis and
interlobular septal thickening.

Both cohorts use the voxel-resolution union of the five TotalSegmentator \citep{wasserthal2023totalsegmentator} lung
lobes (labels 10--14 in \texttt{totalsegmentator\_total\_v2\_7\_0}). A token is
included if any voxel belongs to this union. Applying the same rule prevents
boundary-token eligibility from encoding cohort membership.

\paragraph{Visualization.}

\begin{figure}[tbp]
\centering
\includegraphics[width=\textwidth]{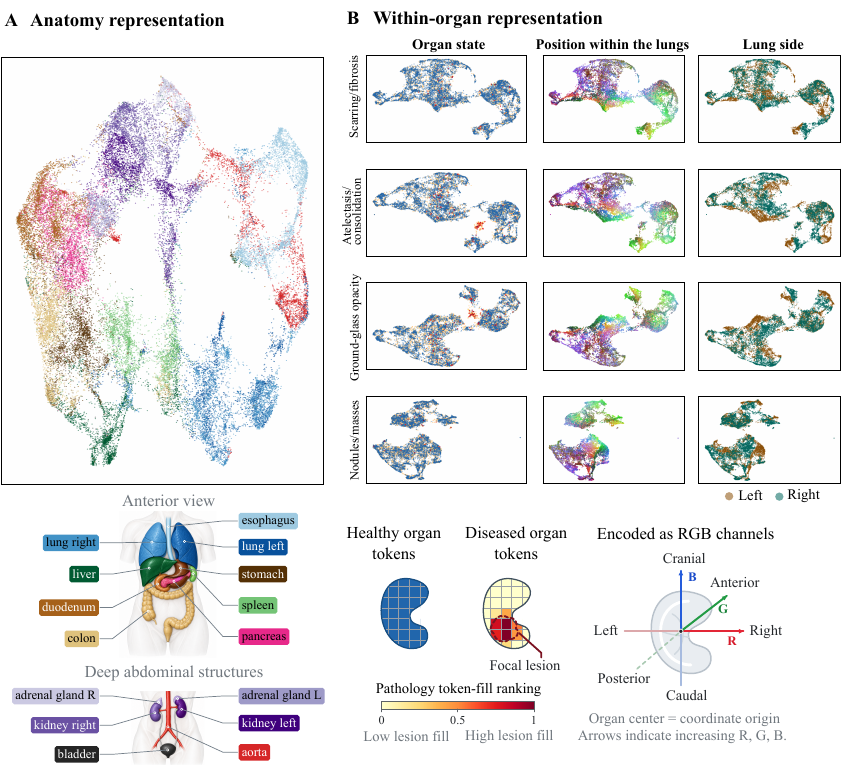}
\caption{Thoracic counterpart to Figure~\ref{fig:umap}.
\textbf{Left:} CT-RATE anatomy tokens colored by 15 structures.
\textbf{Right:} Per pathology family, ReX lung tokens versus report-normal
CT-RATE lung tokens. The first coloring encodes organ state and ranks
pathology fill within the family; its pale end includes diseased-lung tokens
with no annotated pathology. The next two colorings show position in the
whole-lung frame and lung side on the same organ-state embedding. The anatomy
and organ-state panels use different token populations. Report-normal controls
are annotation-negative.  The anatomy legend is shared with Figure~\ref{fig:umap}; abdominal and pelvic structures are only partially covered by chest CT.}
\label{fig:umap-rex}
\end{figure}

Figure~\ref{fig:umap-rex} uses the same visualization conventions
as Figure~\ref{fig:umap} (Appendix~\ref{app:umap-protocol}). Organ-state panels contain the relevant lung-token
populations, while the anatomy panel uses CT-RATE tokens from 15 structures.
For position coloring, voxel coordinates in millimetres are centered on each
scan's whole-lung centroid and divided by the per-axis standard deviation.
The canonical axes map to RGB through $\mathrm{clip}(u/5+0.5,0,1)$.
Because this frame spans both lungs, its left--right channel also carries side
information; the side panel is a consistency check.

\paragraph{Qualitative observations and scope.}
Report-normal and pathology-bearing lung tokens intermix in the UMAPs,
with pathology-overlapping tokens forming a sparse subset and organ-relative
position producing structured color fields. This overlap does not establish
the absence of pathology information or quantify disease discrimination.
Organ-state labels also distinguish the source cohorts, so the visualization
cannot separate disease-related structure from unmeasured cohort differences.

%% file: appendix/scan_level_localization_table.tex
\begin{table}[tbp]
\centering\scriptsize
\setlength{\tabcolsep}{2pt}
\caption{Annotated-cohort MIL and alternative routes. Pathology-token attention
density is the geometric mean of scan ratios to surrounding-organ/lobe
attention; brackets are 95\% group-bootstrap CIs. Ablation is median token deletion effect as a percentage of the fixed
full-sample positive--negative median-logit margin; its intervals are
95\% percentile group-bootstrap CIs from 1{,}000 replicates
(Appendix~\ref{app:routes}). Route/encoder Skill uses identical evaluation rows
but separately cross-fitted routes. Position is the volume-adjusted scan-family
association of largest-pathology typicality with all-pathology-token ablation.
$^*$ denotes matching nominal wild-cluster $p<0.05$
(Appendix~\ref{app:positional}).}

\label{tab:route}
\begin{tabular*}{\textwidth}{@{\extracolsep{\fill}}lcccccc@{}}
\toprule
 & & \shortstack{Pathology-token\\attention density} & \shortstack{Token ablation\\(\% margin)} & \multicolumn{2}{c}{Route / encoder Skill} & \shortstack{Position\\$\rho_{\rm pos}$} \\
\cmidrule(lr){5-6}
Pathology & Skill & & & Acq. & Co-occur. & \\
\midrule
\multicolumn{7}{l}{\textit{ATLAS}} \\
Liver lesion & 0.376 & \metric{2.1$\times$}{[1.7, 2.5]} & \metric{0.45}{[0.28, 0.65]} & 0.84 & 0.17 & $+0.178^{*}$ \\
Kidney lesion & 0.584 & \metric{0.65$\times$}{[0.57, 0.75]} & \metric{0.18}{[0.12, 0.28]} & \best{1.08} & 0.13 & $+0.129^{*}$ \\
Pancreatic lesion & 0.386 & \metric{0.99$\times$}{[0.79, 1.24]} & \metric{1.69}{[1.14, 2.51]} & 0.68 & 0.00 & $+0.006$ \\
\addlinespace
\multicolumn{7}{l}{\textit{ReX}} \\
Scarring/fibrosis & 0.198 & \metric{0.84$\times$}{[0.78, 0.91]} & \metric{$-0.04$}{[-0.22, 0.02]} & 0.06 & \best{1.08} & $+0.032$ \\
Atelectasis/consol. & 0.558 & \metric{\best{9.3$\times$}}{[7.6, 11.7]} & \metric{\best{10.47}}{[6.02, 14.47]} & 0.01 & 0.34 & $-0.036$ \\
Ground-glass opacity & 0.607 & \metric{8.2$\times$}{[6.9, 9.9]} & \metric{2.39}{[1.38, 3.77]} & $-0.02$ & 0.65 & $+0.100$ \\
Nodules/masses & 0.386 & \metric{1.2$\times$}{[1.0, 1.4]} & \metric{0.08}{[0.05, 0.09]} & 0.09 & \best{1.24} & $+0.124^{*}$ \\
\bottomrule
\end{tabular*}
\end{table}

%% file: appendix/recurrence_inference_table.tex
\begin{table}[htbp]
\centering\scriptsize
\caption{Typicality versus scan-family ablation cost or pathology-token attention,
adjusting ranks for largest-pathology volume. $n/G$: rows / case or patient
groups. Nominal conditional wild-cluster $p$ values test the partial coefficients;
no multiplicity adjustment.}
\label{tab:recurrence-inference}
\begin{tabular*}{\textwidth}{@{\extracolsep{\fill}}lccccc@{}}
\toprule
Family & $n/G$ & Necessity $\rho$ & $p$ & Attention $\rho$ & $p$ \\
\midrule
ATLAS kidney & 374/374 & $+0.129$ & 0.01617 & $+0.197$ & 0.00021 \\
ATLAS liver & 261/261 & $+0.178$ & 0.00507 & $+0.106$ & 0.09529 \\
ATLAS pancreas & 197/197 & $+0.006$ & 0.93144 & $-0.083$ & 0.23699 \\
ReX scarring/fibrosis & 230/213 & $+0.032$ & 0.62385 & $-0.355$ & $<0.0001$ \\
ReX atelectasis/consol. & 301/269 & $-0.036$ & 0.54999 & $+0.021$ & 0.72843 \\
ReX ground-glass & 356/345 & $+0.100$ & 0.06187 & $+0.128$ & 0.01917 \\
ReX nodules/masses & 411/382 & $+0.124$ & 0.01948 & $-0.003$ & 0.96224 \\
\bottomrule
\end{tabular*}
\end{table}

%% file: appendix/hidden_target_recoverability.tex
\providecommand{\metric}[2]{\begin{tabular}[c]{@{}c@{}}#1\\[-0.25ex]{\scriptsize #2}\end{tabular}}

\subsection{Hidden-target recovery}
\label{app:e4}

\begin{figure}[p]
\centering
\includegraphics[width=\linewidth]{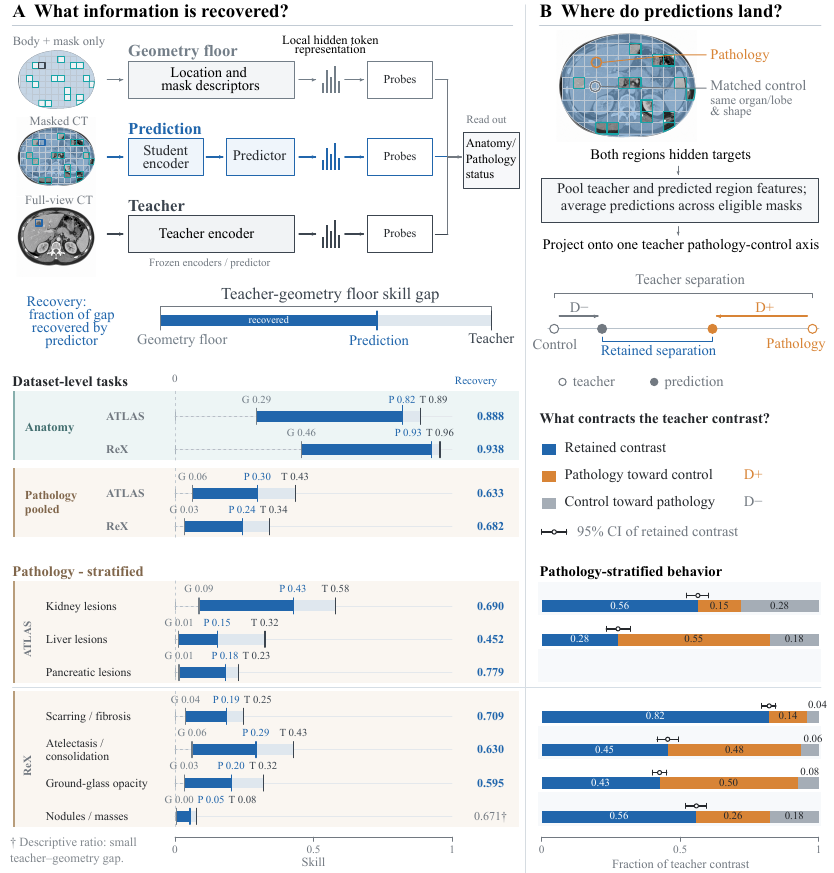}
\caption{Detailed hidden-target recovery and matched-region displacement.
\textbf{A:} Separate probes read geometry (G), prediction (P) and teacher (T).
Bars show absolute Skill for pooled anatomy/pathology tasks in ATLAS and ReX,
then seven pathology categories. Recovery $R_{\mathrm{rec}}=1-Q_{\mathrm{short}}$
is the fraction of the teacher--geometry Skill gap recovered by prediction;
Eq.~\ref{eq:recovery} defines the complementary main-text shortfall.
The nodule/mass ratio is descriptive because the teacher--geometry gap is
below 0.10. \textbf{B:} Both the pathology region
and its matched annotation-negative control are hidden. Pooled region
features are projected onto a cross-fitted teacher pathology--control axis.
Retained contrast $C_{\mathrm{ret}}=1-A_{\mathrm{ctr}}$ and signed
displacements $D_+$ and $D_-$ sum to one; Eq.~\ref{eq:retained-contrast}
defines the main-text attenuation $A_{\mathrm{ctr}}=D_++D_-$.
Whiskers show 95\% intervals for retained contrast.
Pancreatic lesions lack
sufficient eligible matched pairs for B. Schematics illustrate the protocols;
full estimates and uncertainty are reported in this appendix and
Appendix~\ref{app:e5}.}
\label{fig:recovery-displacement-detail}
\end{figure}

Figure~\ref{fig:recovery-displacement-detail} expands the compact main
Figure~\ref{fig:e4-ladder}: it includes the geometry floor, both datasets'
pooled tasks and all family-specific results. Its matched-region panel also
summarizes the separate displacement analysis in Appendix~\ref{app:e5}.

\paragraph{Relation to the main-text metric.}
For task $c$, this appendix reports normalized Skill recovery
\[
R_{\mathrm{rec},c}
=\frac{\mathrm{Skill}_c(\widehat Z)-\mathrm{Skill}_c(\mathcal B)}
       {\mathrm{Skill}_c(Z)-\mathrm{Skill}_c(\mathcal B)}
=1-Q_{\mathrm{short},c},
\]
where $Q_{\mathrm{short},c}$ is the normalized Skill shortfall in
Section~\ref{sec:results-hidden}, Eq.~\ref{eq:recovery}.
The equality uses the same task, geometry floor, evaluation population and
Skill aggregation. Tables and the detailed figure below report
$R_{\mathrm{rec}}$; the main text and its figure report the complementary
$Q_{\mathrm{short}}$, with the task index omitted when clear.

\paragraph{Cohort and token selection.}
Hidden-target recovery uses all eligible annotated source candidates in the expanded extraction,
without requiring a successful translated matched region. The source scopes
contain 768 ATLAS scans with kidney, liver or pancreatic pathologies and 887 ReX
scans with the four pulmonary pathology categories. Each scan has one fixed deployed
spatial/intensity draw and eight production-mask replays. A scan is retained if
at least one eligible pathology or control token is hidden and targeted, yielding
756 ATLAS studies and 887 ReX scans from 813 patients
(Table~\ref{tab:e4-cohorts}). Nine ATLAS scans and one ReX scan have control rows
but no pathology-positive target rows in these replays and remain included.
These are eligible annotated-source cohorts, not all scans in either dataset.

A positive token has nonzero overlap with an included pathology annotation.
Controls are all eligible tokens overlapping the corresponding parent organ in
ATLAS or the lobe(s) assigned to the source pathology in ReX, with zero overlap
with \emph{any} released pathology annotation. ATLAS colon pathologies are excluded
from controls but are not a positive family in this analysis. ReX controls do
not expand to the whole lung and are annotation-negative, not verified healthy.
For pooled pathology status, positives and parent supports are unions over the
included categories; each scan--token--mask row occurs once. Family-specific
analyses instead fit a separate binary probe for each pathology and can share
scans or tokens. The pooled score is calculated from the pooled probe's scores,
not by averaging family scores or recovery ratios.

\begin{table}[t]
\centering
\scriptsize
\setlength{\tabcolsep}{3pt}
\caption{Hidden-target recovery pathology-task cohorts and positive prevalence $\pi$ (\%). Source counts are in-scope annotated candidates; evaluated scans have at least one eligible hidden-target pathology or control row. Groups are ATLAS studies and ReX patients. Rows are token--mask observations, not independent scans; family cohorts overlap. The pooled populations contain 119{,}656/189{,}857 positive and 727{,}909/1{,}932{,}269 negative rows in ATLAS/ReX.}
\label{tab:e4-cohorts}
\begin{tabular}{@{}llrrrrr@{}}
\toprule
Dataset & Scope & Source scans & Eval. scans & Groups & Rows & $\pi$ (\%) \\
\midrule
ATLAS & Pooled & 768 & 756 & 756 & 847{,}565 & 14.12 \\
ATLAS & Kidney lesions & 440 & 430 & 430 & 318{,}093 & 21.72 \\
ATLAS & Liver lesions & 297 & 292 & 292 & 485{,}387 & 7.82 \\
ATLAS & Pancreatic lesions & 224 & 221 & 221 & 58{,}141 & 22.29 \\
ATLAS & Kidney/liver union & 606 & 595 & 595 & 791{,}784 & 13.52 \\
ReX & Pooled & 887 & 887 & 813 & 2{,}122{,}126 & 8.95 \\
ReX & Scarring/fibrosis & 237 & 237 & 220 & 449{,}771 & 7.06 \\
ReX & Atelectasis/consolidation & 306 & 306 & 274 & 531{,}685 & 13.68 \\
ReX & Ground-glass opacity & 363 & 363 & 351 & 768{,}568 & 11.78 \\
ReX & Nodules/masses & 420 & 419 & 390 & 818{,}085 & 2.11 \\
\bottomrule
\end{tabular}
\end{table}

\paragraph{Anatomical targets.}
Anatomy uses the labelled tokens from this pathology/control population plus the
fixed foreground reference sample of up to 48 additional tokens per scan,
selected round-robin across anatomical labels using a dataset/scan-derived
seed. Reference tokens may lie outside the pathology-bearing organ. A token's
class is the structure with the largest nonzero annotated voxel mass. The label sets
contain 15 ATLAS and 13 ReX structures
(Table~\ref{tab:e4-anatomy-prevalences}). Anatomy and pathology status share the
eligible scans and mask replays.
The anatomy tasks contain 1{,}033{,}142 ATLAS and 2{,}316{,}592 ReX rows, versus
847{,}565 and 2{,}122{,}126 for pathology status.

\begin{table}[t]
\centering
\scriptsize
\setlength{\tabcolsep}{3pt}
\caption{Class prevalences (\%) in the pooled anatomy tasks: 1{,}033{,}142 ATLAS and 2{,}316{,}592 ReX token--mask rows, on the same 756/887 eligible scans as their pathology tasks. Structures are listed in label order; horizontal alignment across datasets has no anatomical meaning.}
\label{tab:e4-anatomy-prevalences}
\begin{tabular}{@{}lrlr@{}}
\toprule
ATLAS structure & $\pi$ (\%) & ReX structure & $\pi$ (\%) \\
\midrule
Liver & 38.69 & Left upper lobe & 15.70 \\
Spleen & 3.32 & Left lower lobe & 17.26 \\
Left kidney & 13.53 & Right upper lobe & 12.19 \\
Right kidney & 12.38 & Right middle lobe & 5.42 \\
Pancreas & 4.26 & Right lower lobe & 15.91 \\
Stomach & 4.88 & Trachea & 1.23 \\
Aorta & 2.46 & Oesophagus & 0.61 \\
Left lung & 2.30 & Heart & 4.32 \\
Right lung & 6.64 & Aorta & 2.51 \\
Left adrenal & 0.81 & Liver & 4.97 \\
Right adrenal & 0.68 & Vertebrae & 4.00 \\
Colon & 5.14 & Autochthon & 3.00 \\
Duodenum & 2.93 & Chest-wall bone & 12.87 \\
Oesophagus & 1.09 &  &  \\
Bladder & 0.87 &  &  \\
\bottomrule
\end{tabular}
\end{table}

\paragraph{Information states and probe fitting.}
The prediction and normalized EMA-teacher target are read at the same hidden,
targeted token, using their 768-dimensional representations. A target token's
concealment does not imply concealment of the entire pathology. The geometry
floor uses log1p visible/target counts, integer grid Z/Y/X coordinates, body
occupancy and nearest-visible-token-centre distance in millimetres (nonfinite
values become zero). Seven numeric features receive cubic splines with four
uniformly spaced knots over each training feature's range, without a bias
column, followed by standardization. Mask family is one-hot encoded, ignoring
unknown categories. All transforms fit training rows only.
The floor contains body geometry but no pathology labels or local tissue-intensity
descriptors; distance is token-to-token, not annotation-to-token. Representation
probes use standardized features. Each state has its own $L_2$-regularized
logistic probe; anatomy uses a multiclass classifier and pathology status a binary
classifier.

Five fixed study/patient folds provide out-of-fold probabilities. Within each
outer training partition, one training fold is reserved for selection of
$C\in\{0.01,0.1,1\}$ using the task's primary Skill. Preprocessing is fitted
on the inner fitting partition during selection and on the outer training
partition for the final fit. Only probe fitting is capped at 50{,}000 rows:
one row per group-by-class stratum is guaranteed, remaining capacity is
allocated proportionally, and inverse-inclusion weights restore the training
row distribution. No class balancing is applied. Every eligible held-out row
is scored. Information states share fitting rows, evaluation rows and group
folds within a task; task-specific fitting subsets can differ. Fitting-row
selection and bootstrapping use seed 20260908. All retained fits passed
convergence checks.

\paragraph{Skill and aggregation.}
\label{app:skill_aggregation}
Average precision is the non-interpolated statistic
\begin{equation}
 \mathrm{AP}_c=\sum_j (R_{c,j}-R_{c,j-1})P_{c,j},\qquad
 S_c=\frac{\mathrm{AP}_c-\pi_c}{1-\pi_c},
\label{eq:e4-ap}
\end{equation}
where $P_{c,j}$ and $R_{c,j}$ are precision and recall at successive score
thresholds and $\pi_c$ is the evaluated class prevalence. This is the
\texttt{average\_precision\_score} convention, not trapezoidal PR integration.
Pathology status uses $S_{\mathrm{pathology}}=S_1$ for the positive-pathology class;
the non-pathology Skill is not averaged into it. Anatomy uses
$S_{\mathrm{anatomy}}=K^{-1}\sum_{c=1}^{K}S_c$ with $K=15$ or $13$:
normalization precedes the unweighted class average. Metrics are computed over
concatenated out-of-fold scores, rather than averaged over folds. Pooled pathology
scores weight evaluated rows equally, not categories or patients equally.

Recovery $R_{\mathrm{rec}}$ measures the fraction of a measured
geometry-to-teacher \emph{probe-performance} gap closed by the prediction;
Eq.~\ref{eq:recovery} reports the complementary shortfall.

\paragraph{Uncertainty and ratio reliability.}
We use 1{,}000 bootstrap draws of ATLAS studies or ReX patients, sharing the
expanded dataset's group draws across information states, tasks and family
scopes. AP and prevalence are recomputed with the resampled group weights.
This gives paired anatomy-minus-pathology contrasts despite different task rows.
Percentile 95\% intervals condition on the fitted out-of-fold probabilities;
they do not refit probes or sample additional crops, masks or checkpoints.
Family intervals are pointwise and no multiplicity-adjusted family ranking is
claimed. The reliability rule requires
$G=S(Z)-S(\mathcal B)>0.10$, a positive lower 95\% bound for $G$, and at least
950 bootstrap draws with a positive gap. Ratios failing this rule are reported
descriptively, without clipping or excluding their rows from the pooled task.
The complement is taken on the same bootstrap replicates used for recovery:
an interval $[L,U]$ for $R_{\mathrm{rec}}$ maps to $[1-U,1-L]$ for
$Q_{\mathrm{short}}$, with the same ratio-reliability rule. The paired
anatomy-minus-pathology recovery difference in Table~\ref{tab:e4-contrasts}
equals the pathology-minus-anatomy shortfall difference,
$R_{\mathrm{rec},\mathrm{anatomy}}-R_{\mathrm{rec},\mathrm{pathology}}
=Q_{\mathrm{short},\mathrm{pathology}}-Q_{\mathrm{short},\mathrm{anatomy}}$.
The complement applies to each task's reported ratio, including the pooled
tasks; a pooled ratio is not an average of family-specific ratios.

\paragraph{Pooled and pathology-specific results.}
Tables~\ref{tab:recovery} and \ref{tab:e4-families} give all Skill and
recovery intervals for Figure~\ref{fig:recovery-displacement-detail}A; the main Figure~\ref{fig:e4-anatomy-finding}A shows a compact selection.
The paired anatomy-minus-pathology differences are positive in both datasets
(Table~\ref{tab:e4-contrasts}). Pathology recovery is substantial but varies
across categories. Pancreas has a higher point ratio than kidney (0.779 versus
0.690) alongside much lower prediction/teacher Skills (0.181/0.229 versus
0.427/0.580).

For nodules/masses, $G=0.073$ [0.033, 0.130] fails the point-gap threshold,
although all 1{,}000 ratio draws are valid and the gap interval is positive.
At 2.11\% prevalence, geometry/prediction/teacher AP is 0.026/0.074/0.097;
the corresponding Skills are 0.005/0.054/0.078. The small denominator reflects
limited teacher Skill rather than a strong geometry floor. The family ratio
0.671 [0.543, 0.808] is descriptive; the pooled ReX task retains these pathologies
and passes the reliability rule.

\begin{table}[t]
\centering
\scriptsize
\setlength{\tabcolsep}{3pt}
\caption{Expanded hidden-target recovery pooled results with the spline geometry floor. Values are Skill and recovery, with pointwise 95\% group-bootstrap intervals. Anatomy uses class-macro Skill; pathology status uses positive-class Skill. The four rows correspond to the pooled tasks in Figure~\ref{fig:recovery-displacement-detail}A; only ATLAS anatomy is shown in the main Figure~\ref{fig:e4-anatomy-finding}A.}
\label{tab:recovery}
\begin{tabular}{@{}llcccc@{}}
\toprule
Dataset & Task & Geometry $\mathcal B$ & Prediction $\widehat Z$ & Teacher $Z$ & $R_{\mathrm{rec}}$ \\
\midrule
ATLAS & Anatomy & \metric{0.294}{[0.282, 0.308]} & \metric{0.820}{[0.812, 0.828]} & \metric{0.886}{[0.879, 0.893]} & \metric{0.888}{[0.881, 0.894]} \\
ATLAS & Pathology status & \metric{0.064}{[0.042, 0.093]} & \metric{0.298}{[0.246, 0.349]} & \metric{0.434}{[0.386, 0.482]} & \metric{0.633}{[0.561, 0.686]} \\
ReX & Anatomy & \metric{0.456}{[0.449, 0.464]} & \metric{0.925}{[0.923, 0.928]} & \metric{0.956}{[0.954, 0.958]} & \metric{0.938}{[0.934, 0.941]} \\
ReX & Pathology status & \metric{0.034}{[0.026, 0.043]} & \metric{0.243}{[0.214, 0.275]} & \metric{0.341}{[0.311, 0.374]} & \metric{0.682}{[0.639, 0.724]} \\
\bottomrule
\end{tabular}
\end{table}
\begin{table}[t]
\centering
\scriptsize
\setlength{\tabcolsep}{3pt}
\caption{Pathology-specific hidden-target recovery positive-class Skills, teacher--geometry gap $G$, and recovery, with pointwise 95\% intervals. Each family has its own probe fits. $\dagger$~The nodule/mass gap is below 0.10, so its ratio is descriptive. Family cohorts overlap and these scores are not averaged to obtain the pooled result.}
\label{tab:e4-families}
\begin{tabular}{@{}llcccrc@{}}
\toprule
Dataset & Pathology & Geometry & Prediction & Teacher & $G$ & $R_{\mathrm{rec}}$ \\
\midrule
ATLAS & Kidney lesions & \metric{0.086}{[0.046, 0.132]} & \metric{0.427}{[0.343, 0.495]} & \metric{0.580}{[0.508, 0.639]} & 0.493 & \metric{0.690}{[0.613, 0.748]} \\
ATLAS & Liver lesions & \metric{0.012}{[0.003, 0.023]} & \metric{0.153}{[0.092, 0.228]} & \metric{0.325}{[0.241, 0.411]} & 0.313 & \metric{0.452}{[0.323, 0.560]} \\
ATLAS & Pancreatic lesions & \metric{0.014}{[-0.010, 0.041]} & \metric{0.181}{[0.127, 0.240]} & \metric{0.229}{[0.175, 0.289]} & 0.215 & \metric{0.779}{[0.610, 0.932]} \\
ReX & Scarring/fibrosis & \metric{0.038}{[0.022, 0.058]} & \metric{0.186}{[0.150, 0.229]} & \metric{0.247}{[0.203, 0.294]} & 0.209 & \metric{0.709}{[0.626, 0.794]} \\
ReX & Atelectasis/consolidation & \metric{0.062}{[0.042, 0.087]} & \metric{0.293}{[0.245, 0.336]} & \metric{0.428}{[0.380, 0.473]} & 0.366 & \metric{0.630}{[0.561, 0.696]} \\
ReX & Ground-glass opacity & \metric{0.033}{[0.022, 0.046]} & \metric{0.203}{[0.174, 0.234]} & \metric{0.319}{[0.286, 0.353]} & 0.286 & \metric{0.595}{[0.536, 0.648]} \\
ReX & Nodules/masses & \metric{0.005}{[0.002, 0.009]} & \metric{0.054}{[0.024, 0.098]} & \metric{0.078}{[0.036, 0.134]} & 0.073 & \metric{0.671$^{\dagger}$}{[0.543, 0.808]} \\
\bottomrule
\end{tabular}
\end{table}

\paragraph{Geometry and cohort sensitivities.}
The alternative floor combines standardized geometry with 256 fixed random
Fourier features \citep{rahimi2007random} (Gaussian projection scale 0.35, seed 20260826) and mask-family
indicators.
It changes pooled anatomy/pathology recovery to 0.885/0.635 in ATLAS and
0.933/0.687 in ReX (Table~\ref{tab:e4-rich-floor}); both paired anatomy
advantages remain positive (Table~\ref{tab:e4-contrasts}).

The expanded kidney/liver-only ATLAS analysis has 595 scans and pooled pathology
recovery 0.637 [0.561, 0.693]. Including pancreas gives 756 scans and
0.633 [0.561, 0.686], a paired change of $-0.004$ [$-0.020$, 0.013].

\begin{table}[t]
\centering
\scriptsize
\setlength{\tabcolsep}{3pt}
\caption{Richer geometry-floor sensitivity on identical hidden-target recovery rows and folds. Prediction and teacher scores are unchanged; the alternative floor combines standardized geometry, 256 random Fourier features and mask-family indicators.}
\label{tab:e4-rich-floor}
\begin{tabular}{@{}llcc@{}}
\toprule
Dataset & Task & Geometry Skill & $R_{\mathrm{rec}}$ \\
\midrule
ATLAS & Anatomy & \metric{0.309}{[0.296, 0.322]} & \metric{0.885}{[0.878, 0.892]} \\
ATLAS & Pathology status & \metric{0.063}{[0.039, 0.090]} & \metric{0.635}{[0.570, 0.689]} \\
ReX & Anatomy & \metric{0.493}{[0.486, 0.501]} & \metric{0.933}{[0.929, 0.937]} \\
ReX & Pathology status & \metric{0.029}{[0.022, 0.037]} & \metric{0.687}{[0.642, 0.727]} \\
\bottomrule
\end{tabular}
\end{table}
\begin{table}[t]
\centering
\scriptsize
\setlength{\tabcolsep}{3pt}
\caption{Paired anatomy-minus-pathology recovery differences. Shared study/patient bootstrap draws pair the uncertainty, while anatomy and pathology status retain their individual token populations and score aggregation.}
\label{tab:e4-contrasts}
\begin{tabular}{@{}llc@{}}
\toprule
Dataset & Geometry floor & $R_{\mathrm{rec},\mathrm{anatomy}}-R_{\mathrm{rec},\mathrm{pathology}}$ \\
\midrule
ATLAS & Spline & \metric{0.255}{[0.202, 0.325]} \\
ATLAS & Fourier & \metric{0.251}{[0.196, 0.316]} \\
ReX & Spline & \metric{0.256}{[0.214, 0.299]} \\
ReX & Fourier & \metric{0.246}{[0.205, 0.290]} \\
\bottomrule
\end{tabular}
\end{table}

%% file: appendix/matched_region_displacement.tex
\providecommand{\eFiveMetric}[2]{\begin{tabular}[c]{@{}c@{}}#1\\[-0.25ex]{\scriptsize #2}\end{tabular}}

\subsection{Matched-region contrast, displacement and robustness}
\label{app:e5}

Figure~\ref{fig:recovery-displacement-detail}B shows the full family-wise
contrast decomposition; the compact main Figure~\ref{fig:e5-shrinkage}B
highlights atelectasis/consolidation. This appendix reports retained
contrast $C_{\mathrm{ret}}$, whereas Section~\ref{sec:results-feature}
reports fractional contrast attenuation
$A_{\mathrm{ctr}}=1-C_{\mathrm{ret}}$ (Eq.~\ref{eq:retained-contrast}).
The signed displacements $D_+$ and $D_-$ have the same definitions in both
presentations. Main-text percentages are 100 times the corresponding fractions.

\paragraph{Cohort, replay and feature pooling.}
Matched-region displacement retains the original ATLAS kidney/liver and ReX scarring/fibrosis,
atelectasis/consolidation, ground-glass opacity and nodule/mass categories; it does
not use hidden-target recovery's expanded organ/lobe-wide cohort.
Pancreatic lesions are excluded because too few eligible pathology--control
pairs are available. Each pathology is paired with its
frozen exact-shape translated control in the same organ or lung lobe, excluding
released pathology annotations. Translation uses 24\,mm grid steps, preserving
footprint and sub-token fill. Among at most 256 candidates, the original match
minimizes $(d+|\Delta b|)/24-0.05f$, for centroid displacement $d$ and
parent-boundary-distance difference $\Delta b$ in millimetres and anchor parent
fill $f$; token-row ordering resolves ties. ReX controls are unannotated rather than verified
healthy tissue. All six categories pass the separate normalized-teacher sensitivity
gate in Table~\ref{tab:e3-teacher}.

One deployed spatial/intensity draw and 64 production-mask draws are fixed per
scan. A pair contributes on a draw when neither region has any visible context
overlap and both have nonzero targeted voxel mass. The 768-dimensional teacher
feature is pooled over the represented region with voxel-overlap weights; the
predicted feature is pooled with the same type of weights over its targeted
support, then averaged equally over retained mask draws for that role and pair.
The primary analysis retains complete pairs. Thus events are first averaged
within pairs; scans with more retained draws do not acquire extra weight solely
from their draw count. The slope subsequently weights pairs by teacher-contrast
energy, rather than assigning equal weight to patients.

\paragraph{Cross-fitted directions.}
A linear teacher classifier supplies a pathology-sensitive direction without
the held-out patient/study fold. Training-fold standardization precedes $L_2$ logistic fitting (LBFGS,
maximum 3{,}000 iterations, no class weighting). Nested grouped validation
selects $C\in\{0.01,0.1,1,10\}$ by mean AUROC, breaking ties toward $C=1$.
ReX alternative-control sensitivities fit binary matched-control gates. Dividing the standardized coefficient by its training
feature scale and normalizing to unit length gives a direction in raw normalized
teacher-feature space. Each pair is projected with its held-out-fold direction,
which is held fixed across the control comparisons specified below.

\paragraph{Shrinkage and signed displacement.}
For projected teacher values $z_p^+,z_p^-$ and mask-averaged predictions
$\widehat z_p^+,\widehat z_p^-$, write $x_p=z_p^+-z_p^-$ and
$y_p=\widehat z_p^+-\widehat z_p^-$. The origin-constrained OLS coefficient is
$C_{\mathrm{ret}}=\sum_p x_py_p/\sum_p x_p^2$. Its two displacements are
\begin{equation}
 D_+=\frac{\sum_p x_p(z_p^+-\widehat z_p^+)}{\sum_p x_p^2},\qquad
 D_-=\frac{\sum_p x_p(\widehat z_p^--z_p^-)}{\sum_p x_p^2}.
 \label{eq:displacement}
\end{equation}
The identity $y_p=x_p-(z_p^+-\widehat z_p^+)-(\widehat z_p^--z_p^-)$ gives
$C_{\mathrm{ret}}=1-D_+-D_-$, hence
$A_{\mathrm{ctr}}=1-C_{\mathrm{ret}}=D_++D_-$ for the same pairs,
axis and weights. The displacements and their asymmetry $D_+-D_-$ are
unchanged by this re-expression. These are $x_p^2$-weighted signed
fractions of each teacher gap, not unweighted means or absolute feature errors.
The identities hold before rounding; displayed components can differ from
the displayed total in the last reported digit.
Gaps with $|x_p|\leq10^{-9}$ are omitted; the only affected primary pair is
one ground-glass pair with an exactly zero gap, which contributes zero
to every relevant sum.

Values $0<C_{\mathrm{ret}}<1$ indicate attenuated aggregate contrast;
negative values indicate reversal and values above one amplification. Unit slope
does not imply pairwise equality, and zero slope need not imply equal predictions
for the two roles. Likewise, $D_+=0$ indicates no aggregate directed displacement,
not zero absolute error, and $D_+=1$ indicates a weighted displacement equal to
the teacher gap.

Table~\ref{tab:e5-primary} contains the full values underlying
Figures~\ref{fig:e5-shrinkage}B and~\ref{fig:recovery-displacement-detail}B.
The detailed stack total is identically one, including in each bootstrap
replicate; its components are correlated. Whiskers in the detailed figure
show only the retained-contrast coefficient interval. Component intervals are reported separately here.
An interval $[L,U]$ for $C_{\mathrm{ret}}$ corresponds to
$[1-U,1-L]$ for $A_{\mathrm{ctr}}$. Equivalently, attenuation is the
sum $D_++D_-$ within each paired bootstrap replicate. Multiply the transformed
estimate and interval bounds by 100 to express them as percentages.
\begin{table}[tbp]
\centering\scriptsize
\setlength{\tabcolsep}{3pt}
\caption{Primary matched-region displacement estimates and pointwise 95\% group-bootstrap intervals. $N/G$: region pairs / patient or study groups. Signed displacement is measured in teacher-contrast units.}
\label{tab:e5-primary}
\begin{tabular}{@{}lccccc@{}}
\toprule
Family & $N/G$ & $C_{\mathrm{ret}}$ & $D_+$ & $D_-$ & $D_+-D_-$ \\
\midrule
ATLAS kidney & 671/280 & \eFiveMetric{0.564}{[0.523, 0.602]} & \eFiveMetric{0.154}{[0.119, 0.191]} & \eFiveMetric{0.282}{[0.247, 0.321]} & \eFiveMetric{-0.128}{[-0.192, -0.069]} \\
ATLAS liver & 808/261 & \eFiveMetric{0.276}{[0.233, 0.319]} & \eFiveMetric{0.548}{[0.497, 0.595]} & \eFiveMetric{0.177}{[0.147, 0.208]} & \eFiveMetric{0.371}{[0.298, 0.436]} \\
ReX scarring/fibrosis & 400/185 & \eFiveMetric{0.821}{[0.795, 0.846]} & \eFiveMetric{0.137}{[0.112, 0.166]} & \eFiveMetric{0.042}{[0.023, 0.060]} & \eFiveMetric{0.095}{[0.060, 0.137]} \\
ReX atelectasis/consol. & 457/189 & \eFiveMetric{0.455}{[0.416, 0.493]} & \eFiveMetric{0.482}{[0.439, 0.523]} & \eFiveMetric{0.063}{[0.041, 0.086]} & \eFiveMetric{0.419}{[0.364, 0.471]} \\
ReX ground-glass & 723/264 & \eFiveMetric{0.426}{[0.400, 0.451]} & \eFiveMetric{0.499}{[0.468, 0.529]} & \eFiveMetric{0.076}{[0.054, 0.095]} & \eFiveMetric{0.423}{[0.378, 0.467]} \\
ReX nodules/masses & 942/379 & \eFiveMetric{0.558}{[0.520, 0.594]} & \eFiveMetric{0.265}{[0.230, 0.299]} & \eFiveMetric{0.177}{[0.149, 0.206]} & \eFiveMetric{0.088}{[0.036, 0.138]} \\
\bottomrule
\end{tabular}
\end{table}

\paragraph{Generic-direction calibration.}
For each family, 4,096 independent sets of isotropic unit vectors in 768 dimensions
are applied to the same pair contrasts, with one direction per outer fold in
each set. The fixed random bank is shared across families. Each direction-set
coefficient is calculated separately; we report its mean, central 95\% range
and the paired difference between the pathology coefficient and the random mean
(Table~\ref{tab:e5-random}). Direction percentiles describe variation between
axes on the observed data, not sampling CIs or exchangeable null-test $p$-values.
In particular, atelectasis is below the generic mean but around its 9.7th
percentile, within the central random-axis range.

Coefficients are averaged after projection and slope estimation. Averaging
direction vectors first would instead produce one effective direction per fold
and would not estimate the mean of individual slopes.
Because attenuation is one minus retained contrast, its
pathology-minus-reference difference is the negative of the corresponding
retention difference. Thus negative differences in
Tables~\ref{tab:e5-random} and~\ref{tab:e5-generic} indicate greater
attenuation along the pathology direction, and positive differences indicate
less. The random-axis mean and the moment reference $g$ remain distinct
references; each is complemented separately.

\begin{table}[tbp]
\centering\scriptsize
\setlength{\tabcolsep}{3pt}
\caption{Generic calibration using 4,096 random-axis sets. Bracketed intervals under means and differences are group-bootstrap CIs; the central axis range describes variation between directions. Rank is the percentage of sampled coefficients below the pathology coefficient, not a $p$-value.}
\label{tab:e5-random}
\begin{tabular}{@{}lcccc@{}}
\toprule
Family & Random mean & Central 95\% of axes & Pathology minus mean & Rank (\%) \\
\midrule
ATLAS kidney & \eFiveMetric{0.480}{[0.467, 0.493]} & [0.406, 0.558] & \eFiveMetric{0.083}{[0.046, 0.118]} & 98.2 \\
ATLAS liver & \eFiveMetric{0.443}{[0.426, 0.459]} & [0.370, 0.520] & \eFiveMetric{-0.167}{[-0.202, -0.131]} & 0.0 \\
ReX scarring/fibrosis & \eFiveMetric{0.643}{[0.624, 0.660]} & [0.551, 0.738] & \eFiveMetric{0.178}{[0.158, 0.198]} & 100.0 \\
ReX atelectasis/consol. & \eFiveMetric{0.525}{[0.501, 0.549]} & [0.413, 0.634] & \eFiveMetric{-0.070}{[-0.097, -0.041]} & 9.7 \\
ReX ground-glass & \eFiveMetric{0.564}{[0.550, 0.579]} & [0.461, 0.663] & \eFiveMetric{-0.139}{[-0.159, -0.119]} & 0.5 \\
ReX nodules/masses & \eFiveMetric{0.572}{[0.560, 0.585]} & [0.506, 0.646] & \eFiveMetric{-0.014}{[-0.047, 0.019]} & 36.1 \\
\bottomrule
\end{tabular}
\end{table}

An exact isotropic moment reference provides a second calibration:
\begin{equation}
 g=\frac{\sum_p\Delta T_p^\top\Delta P_p}{\sum_p\|\Delta T_p\|_2^2}.
 \label{eq:e5-generic}
\end{equation}
Here $\Delta T_p$ and $\Delta P_p$ are the unprojected teacher and predicted
pathology-minus-control feature vectors. For isotropic unit $u$ in $d$ dimensions,
$\mathbb E[(u^\top a)(u^\top b)]=a^\top b/d$. Thus $g$ is the ratio of
expected projection numerator and denominator, equivalently an energy-weighted
generic slope; it is not the expectation of their ratio. Paired intervals and the random-bank
mean give the same primary directional conclusions (Table~\ref{tab:e5-generic}).

\begin{table}[tbp]
\centering\scriptsize
\setlength{\tabcolsep}{3pt}
\caption{Exact isotropic moment reference $g$, paired differences and six-family Holm-adjusted approximate bootstrap $p$-values. The repeated 0.012 values reflect finite bootstrap resolution. Energy ratio is pathology-axis teacher-contrast energy divided by mean random-axis energy.}
\label{tab:e5-generic}
\begin{tabular}{@{}lcccc@{}}
\toprule
Family & $g$ & $C_{\mathrm{ret}}-g$ & Adj.\ $p$ & Energy ratio \\
\midrule
ATLAS kidney & \eFiveMetric{0.483}{[0.469, 0.496]} & \eFiveMetric{0.081}{[0.044, 0.116]} & 0.012 & 1.16 \\
ATLAS liver & \eFiveMetric{0.445}{[0.427, 0.461]} & \eFiveMetric{-0.169}{[-0.204, -0.133]} & 0.012 & 2.74 \\
ReX scarring/fibrosis & \eFiveMetric{0.649}{[0.630, 0.667]} & \eFiveMetric{0.173}{[0.153, 0.191]} & 0.012 & 19.89 \\
ReX atelectasis/consol. & \eFiveMetric{0.526}{[0.500, 0.550]} & \eFiveMetric{-0.071}{[-0.097, -0.042]} & 0.012 & 13.42 \\
ReX ground-glass & \eFiveMetric{0.565}{[0.550, 0.581]} & \eFiveMetric{-0.140}{[-0.160, -0.121]} & 0.012 & 8.54 \\
ReX nodules/masses & \eFiveMetric{0.575}{[0.562, 0.588]} & \eFiveMetric{-0.016}{[-0.050, 0.016]} & 0.378 & 0.97 \\
\bottomrule
\end{tabular}
\end{table}

\paragraph{Control selection and teacher-axis sensitivity.}
There are four controls in total: the original nearest match and three
alternatives. Alternative voxel masks preserve shape, voxel count and
boundary-touch status, remain inside the parent organ/lobe, avoid all released
pathologies and do not overlap one another. Eligible translations lie within the
frozen geometry-score caliper of 2.0 above the original match and are selected in
stable-hash order without using model outcomes. The three axes are fitted against
the original control on the robustness subset, alternative~1, and the mean of
available control features per pair (consensus), before fitting a direction. Eligible pair counts differ across controls;
the common-pair comparison below holds the evaluated pairs and masks fixed.
Intervals for these control/axis comparisons are pointwise, not simultaneous
confidence bounds.

All alternative-control and original-to-alternative teacher gates pass.
The teacher-only sensitivity requires token-disjoint original and alternative-1
controls, not disjoint pathology and all four controls. Binary original,
alternative and consensus gates are refitted on the restricted population
(at least 100 pairs and 10 per fold). The statistic divides mean held-out
pathology-minus-two-control-mean score by RMS within-pair control spread.
With gates fixed, group-bootstrap intervals are positive in all six categories
(Table~\ref{tab:teacher-token-disjoint}). This is a teacher diagnostic, not
a predictor-displacement or context-intervention endpoint. Axis alignment is imperfect, and control generators are
partly distinguishable from teacher features (Table~\ref{tab:e5-teacher-axes}).
All four ReX control-generator intervals exclude 0.5.
\input{appendix/teacher_token_disjoint_table.tex}
\begin{table}[tbp]
\centering\scriptsize
\setlength{\tabcolsep}{3pt}
\caption{Teacher-axis sensitivity on the subset with the required alternative control. Columns report held-out pathology/control AUROC, original--alternative direction cosine (median across folds), and the AUROC for distinguishing alternative-1 from original control tissue.}
\label{tab:e5-teacher-axes}
\begin{tabular}{@{}lcccccc@{}}
\toprule
Family & Pairs & Original & Alt.\ 1 & Consensus & Cosine & Control generator \\
\midrule
ATLAS kidney & 498 & 0.665 & 0.661 & 0.688 & 0.515 & \eFiveMetric{0.525}{[0.491, 0.560]} \\
ATLAS liver & 756 & 0.710 & 0.742 & 0.773 & 0.590 & \eFiveMetric{0.508}{[0.483, 0.535]} \\
ReX scarring/fibrosis & 361 & 0.939 & 0.960 & 0.969 & 0.681 & \eFiveMetric{0.661}{[0.631, 0.693]} \\
ReX atelectasis/consol. & 388 & 0.907 & 0.942 & 0.957 & 0.677 & \eFiveMetric{0.566}{[0.532, 0.601]} \\
ReX ground-glass & 636 & 0.879 & 0.915 & 0.933 & 0.692 & \eFiveMetric{0.584}{[0.555, 0.613]} \\
ReX nodules/masses & 930 & 0.644 & 0.745 & 0.802 & 0.511 & \eFiveMetric{0.605}{[0.579, 0.629]} \\
\bottomrule
\end{tabular}
\end{table}

\paragraph{Identical-pair, identical-mask control comparison.}
The common-pair sensitivity restricts to primary pair IDs and intersects
mask draws qualifying for all four controls. Consequently each control uses the
same pairs, folds and draws within a family. Pathology teacher and predicted
features are exactly identical across alternatives. All three axes remain fixed;
contrasts between controls on one axis therefore isolate control choice within
this restricted population and masking condition.

All 72 common-mask coefficients and their pointwise upper CIs are below one
(largest upper limit 0.901). All 12 asymmetry CIs per family retain their sign for
kidney, liver, scarring, atelectasis and ground-glass. Nodule asymmetry is less
stable: three consensus-axis alternative-control intervals cross zero, including
$-0.006$ $[-0.036,0.025]$ for alternative~2. Exact magnitudes change: on the
consensus axis, switching from the original control to alternatives raises the
coefficient by 0.082--0.103 for liver, 0.059--0.068 for ground-glass and
0.122--0.145 for nodules, with paired pointwise intervals excluding zero.
Table~\ref{tab:e5-controls} reports ranges across the tested combinations.
Pathology-minus-$g$ intervals remain below zero in every combination for liver,
atelectasis and ground-glass and above zero for scarring; kidney points remain
positive but three intervals cross zero. Nodules vary from no clear difference
to greater preservation than generic.
\begin{table}[tbp]
\centering\scriptsize
\setlength{\tabcolsep}{3pt}
\caption{Common-pair, common-mask sensitivity: ranges across four controls crossed with three fixed teacher axes. Entries are min--max point estimates, not confidence intervals. $N/G$ is identical across all 12 cells within a family.}
\label{tab:e5-controls}
\begin{tabular}{@{}lccccc@{}}
\toprule
Family & $N/G$ & $C_{\mathrm{ret}}$ & $D_+$ & $D_-$ & $D_+-D_-$ \\
\midrule
ATLAS kidney & 298/172 & $[0.514, 0.621]$ & $[0.100, 0.175]$ & $[0.270, 0.320]$ & $[-0.203, -0.136]$ \\
ATLAS liver & 683/231 & $[0.285, 0.391]$ & $[0.431, 0.517]$ & $[0.164, 0.211]$ & $[0.253, 0.319]$ \\
ReX scarring/fibrosis & 273/141 & $[0.832, 0.875]$ & $[0.095, 0.127]$ & $[0.031, 0.048]$ & $[0.058, 0.090]$ \\
ReX atelectasis/consol. & 289/131 & $[0.435, 0.502]$ & $[0.421, 0.505]$ & $[0.058, 0.101]$ & $[0.324, 0.445]$ \\
ReX ground-glass & 513/208 & $[0.391, 0.500]$ & $[0.442, 0.533]$ & $[0.055, 0.093]$ & $[0.371, 0.458]$ \\
ReX nodules/masses & 911/374 & $[0.546, 0.808]$ & $[0.105, 0.269]$ & $[0.083, 0.185]$ & $[-0.006, 0.083]$ \\
\bottomrule
\end{tabular}
\end{table}

\paragraph{Distance and production-mask strata.}
Distance strata select role-specific events by the nearest visible
context-token distance, then average within role and retain pairs with both
roles in that bin. The role averages need not use the same draws, and eligible
pairs change across bins. All six farthest-bin coefficients are
below their nearest-bin values; five sequences decrease monotonically. Scarring
has a small final increase from 0.786 to 0.802
(Table~\ref{tab:e5-distance}). Thus the data support a broad distance trend. The three production
mask families also yield attenuated coefficients
(Table~\ref{tab:e5-masks}).
\begin{table}[tbp]
\centering\scriptsize
\setlength{\tabcolsep}{3pt}
\caption{Distance-stratified coefficients with 95\% group-bootstrap intervals and pair counts. Distance is the region-boundary-to-nearest-visible-token distance in millimetres; bins are left-closed and right-open. Roles are filtered separately before pair-level averaging.}
\label{tab:e5-distance}
\begin{tabular}{@{}lcccc@{}}
\toprule
Family & $[0,24)$ & $[24,48)$ & $[48,96)$ & $[96,\infty)$ \\
\midrule
ATLAS kidney & \eFiveMetric{0.674}{[0.633, 0.712]; $n=671$} & \eFiveMetric{0.533}{[0.490, 0.574]; $n=671$} & \eFiveMetric{0.440}{[0.392, 0.491]; $n=669$} & \eFiveMetric{0.407}{[0.350, 0.459]; $n=486$} \\
ATLAS liver & \eFiveMetric{0.326}{[0.283, 0.371]; $n=808$} & \eFiveMetric{0.263}{[0.222, 0.309]; $n=805$} & \eFiveMetric{0.222}{[0.176, 0.268]; $n=803$} & \eFiveMetric{0.166}{[0.109, 0.232]; $n=614$} \\
ReX scarring/fibrosis & \eFiveMetric{0.849}{[0.826, 0.871]; $n=400$} & \eFiveMetric{0.810}{[0.782, 0.838]; $n=400$} & \eFiveMetric{0.786}{[0.757, 0.814]; $n=396$} & \eFiveMetric{0.802}{[0.761, 0.843]; $n=231$} \\
ReX atelectasis/consol. & \eFiveMetric{0.484}{[0.444, 0.525]; $n=457$} & \eFiveMetric{0.434}{[0.394, 0.478]; $n=457$} & \eFiveMetric{0.421}{[0.383, 0.463]; $n=456$} & \eFiveMetric{0.420}{[0.369, 0.470]; $n=194$} \\
ReX ground-glass & \eFiveMetric{0.448}{[0.423, 0.475]; $n=723$} & \eFiveMetric{0.411}{[0.385, 0.437]; $n=723$} & \eFiveMetric{0.395}{[0.371, 0.421]; $n=719$} & \eFiveMetric{0.387}{[0.352, 0.425]; $n=324$} \\
ReX nodules/masses & \eFiveMetric{0.606}{[0.567, 0.644]; $n=942$} & \eFiveMetric{0.547}{[0.505, 0.587]; $n=942$} & \eFiveMetric{0.497}{[0.452, 0.539]; $n=942$} & \eFiveMetric{0.471}{[0.424, 0.518]; $n=633$} \\
\bottomrule
\end{tabular}
\end{table}
\begin{table}[tbp]
\centering\scriptsize
\setlength{\tabcolsep}{3pt}
\caption{Coefficients and 95\% group-bootstrap intervals within each production mask family. These are conditional strata, not interventions on the training mask distribution.}
\label{tab:e5-masks}
\begin{tabular}{@{}lccc@{}}
\toprule
Family & Multi-block target & Multi-scale blocks & Single-crop context \\
\midrule
ATLAS kidney & \eFiveMetric{0.589}{[0.548, 0.627]} & \eFiveMetric{0.586}{[0.544, 0.624]} & \eFiveMetric{0.510}{[0.464, 0.551]} \\
ATLAS liver & \eFiveMetric{0.294}{[0.251, 0.340]} & \eFiveMetric{0.290}{[0.247, 0.338]} & \eFiveMetric{0.245}{[0.198, 0.296]} \\
ReX scarring/fibrosis & \eFiveMetric{0.833}{[0.808, 0.858]} & \eFiveMetric{0.834}{[0.808, 0.857]} & \eFiveMetric{0.796}{[0.769, 0.825]} \\
ReX atelectasis/consol. & \eFiveMetric{0.463}{[0.423, 0.503]} & \eFiveMetric{0.458}{[0.418, 0.498]} & \eFiveMetric{0.442}{[0.403, 0.481]} \\
ReX ground-glass & \eFiveMetric{0.431}{[0.405, 0.456]} & \eFiveMetric{0.427}{[0.400, 0.453]} & \eFiveMetric{0.417}{[0.390, 0.442]} \\
ReX nodules/masses & \eFiveMetric{0.574}{[0.536, 0.611]} & \eFiveMetric{0.570}{[0.531, 0.608]} & \eFiveMetric{0.526}{[0.486, 0.566]} \\
\bottomrule
\end{tabular}
\end{table}

\paragraph{Uncertainty and influence.}
Primary intervals and generic-reference comparisons use 1,000 paired bootstrap
resamples of ATLAS released case/study groups or ReX patients, retaining every
pair within a resampled group.
Intervals are conditional on saved crops, masks, trained axes and checkpoint;
they do not include refitting or pretraining uncertainty. Holm correction \citep{holm1979simple} covers
the six primary pathology-minus-$g$ comparisons; add-one two-sided bootstrap tail
counts avoid zero $p$-values. The first five adjusted values of about 0.012 reflect
finite resampling resolution, while nodules give 0.378. All control-cell intervals
are pointwise and descriptive; the control comparisons are sensitivity analyses.

Eight independent blocks of 512 random-axis sets give mean Monte Carlo standard
errors below 0.0008 and percentile-endpoint errors below 0.003. No patient/study
contributes more than 4.1\% of primary teacher-contrast energy; removing any one
group changes a primary coefficient by less than 0.009
(Table~\ref{tab:e5-stability}).

\begin{table}[tbp]
\centering\scriptsize
\setlength{\tabcolsep}{3pt}
\caption{Primary stability diagnostics. Effective groups and largest group weight use squared teacher-gap weights. Leave-one-out ranges delete one patient/study group. Monte Carlo standard errors (MCSE) use eight independent blocks of 512 random-axis sets.}
\label{tab:e5-stability}
\begin{tabular}{@{}lccccc@{}}
\toprule
Family & Effective groups & Max.\ weight (\%) & Leave-one-out $C_{\mathrm{ret}}$ & Mean MCSE & Max.\ tail MCSE \\
\midrule
ATLAS kidney & 105.5 & 2.71 & [0.559, 0.570] & 0.0008 & 0.0021 \\
ATLAS liver & 79.6 & 3.05 & [0.268, 0.283] & 0.0005 & 0.0020 \\
ReX scarring/fibrosis & 92.9 & 3.88 & [0.818, 0.824] & 0.0007 & 0.0028 \\
ReX atelectasis/consol. & 88.0 & 3.88 & [0.447, 0.462] & 0.0008 & 0.0030 \\
ReX ground-glass & 105.6 & 2.82 & [0.423, 0.429] & 0.0007 & 0.0022 \\
ReX nodules/masses & 123.2 & 4.03 & [0.552, 0.563] & 0.0004 & 0.0017 \\
\bottomrule
\end{tabular}
\end{table}

\paragraph{Family-specific exceptions.}
Kidney and scarring/fibrosis retain more contrast than the mean generic
direction, but their retained-contrast coefficients, 0.564 and 0.821, remain
below one. Kidney shows greater control-side than pathology-side displacement
($D_+-D_-=-0.128$, 95\% CI $[-0.192,-0.069]$). These measurements establish
heterogeneity in contrast attenuation and displacement asymmetry.

%% file: appendix/teacher_token_disjoint_table.tex
\begin{table}[htbp]
\centering\scriptsize
\caption{Teacher sensitivity with token-disjoint original/alternative-1 controls.
Pathology excess is in RMS control-spread units, with fixed-gate 95\%
case/patient-bootstrap intervals. $N/G$: pairs / groups.}
\label{tab:teacher-token-disjoint}
\begin{tabular}{@{}lcc@{}}
\toprule
Family & $N/G$ & Normalized pathology excess \\
\midrule
ATLAS kidney & 446/222 & 0.953 [0.763, 1.142] \\
ATLAS liver & 674/226 & 1.369 [1.134, 1.631] \\
ReX scarring/fibrosis & 254/136 & 3.970 [3.516, 4.477] \\
ReX atelectasis/consol. & 280/138 & 2.814 [2.445, 3.283] \\
ReX ground-glass & 498/217 & 2.758 [2.456, 3.080] \\
ReX nodules/masses & 881/367 & 0.942 [0.836, 1.055] \\
\bottomrule
\end{tabular}
\end{table}

%% file: appendix/visible_context_interventions.tex
\providecommand{\eSixMetric}[2]{\begin{tabular}[c]{@{}c@{}}#1\\[-0.25ex]{\scriptsize #2}\end{tabular}}

\subsection{Visible-region interventions and robustness}
\label{app:e6}

\paragraph{Target definitions.}
We evaluate replacement effects on each region's own clean target neighbourhood
and on the intersection shared by the pathology and control regions. The
shared-target comparison uses identical targets and loss weights for both
interventions. Both target definitions, their common-event comparison and
endpoint-specific sensitivities are reported below; the reporting correction
covers both target definitions.
Donor/equal-norm replacement, dose response and available-cohort alternative
controls cover both target endpoints. Token-disjoint, and covariate-adjusted sensitivities cover own
neighbourhood endpoints.

\paragraph{Cohort and production replay.}
The source cohort contains 1,994 pathology--control pairs: 143 kidney, 417 liver, 281 scarring/fibrosis, 327
atelectasis/consolidation, 505 ground-glass and 321 nodule/mass pairs. It contains
279 ATLAS scans and 625 ReX scans from 586 ReX patients. Pancreatic lesions are excluded because too few eligible pathology--control
pairs are available; the expanded hidden-target recovery cohort is not added. Each pathology retains its original exact-shape
translated control within the parent organ or dominant lung lobe, together with
up to three eligible alternative controls. Region matching
is at voxel scale; different translated regions may overlap the same token.

We restore the deployed checkpoint, crop/augmentation draw, token grid and 64
production masks per scan. Cached full-view EMA-teacher targets and final
masked-student context activations are retained at their native precision.
The token pitch is 24\,mm and feature width is 768. Predictor interventions are evaluated in FP32 with TF32 disabled, using the
same cached student/teacher tensors. No encoder, predictor, teacher or
diagnostic direction is refitted. Eligibility requires both original
pair members to contain visible token mass in the same mask event. Additional
controls are evaluated only when that original event and the additional control
qualify. Table~\ref{tab:e6-support} reports retained support after requiring at
least eight eligible events per pair. Counts are pathology/control pairs, not
independent patients; the bootstrap preserves the study/patient clusters.
\begin{table}[tbp]
\centering\scriptsize
\setlength{\tabcolsep}{2pt}
\caption{Visible-context intervention source and retained support. O: own neighbourhoods; S: shared targets. Groups are ATLAS released studies/cases or ReX patients. Each view requires eight events per pair. Kidney is below the 100-pair threshold.}
\label{tab:e6-support}
\begin{tabular}{@{}lrrrrrrr@{}}
\toprule
Family & Source pairs & O pairs & O groups & O events & S pairs & S groups & S events \\
\midrule
ATLAS kidney & 143 & 62 & 50 & 573 & 60 & 48 & 555 \\
ATLAS liver & 417 & 405 & 206 & 4947 & 402 & 206 & 4900 \\
ReX scarring/fibrosis & 281 & 243 & 141 & 2800 & 242 & 141 & 2784 \\
\shortstack[l]{ReX atelectasis/\\consolidation} & 327 & 282 & 149 & 3377 & 280 & 149 & 3353 \\
ReX ground glass & 505 & 447 & 219 & 5187 & 444 & 218 & 5146 \\
ReX nodules/masses & 321 & 290 & 205 & 2936 & 288 & 204 & 2918 \\
\bottomrule
\end{tabular}
\end{table}

\paragraph{Clean tokens, donors and references.}
Parent membership requires nonzero overlap with the organ/lobe mask, rather
than complete token purity. Kidney uses the bilateral kidney
union; ReX retains the pathology's assigned dominant lobe. A token is excluded
from the clean population if it overlaps any released pathology annotation in the
scan, including unmatched and out-of-scope annotations, or any available
matched control for the current source pathology. The same excluded population
is used across that pathology's control comparisons. This does not exclude
unannotated disease.

For each source/mask event, the eligible visible clean tokens are deterministically
permuted. The first $\max(8,\lfloor K/2\rfloor)$ form the donor pool and the
remainder form a separate reference pool, where $K$ is the eligible count.
At least eight donors and enough reference tokens for the larger original
region are required. Otherwise the event is excluded; there is no global-context
fallback. The replacement $\bar h$ is the unweighted mean of the original masked
student activations in the donor pool, computed once per event and shared
between roles. It excludes both regions, all disallowed annotations/controls
and all target tokens. Reference tokens do not contribute to their own
replacement mean. Eight reference permutations are generated without replacement
within a draw. For a region containing $k$ visible tokens, the first $k$ entries
are used; equal-size regions therefore use coupled reference sets. Reference
selection matches count and parent tissue, not every spatial property.

\paragraph{Target sets and fixed loss.}
Let $R_\pm$ contain visible tokens overlapping the pathology/control and let
$c_\pm$ be the unweighted mean of their token-centre coordinates. The role-specific
set $N_\pm$ contains clean same-parent target tokens whose centres are 24--96\,mm
from $c_\pm$, inclusively. These are centre distances, distinct from the
boundary-to-token-box distances used in mask-exposure analysis. Each $N_\pm$ must contain at least
eight tokens. The shared readout uses $N_\cap=N_+\cap N_-$ and additionally
requires $|N_\cap|\geq8$. The original role-specific readout uses $N_+$ for the
pathology and $N_-$ for the control. Shared targets equalize the evaluated tokens
and baseline loss but need not have equal distances to the two regions.

For foreground occupancy $o_j$ and teacher target $t_j$, the loss is
\begin{equation}
 L_N(h)=\frac{\sum_{j\in N}(0.1+0.9o_j)
                 \|f(h)_j-t_j\|_1/768}
                {\sum_{j\in N}(0.1+0.9o_j)}.
 \label{eq:e6-loss-app}
\end{equation}
Replacement modifies only selected final student activation values. Token
identities, positions, masks, teacher targets and target weights are held fixed.
Every intervention is formed from the unmodified context rather than from a
previously perturbed arm. For role $r$, use $N_r=N_\cap$ in the shared analysis
and that role's own set in the own-neighbourhood analysis, and define
\begin{align}
 A_r&=L_{N_r}(h^{R_r\leftarrow\bar h})-L_{N_r}(h),\nonumber\\
 G_r&=\frac{1}{8}\sum_{b=1}^{8}
       \{L_{N_r}(h^{Q_{r,b}\leftarrow\bar h})-L_{N_r}(h)\},\nonumber\\
 d&=(A_+-G_+)-(A_--G_-),\qquad d_{\mathrm{raw}}=A_+-A_-.
 \label{eq:e6-event-app}
\end{align}
Here $Q_{r,b}$ is the $b$th count-matched reference set. The baseline cancels
within each adjusted role contrast, even in the own-neighbourhood analysis.
The random-reference subtraction is a defined calibration.

\paragraph{Aggregation, effect scales and uncertainty.}
Reference draws are averaged within event, eligible events within pair, and
pairs equally. At least eight events per pair and 100 pairs per family are
required for supported family inference. Filtering and event-count requirements
are reapplied within each sensitivity. Kidney falls below family support and is
reported descriptively without an inferential $p$-value.

Writing $\langle\cdot\rangle$ for this hierarchical average, the reported
$\Delta=\langle d\rangle$ and $R_{\mathrm{ctrl}}=-100\Delta/\langle A_-\rangle$
use the reference-adjusted numerator and the natural control replacement cost
as denominator. Positive $R_{\mathrm{ctrl}}$ denotes a negative adjusted
contrast, measured in units of that cost. It need not lie between zero and 100.
The literal raw reduction is instead
$R_{\mathrm{raw}}=-100\langle d_{\mathrm{raw}}\rangle/\langle A_-\rangle$.
The baseline scale is $100\Delta/\langle B\rangle$, where
$B=\{L_{N_+}(h)+L_{N_-}(h)\}/2$ for own neighbourhoods and $B=L_{N_\cap}(h)$
for shared targets.
Table~\ref{tab:context-main} reports both normalizations in the
control-minus-pathology orientation: $R_{\mathrm{ctrl}}$ as defined here and
the sign-reversed baseline scale $-100\Delta/\langle B\rangle$, with interval
$[-u,-l]$ when the interval for $100\Delta/\langle B\rangle$ is $[l,u]$.
Positive main-text values therefore denote a higher control replacement cost.
The appendix tables report $\Delta$ and $100\Delta/\langle B\rangle$ in the
pathology-minus-control orientation, where the same result is negative, whereas
$R_{\mathrm{ctrl}}$ has the same sign in both. These are ratios of family means, not averages of
individual-event ratios. Table~\ref{tab:e6-costs} distinguishes the raw and
adjusted scales.

We draw 10,000 study/patient bootstrap resamples, preserving all pairs and their
nested events within a sampled group and recomputing the equal-pair mean.
Numerator and denominator use the same resample. All control-cost denominators
remain positive across the 10,000 resamples for every family/target definition.
Intervals are pointwise 2.5th--97.5th percentiles, conditional on the checkpoint,
fixed crop/masks, diagnostic directions and sampled reference sets. Two-sided
bootstrap sign-tail $p$-values use
$\min\{1,2[1+\min(\#\{\Delta^*\leq0\},\#\{\Delta^*\geq0\})]/10001\}$;
these are approximate, not exact randomization tests. The joint reporting
correction uses Holm \citep{holm1979simple}  across four ATLAS contrasts and Benjamini--Hochberg \citep{benjamini1995fdr}
across eight ReX contrasts, spanning both target definitions. Unsupported kidney entries contribute $p=1$ to these categories but are
not displayed as tests. The same zero-contrast test accompanies the normalized
scale because its denominator is positive. Pointwise intervals need not agree
with multiplicity-adjusted decisions: the shared nodule interval excludes zero,
but its jointly adjusted $p=0.0656$ does not pass 0.05.
\begin{table}[tbp]
\centering\scriptsize
\setlength{\tabcolsep}{2pt}
\caption{Both target definitions with the joint reporting correction spanning four ATLAS and eight ReX contrasts. Intervals are pointwise; kidney is descriptive. Baseline percentage uses the signed adjusted contrast and therefore has the opposite sign to $R_{\mathrm{ctrl}}$.}
\label{tab:e6-endpoints}
\begin{tabular}{@{}llrrrrr@{}}
\toprule
Family & Targets & $n$ & $\Delta\times10^3$ & $R_{\mathrm{ctrl}}$ (\%) & $100\Delta/\langle B\rangle$ & Adj.\ $p$ \\
\midrule
ATLAS kidney & Own & 62 & \eSixMetric{$-1.209$}{$[-2.836, 0.440]$} & \eSixMetric{$9.5$}{$[-3.6, 21.3]$} & \eSixMetric{$-0.335$}{$[-0.792, 0.121]$} & --- \\
ATLAS kidney & Shared & 60 & \eSixMetric{$-1.525$}{$[-3.270, 0.252]$} & \eSixMetric{$11.6$}{$[-2.0, 24.0]$} & \eSixMetric{$-0.423$}{$[-0.913, 0.068]$} & --- \\
ATLAS liver & Own & 405 & \eSixMetric{$-0.977$}{$[-1.726, -0.302]$} & \eSixMetric{$6.7$}{$[2.2, 11.2]$} & \eSixMetric{$-0.273$}{$[-0.481, -0.085]$} & $0.0114$ \\
ATLAS liver & Shared & 402 & \eSixMetric{$-2.124$}{$[-3.052, -1.273]$} & \eSixMetric{$12.8$}{$[8.1, 17.3]$} & \eSixMetric{$-0.599$}{$[-0.860, -0.362]$} & $<0.001$ \\
ReX scarring/fibrosis & Own & 243 & \eSixMetric{$-1.176$}{$[-2.410, 0.024]$} & \eSixMetric{$5.3$}{$[-0.1, 10.4]$} & \eSixMetric{$-0.352$}{$[-0.719, 0.007]$} & $0.0720$ \\
ReX scarring/fibrosis & Shared & 242 & \eSixMetric{$-8.064$}{$[-9.900, -6.346]$} & \eSixMetric{$27.3$}{$[22.4, 32.0]$} & \eSixMetric{$-2.433$}{$[-2.989, -1.914]$} & $<0.001$ \\
\shortstack[l]{ReX atelectasis/\\consolidation} & Own & 282 & \eSixMetric{$-1.743$}{$[-2.994, -0.503]$} & \eSixMetric{$6.7$}{$[2.0, 11.2]$} & \eSixMetric{$-0.506$}{$[-0.872, -0.146]$} & $0.0116$ \\
\shortstack[l]{ReX atelectasis/\\consolidation} & Shared & 280 & \eSixMetric{$-5.535$}{$[-7.192, -3.890]$} & \eSixMetric{$17.6$}{$[12.8, 22.0]$} & \eSixMetric{$-1.616$}{$[-2.103, -1.136]$} & $<0.001$ \\
ReX ground glass & Own & 447 & \eSixMetric{$-0.545$}{$[-1.456, 0.364]$} & \eSixMetric{$2.1$}{$[-1.5, 5.6]$} & \eSixMetric{$-0.162$}{$[-0.433, 0.109]$} & $0.2795$ \\
ReX ground glass & Shared & 444 & \eSixMetric{$-3.847$}{$[-4.935, -2.713]$} & \eSixMetric{$12.6$}{$[9.0, 15.9]$} & \eSixMetric{$-1.152$}{$[-1.475, -0.812]$} & $<0.001$ \\
ReX nodules/masses & Own & 290 & \eSixMetric{$-0.191$}{$[-0.968, 0.586]$} & \eSixMetric{$1.3$}{$[-4.3, 6.6]$} & \eSixMetric{$-0.058$}{$[-0.293, 0.178]$} & $0.6427$ \\
ReX nodules/masses & Shared & 288 & \eSixMetric{$-0.975$}{$[-1.937, -0.042]$} & \eSixMetric{$6.0$}{$[0.3, 11.4]$} & \eSixMetric{$-0.297$}{$[-0.589, -0.013]$} & $0.0656$ \\
\bottomrule
\end{tabular}
\end{table}

Own-neighbourhood estimates retain corrected support for liver and atelectasis;
shared-target estimates retain support for liver, scarring, atelectasis and
ground glass under the joint correction spanning both target definitions
(Table~\ref{tab:e6-endpoints}).
\begin{table}[tbp]
\centering\scriptsize
\setlength{\tabcolsep}{2pt}
\caption{Absolute baseline loss, role-specific mean replacement costs, raw contrasts and raw relative reduction. $A_+$ and $A_-$ are unadjusted costs. Raw contrast/reduction intervals are pointwise and are not the tests of reference-adjusted $\Delta$.}
\label{tab:e6-costs}
\begin{tabular}{@{}llrrrrr@{}}
\toprule
Family & Targets & $\langle B\rangle$ & $\langle A_+\rangle\!\times\!10^3$ & $\langle A_-\rangle\!\times\!10^3$ & $\Delta_{\mathrm{raw}}\!\times\!10^3$ & $R_{\mathrm{raw}}$ (\%) \\
\midrule
ATLAS kidney & Own & $0.361$ & $11.174$ & $12.740$ & \eSixMetric{$-1.566$}{$[-3.163, 0.032]$} & \eSixMetric{$12.3$}{$[-0.3, 23.7]$} \\
ATLAS kidney & Shared & $0.360$ & $11.363$ & $13.193$ & \eSixMetric{$-1.830$}{$[-3.547, -0.154]$} & \eSixMetric{$13.9$}{$[1.2, 25.8]$} \\
ATLAS liver & Own & $0.358$ & $13.520$ & $14.575$ & \eSixMetric{$-1.056$}{$[-1.836, -0.367]$} & \eSixMetric{$7.2$}{$[2.7, 12.0]$} \\
ATLAS liver & Shared & $0.354$ & $14.631$ & $16.605$ & \eSixMetric{$-1.974$}{$[-2.900, -1.136]$} & \eSixMetric{$11.9$}{$[7.2, 16.5]$} \\
ReX scarring/fibrosis & Own & $0.334$ & $21.717$ & $22.362$ & \eSixMetric{$-0.645$}{$[-1.834, 0.534]$} & \eSixMetric{$2.9$}{$[-2.5, 8.0]$} \\
ReX scarring/fibrosis & Shared & $0.331$ & $22.176$ & $29.549$ & \eSixMetric{$-7.373$}{$[-9.157, -5.703]$} & \eSixMetric{$25.0$}{$[20.1, 29.6]$} \\
\shortstack[l]{ReX atelectasis/\\consolidation} & Own & $0.344$ & $24.852$ & $26.163$ & \eSixMetric{$-1.311$}{$[-2.656, 0.006]$} & \eSixMetric{$5.0$}{$[-0.0, 9.8]$} \\
\shortstack[l]{ReX atelectasis/\\consolidation} & Shared & $0.342$ & $26.280$ & $31.440$ & \eSixMetric{$-5.160$}{$[-6.996, -3.379]$} & \eSixMetric{$16.4$}{$[11.2, 21.4]$} \\
ReX ground glass & Own & $0.336$ & $25.643$ & $25.665$ & \eSixMetric{$-0.022$}{$[-1.034, 0.996]$} & \eSixMetric{$0.1$}{$[-3.9, 3.9]$} \\
ReX ground glass & Shared & $0.334$ & $27.180$ & $30.560$ & \eSixMetric{$-3.380$}{$[-4.562, -2.216]$} & \eSixMetric{$11.1$}{$[7.4, 14.7]$} \\
ReX nodules/masses & Own & $0.330$ & $14.232$ & $14.294$ & \eSixMetric{$-0.062$}{$[-0.920, 0.776]$} & \eSixMetric{$0.4$}{$[-5.6, 6.2]$} \\
ReX nodules/masses & Shared & $0.328$ & $15.551$ & $16.352$ & \eSixMetric{$-0.802$}{$[-1.815, 0.178]$} & \eSixMetric{$4.9$}{$[-1.2, 10.6]$} \\
\bottomrule
\end{tabular}
\end{table}

Both roles have positive mean full-replacement costs. For shared-target liver,
$R_{\mathrm{raw}}=11.9\%$, whereas $R_{\mathrm{ctrl}}=12.8\%$ after the
reference correction. The corresponding adjusted difference is $-0.60\%$ of
baseline local loss. For shared scarring, the raw reduction is $25.0\%$ and the
adjusted contrast is equivalent to $27.3\%$ of control replacement cost.

\paragraph{Student sensitivity before intervention.}
Frozen cross-fitted linear pathology/control directions were originally fitted on
full-context final student features, with standardization and regularization
selected inside the training folds. Undoing the training standardization and
normalizing the coefficient yields a unit direction in raw student-feature
space. We apply each held-out-fold direction to the actual pre-replacement
masked-context features. A region is pooled with its visible voxel-overlap
weights; scores are averaged over eligible events per pair. Each pair supplies
one pathology and one control score, giving prevalence 0.5. The diagnostic reports
AUROC, positive-class AP and Skill $2\mathrm{AP}-1$, with 2,000 group-bootstrap
AUROC intervals. It requires at least 100 pairs, at least ten pairs per outer
fold, AUROC above 0.60 and lower interval bound above 0.55. Shared-target support
is assessed separately using the same frozen directions and criteria.
\begin{table}[tbp]
\centering\scriptsize
\setlength{\tabcolsep}{2pt}
\caption{Actual pre-replacement masked-student diagnostic on own (O) and shared (S) event support, using unchanged held-out directions. Each pair supplies one pathology and one control (prevalence 0.5). Shared positive-class AP and Skill are point estimates; AUROC intervals use 2,000 group resamples.}
\label{tab:e6-student}
\begin{tabular}{@{}llrrrrl@{}}
\toprule
Family & Pairs O/S & AUROC O & AUROC S & AP S & Skill S & Gate S \\
\midrule
ATLAS kidney & 62/60 & \eSixMetric{$0.681$}{$[0.599, 0.767]$} & \eSixMetric{$0.679$}{$[0.588, 0.771]$} & $0.677$ & $0.353$ & Support fail \\
ATLAS liver & 405/402 & \eSixMetric{$0.801$}{$[0.772, 0.835]$} & \eSixMetric{$0.801$}{$[0.770, 0.834]$} & $0.819$ & $0.638$ & Pass \\
ReX scarring/fibrosis & 243/242 & \eSixMetric{$0.950$}{$[0.931, 0.968]$} & \eSixMetric{$0.950$}{$[0.930, 0.967]$} & $0.946$ & $0.893$ & Pass \\
\shortstack[l]{ReX atelectasis/\\consolidation} & 282/280 & \eSixMetric{$0.930$}{$[0.902, 0.955]$} & \eSixMetric{$0.931$}{$[0.903, 0.956]$} & $0.929$ & $0.858$ & Pass \\
ReX ground glass & 447/444 & \eSixMetric{$0.909$}{$[0.885, 0.929]$} & \eSixMetric{$0.908$}{$[0.885, 0.928]$} & $0.911$ & $0.823$ & Pass \\
ReX nodules/masses & 290/288 & \eSixMetric{$0.684$}{$[0.649, 0.720]$} & \eSixMetric{$0.681$}{$[0.645, 0.717]$} & $0.694$ & $0.388$ & Pass \\
\bottomrule
\end{tabular}
\end{table}

Liver and all four ReX categories pass in both cohorts. Kidney has detectable
separation but fails total/fold support. Thus a small relative effect does not
simply imply an inability of the student to distinguish the regions. This gate
does not identify what the predictor uses, and replacement after encoding leaves
information already propagated through student self-attention available in other
tokens. It differs from the full-view EMA-teacher gate in
Table~\ref{tab:e3-teacher}.

\paragraph{Shared versus own targets on common events.}
The same interventions are evaluated on each role's neighbourhood and on their
intersection. Restricting the own-neighbourhood analysis to the shared-target
eligible events leaves its estimates close to the own-neighbourhood values
(Table~\ref{tab:e6-targets}).
\begin{table}[tbp]
\centering\scriptsize
\setlength{\tabcolsep}{3pt}
\caption{Own and shared target readouts on identical eligible pair/mask events. Values are adjusted contrasts in $10^{-3}$ loss units with pointwise intervals. The events and interventions are shared; evaluated targets differ.}
\label{tab:e6-targets}
\begin{tabular}{@{}lrrrr@{}}
\toprule
Family & Pairs & Events & Own targets & Shared targets \\
\midrule
ATLAS kidney & 60 & 555 & \eSixMetric{$-1.147$}{$[-2.764, 0.487]$} & \eSixMetric{$-1.525$}{$[-3.270, 0.252]$} \\
ATLAS liver & 402 & 4900 & \eSixMetric{$-0.994$}{$[-1.726, -0.327]$} & \eSixMetric{$-2.124$}{$[-3.052, -1.273]$} \\
ReX scarring/fibrosis & 242 & 2784 & \eSixMetric{$-1.131$}{$[-2.369, 0.069]$} & \eSixMetric{$-8.064$}{$[-9.900, -6.346]$} \\
\shortstack[l]{ReX atelectasis/\\consolidation} & 280 & 3353 & \eSixMetric{$-1.742$}{$[-2.995, -0.502]$} & \eSixMetric{$-5.535$}{$[-7.192, -3.890]$} \\
ReX ground glass & 444 & 5146 & \eSixMetric{$-0.457$}{$[-1.337, 0.465]$} & \eSixMetric{$-3.847$}{$[-4.935, -2.713]$} \\
ReX nodules/masses & 288 & 2918 & \eSixMetric{$-0.182$}{$[-0.949, 0.588]$} & \eSixMetric{$-0.975$}{$[-1.937, -0.042]$} \\
\bottomrule
\end{tabular}
\end{table}

\paragraph{Replacement and magnitude sensitivities.}
Empirical-donor replacement uses four samples of actual eligible donor
activations, with replacement, instead of the donor mean. Samples are coupled
across roles and corresponding references; each role receives as many values as
its visible token count. The original donor/reference separation is retained.
Table~\ref{tab:e6-replacements} reports this alternative finite intervention
with its own reference-adjusted contrast. It preserves negative local contrasts with pointwise intervals below
zero for liver, scarring, atelectasis and ground glass. The shared-target version
also gives a negative nodule interval, but this is an auxiliary comparison.

For each mean-replacement arm we record the Frobenius change norm, per-token
norms and per-feature RMS. The latter equals the Frobenius norm divided by
$\sqrt{768|R_r|}$. Pathology RMS is slightly larger than control RMS for liver and
all ReX categories; whole-region norms additionally reflect different token counts.
Table~\ref{tab:e6-magnitudes} reports the paired magnitude differences; average
norm similarity does not establish event-level matching. An equal-norm
sensitivity sets $c=\min(\|\bar h-h_{R_+}\|_F,\|\bar h-h_{R_-}\|_F)$ and
moves each role along its own mean-replacement direction by
$\alpha_r=c/\|\bar h-h_{R_r}\|_F$. The realized mean relative norm mismatch is
approximately $2\times10^{-8}$. This sensitivity reports raw effects; no reference
subtraction is implied for these scaled interventions.
\begin{table}[tbp]
\centering\scriptsize
\setlength{\tabcolsep}{3pt}
\caption{Replacement sensitivities in $10^{-3}$ loss units, with pointwise intervals. Donor: reference-adjusted contrast using sampled donor activations. Equal norm: raw contrast after scaling to a common Frobenius norm. Both retain the respective own/shared support in Table~\ref{tab:e6-support}.}
\label{tab:e6-replacements}
\begin{tabular}{@{}lrrrr@{}}
\toprule
Family & Donor own & Donor shared & Equal norm own & Equal norm shared \\
\midrule
ATLAS kidney & \eSixMetric{$-0.351$}{$[-1.473, 0.807]$} & \eSixMetric{$-0.523$}{$[-1.773, 0.742]$} & \eSixMetric{$-0.614$}{$[-1.921, 0.732]$} & \eSixMetric{$-0.791$}{$[-2.167, 0.613]$} \\
ATLAS liver & \eSixMetric{$-0.944$}{$[-1.441, -0.487]$} & \eSixMetric{$-1.438$}{$[-1.994, -0.927]$} & \eSixMetric{$-0.756$}{$[-1.361, -0.196]$} & \eSixMetric{$-1.404$}{$[-2.094, -0.736]$} \\
ReX scarring/fibrosis & \eSixMetric{$-1.017$}{$[-1.690, -0.385]$} & \eSixMetric{$-4.316$}{$[-5.249, -3.427]$} & \eSixMetric{$-0.979$}{$[-2.011, 0.021]$} & \eSixMetric{$-6.257$}{$[-7.839, -4.752]$} \\
\shortstack[l]{ReX atelectasis/\\consolidation} & \eSixMetric{$-1.293$}{$[-1.931, -0.664]$} & \eSixMetric{$-3.087$}{$[-3.934, -2.288]$} & \eSixMetric{$-0.875$}{$[-1.955, 0.141]$} & \eSixMetric{$-3.969$}{$[-5.281, -2.642]$} \\
ReX ground glass & \eSixMetric{$-0.746$}{$[-1.178, -0.303]$} & \eSixMetric{$-2.178$}{$[-2.737, -1.626]$} & \eSixMetric{$-0.155$}{$[-0.960, 0.662]$} & \eSixMetric{$-2.758$}{$[-3.709, -1.763]$} \\
ReX nodules/masses & \eSixMetric{$-0.253$}{$[-0.632, 0.130]$} & \eSixMetric{$-0.561$}{$[-1.035, -0.098]$} & \eSixMetric{$0.112$}{$[-0.582, 0.837]$} & \eSixMetric{$-0.522$}{$[-1.377, 0.311]$} \\
\bottomrule
\end{tabular}
\end{table}
\begin{table}[tbp]
\centering\scriptsize
\setlength{\tabcolsep}{3pt}
\caption{Pathology-minus-control perturbation differences on the own-neighbourhood cohort. RMS columns are scaled by $10^3$; Frobenius differences are unscaled. These are activation-change norms, not prediction loss. Magnitudes are identical between BF16 and FP32 replay.}
\label{tab:e6-magnitudes}
\begin{tabular}{@{}lrrr@{}}
\toprule
Family & Region RMS $\times10^3$ & Region Frobenius & Reference RMS $\times10^3$ \\
\midrule
ATLAS kidney & \eSixMetric{$-1.250$}{$[-10.811, 9.199]$} & \eSixMetric{$-0.595$}{$[-1.577, 0.486]$} & \eSixMetric{$-0.454$}{$[-1.514, 0.615]$} \\
ATLAS liver & \eSixMetric{$12.203$}{$[8.566, 15.813]$} & \eSixMetric{$0.651$}{$[0.160, 1.157]$} & \eSixMetric{$0.001$}{$[-0.320, 0.320]$} \\
ReX scarring/fibrosis & \eSixMetric{$24.470$}{$[19.295, 29.904]$} & \eSixMetric{$0.692$}{$[-0.099, 1.493]$} & \eSixMetric{$-0.150$}{$[-0.612, 0.289]$} \\
\shortstack[l]{ReX atelectasis/\\consolidation} & \eSixMetric{$16.475$}{$[12.499, 20.517]$} & \eSixMetric{$0.348$}{$[-0.539, 1.204]$} & \eSixMetric{$0.039$}{$[-0.323, 0.432]$} \\
ReX ground glass & \eSixMetric{$9.765$}{$[6.678, 12.899]$} & \eSixMetric{$0.140$}{$[-0.445, 0.713]$} & \eSixMetric{$-0.060$}{$[-0.353, 0.224]$} \\
ReX nodules/masses & \eSixMetric{$3.867$}{$[0.280, 7.464]$} & \eSixMetric{$0.002$}{$[-0.520, 0.505]$} & \eSixMetric{$0.118$}{$[-0.283, 0.524]$} \\
\bottomrule
\end{tabular}
\end{table}

\paragraph{Dose response.}
For $\alpha\in\{0.25,0.5,1\}$, intervened activations are
$h'_j=(1-\alpha)h_j+\alpha\bar h$, leaving the other context activations intact.
The quarter- and half-dose results are raw contrasts on the same eligible events
as full replacement. Liver and all ReX categories have positive point estimates
at both partial doses under both target definitions, whereas full replacement
gives negative point estimates. With own neighbourhoods, all ten of those
partial-dose intervals exclude zero. With shared targets, both scarring intervals
and the half-dose nodule interval include zero; the remaining positive intervals
exclude zero (Table~\ref{tab:e6-dose}). These pointwise sensitivities demonstrate
nonlinear relative responses.
\begin{table}[tbp]
\centering\scriptsize
\setlength{\tabcolsep}{3pt}
\caption{Raw pathology-minus-control contrasts under partial and full mean replacement, in $10^{-3}$ loss units with pointwise intervals. Each row retains the same events across doses. No reference adjustment is applied to the partial-dose contrasts.}
\label{tab:e6-dose}
\begin{tabular}{@{}llrrrr@{}}
\toprule
Family & Targets & Pairs & $\alpha=0.25$ & $\alpha=0.5$ & $\alpha=1$ \\
\midrule
ATLAS kidney & Own & 62 & \eSixMetric{$-0.123$}{$[-0.244, -0.010]$} & \eSixMetric{$-0.272$}{$[-0.585, 0.030]$} & \eSixMetric{$-1.566$}{$[-3.157, 0.048]$} \\
ATLAS kidney & Shared & 60 & \eSixMetric{$-0.094$}{$[-0.211, 0.016]$} & \eSixMetric{$-0.257$}{$[-0.571, 0.063]$} & \eSixMetric{$-1.830$}{$[-3.547, -0.154]$} \\
ATLAS liver & Own & 405 & \eSixMetric{$0.108$}{$[0.062, 0.156]$} & \eSixMetric{$0.314$}{$[0.178, 0.454]$} & \eSixMetric{$-1.056$}{$[-1.826, -0.342]$} \\
ATLAS liver & Shared & 402 & \eSixMetric{$0.103$}{$[0.058, 0.149]$} & \eSixMetric{$0.216$}{$[0.074, 0.355]$} & \eSixMetric{$-1.974$}{$[-2.900, -1.136]$} \\
ReX scarring/fibrosis & Own & 243 & \eSixMetric{$0.139$}{$[0.081, 0.199]$} & \eSixMetric{$0.862$}{$[0.628, 1.107]$} & \eSixMetric{$-0.645$}{$[-1.807, 0.519]$} \\
ReX scarring/fibrosis & Shared & 242 & \eSixMetric{$0.057$}{$[-0.009, 0.125]$} & \eSixMetric{$0.155$}{$[-0.111, 0.420]$} & \eSixMetric{$-7.373$}{$[-9.157, -5.703]$} \\
\shortstack[l]{ReX atelectasis/\\consolidation} & Own & 282 & \eSixMetric{$0.239$}{$[0.156, 0.323]$} & \eSixMetric{$0.774$}{$[0.470, 1.072]$} & \eSixMetric{$-1.311$}{$[-2.663, 0.045]$} \\
\shortstack[l]{ReX atelectasis/\\consolidation} & Shared & 280 & \eSixMetric{$0.213$}{$[0.122, 0.310]$} & \eSixMetric{$0.393$}{$[0.056, 0.738]$} & \eSixMetric{$-5.160$}{$[-6.996, -3.379]$} \\
ReX ground glass & Own & 447 & \eSixMetric{$0.189$}{$[0.128, 0.252]$} & \eSixMetric{$0.795$}{$[0.554, 1.042]$} & \eSixMetric{$-0.022$}{$[-1.032, 0.981]$} \\
ReX ground glass & Shared & 444 & \eSixMetric{$0.174$}{$[0.110, 0.240]$} & \eSixMetric{$0.510$}{$[0.251, 0.777]$} & \eSixMetric{$-3.380$}{$[-4.562, -2.216]$} \\
ReX nodules/masses & Own & 290 & \eSixMetric{$0.048$}{$[0.012, 0.083]$} & \eSixMetric{$0.197$}{$[0.053, 0.340]$} & \eSixMetric{$-0.062$}{$[-0.912, 0.741]$} \\
ReX nodules/masses & Shared & 288 & \eSixMetric{$0.043$}{$[0.004, 0.083]$} & \eSixMetric{$0.140$}{$[-0.043, 0.306]$} & \eSixMetric{$-0.802$}{$[-1.815, 0.178]$} \\
\bottomrule
\end{tabular}
\end{table}

\paragraph{Alternative controls and support.}
The original plus three additional controls were evaluated without refitting.
Each control uses the same source-specific clean donor/reference population but
must itself have visible tokens and enough qualifying targets. Each control's
own-neighbourhood point estimate is negative on its available cohort; sample
counts differ substantially (Table~\ref{tab:e6-controls}).
\begin{table}[tbp]
\centering\scriptsize
\setlength{\tabcolsep}{3pt}
\caption{Control-specific adjusted contrasts ($\times10^3$), retained pairs and pointwise intervals. Control 0 repeats the primary reporting intervals from
Table~\ref{tab:e6-endpoints}. Each cell uses its own eligible cohort, so columns do not isolate control choice. Entries with fewer than 100 pairs are descriptive.}
\label{tab:e6-controls}
\begin{tabular}{@{}llrrrr@{}}
\toprule
Family & Targets & Control 0 & Control 1 & Control 2 & Control 3 \\
\midrule
ATLAS kidney & Own & \eSixMetric{$-1.209$ ($n=62$)}{$[-2.836, 0.440]$} & \eSixMetric{$-2.327$ ($n=18$)}{$[-4.764, 0.039]$} & \eSixMetric{$-4.534$ ($n=13$)}{$[-7.558, -1.773]$} & \eSixMetric{$-3.506$ ($n=9$)}{$[-9.077, 1.871]$} \\
ATLAS kidney & Shared & \eSixMetric{$-1.525$ ($n=60$)}{$[-3.270, 0.252]$} & \eSixMetric{$-2.596$ ($n=17$)}{$[-5.109, -0.250]$} & \eSixMetric{$-4.889$ ($n=12$)}{$[-8.193, -1.789]$} & \eSixMetric{$-2.605$ ($n=7$)}{$[-5.459, 0.547]$} \\
ATLAS liver & Own & \eSixMetric{$-0.977$ ($n=405$)}{$[-1.726, -0.302]$} & \eSixMetric{$-1.678$ ($n=178$)}{$[-3.213, -0.343]$} & \eSixMetric{$-1.721$ ($n=165$)}{$[-3.208, -0.469]$} & \eSixMetric{$-2.323$ ($n=146$)}{$[-3.495, -1.276]$} \\
ATLAS liver & Shared & \eSixMetric{$-2.124$ ($n=402$)}{$[-3.052, -1.273]$} & \eSixMetric{$-1.836$ ($n=173$)}{$[-3.335, -0.503]$} & \eSixMetric{$-2.294$ ($n=155$)}{$[-3.564, -1.132]$} & \eSixMetric{$-3.875$ ($n=143$)}{$[-5.463, -2.458]$} \\
ReX scarring/fibrosis & Own & \eSixMetric{$-1.176$ ($n=243$)}{$[-2.410, 0.024]$} & \eSixMetric{$-3.339$ ($n=116$)}{$[-5.481, -1.311]$} & \eSixMetric{$-2.882$ ($n=95$)}{$[-5.084, -0.756]$} & \eSixMetric{$-2.955$ ($n=79$)}{$[-5.588, -0.454]$} \\
ReX scarring/fibrosis & Shared & \eSixMetric{$-8.064$ ($n=242$)}{$[-9.900, -6.346]$} & \eSixMetric{$-8.734$ ($n=115$)}{$[-11.271, -6.180]$} & \eSixMetric{$-9.456$ ($n=94$)}{$[-11.926, -7.061]$} & \eSixMetric{$-9.939$ ($n=76$)}{$[-12.515, -7.454]$} \\
\shortstack[l]{ReX atelectasis/\\consolidation} & Own & \eSixMetric{$-1.743$ ($n=282$)}{$[-2.994, -0.503]$} & \eSixMetric{$-2.373$ ($n=136$)}{$[-4.803, -0.028]$} & \eSixMetric{$-3.024$ ($n=97$)}{$[-5.121, -0.874]$} & \eSixMetric{$-0.486$ ($n=77$)}{$[-2.554, 1.855]$} \\
\shortstack[l]{ReX atelectasis/\\consolidation} & Shared & \eSixMetric{$-5.535$ ($n=280$)}{$[-7.192, -3.890]$} & \eSixMetric{$-3.718$ ($n=125$)}{$[-6.089, -1.154]$} & \eSixMetric{$-5.667$ ($n=95$)}{$[-7.586, -3.752]$} & \eSixMetric{$-3.047$ ($n=77$)}{$[-5.402, -0.548]$} \\
ReX ground glass & Own & \eSixMetric{$-0.545$ ($n=447$)}{$[-1.456, 0.364]$} & \eSixMetric{$-0.797$ ($n=200$)}{$[-2.750, 1.161]$} & \eSixMetric{$-2.284$ ($n=169$)}{$[-3.976, -0.576]$} & \eSixMetric{$-1.408$ ($n=159$)}{$[-3.404, 0.564]$} \\
ReX ground glass & Shared & \eSixMetric{$-3.847$ ($n=444$)}{$[-4.935, -2.713]$} & \eSixMetric{$-3.567$ ($n=192$)}{$[-5.971, -1.243]$} & \eSixMetric{$-4.293$ ($n=163$)}{$[-6.403, -2.310]$} & \eSixMetric{$-5.872$ ($n=152$)}{$[-8.339, -3.560]$} \\
ReX nodules/masses & Own & \eSixMetric{$-0.191$ ($n=290$)}{$[-0.968, 0.586]$} & \eSixMetric{$-1.456$ ($n=83$)}{$[-2.935, 0.085]$} & \eSixMetric{$-0.297$ ($n=85$)}{$[-1.677, 1.105]$} & \eSixMetric{$-0.554$ ($n=91$)}{$[-1.991, 0.831]$} \\
ReX nodules/masses & Shared & \eSixMetric{$-0.975$ ($n=288$)}{$[-1.937, -0.042]$} & \eSixMetric{$-2.909$ ($n=82$)}{$[-4.830, -0.879]$} & \eSixMetric{$-1.994$ ($n=84$)}{$[-3.926, -0.146]$} & \eSixMetric{$-1.716$ ($n=88$)}{$[-3.506, 0.048]$} \\
\bottomrule
\end{tabular}
\end{table}

\paragraph{Own-neighbourhood target token overlap and descriptive covariate adjustment.}
In the own-neighbourhood target case, two voxel-disjoint regions may share a 24\,mm token. Removing such events and
reapplying support requirements retains negative own-neighbourhood point estimates
in all categories, but only liver retains a pointwise interval strictly below zero (kidney: 38 pairs, descriptive).

A separate own-neighbourhood regression uses pair means of the adjusted
contrast, log visible-token-count ratio, log target-count ratio, target-distance
difference divided by 24\,mm, log per-feature RMS ratio and two centred mask-family
proportions. The reported intercept evaluates equal count/distance/RMS covariates
at the observed mask mix, using 2,000 group-bootstrap fits. All design matrices
have rank seven. The intercept becomes uncertain for kidney, liver, scarring and
atelectasis, and positive with pointwise intervals above zero for ground glass
and nodules (Table~\ref{tab:e6-adjustments}).
This regression targets a different estimand from the primary contrast. Adjusting for token counts and target distances removes design imbalances between the two regions, but the per-feature RMS ratio is partly a consequence of region content: for liver and all ReX categories, pathology activations lie farther from the donor mean than control activations (Table~\ref{tab:e6-magnitudes}), so adjusting for magnitude can absorb part of the pathology--control difference itself. Few pairs have near-equal perturbation norms, so the equal-covariate intercept relies on the additive linear specification to extrapolate, although the dose response is nonlinear in perturbation strength (Table~\ref{tab:e6-dose}). The intercept is reported with pointwise intervals and no multiplicity correction. We therefore treat it as a descriptive robustness check.
\begin{table}[tbp]
\centering\scriptsize
\setlength{\tabcolsep}{3pt}
\caption{Own-neighbourhood sensitivities: token-disjoint events and the descriptive covariate-adjusted intercept. Adjusted contrasts are scaled by $10^3$ with pointwise intervals. The regression retains original own-neighbourhood support; its intercept is a model-based contrast at equal count, distance and perturbation-magnitude covariates, without multiplicity correction. Kidney token-disjoint support is below 100 pairs.}
\label{tab:e6-adjustments}
\begin{tabular}{@{}lrr@{}}
\toprule
Family & Token disjoint & Regression intercept \\
\midrule
ATLAS kidney & \eSixMetric{$-2.419$ ($n=38$)}{$[-4.754, 0.036]$} & \eSixMetric{$-0.042$}{$[-1.793, 1.724]$} \\
ATLAS liver & \eSixMetric{$-1.209$ ($n=300$)}{$[-2.143, -0.340]$} & \eSixMetric{$-0.262$}{$[-0.833, 0.308]$} \\
ReX scarring/fibrosis & \eSixMetric{$-1.326$ ($n=201$)}{$[-2.741, 0.039]$} & \eSixMetric{$0.533$}{$[-1.391, 2.786]$} \\
\shortstack[l]{ReX atelectasis/\\consolidation} & \eSixMetric{$-1.297$ ($n=224$)}{$[-2.778, 0.124]$} & \eSixMetric{$0.637$}{$[-1.240, 2.360]$} \\
ReX ground glass & \eSixMetric{$-0.586$ ($n=328$)}{$[-1.712, 0.519]$} & \eSixMetric{$1.210$}{$[0.207, 2.288]$} \\
ReX nodules/masses & \eSixMetric{$-0.080$ ($n=192$)}{$[-1.156, 0.983]$} & \eSixMetric{$0.758$}{$[0.064, 1.500]$} \\
\bottomrule
\end{tabular}
\end{table}

%% file: main.bbl
\begin{thebibliography}{42}
\providecommand{\natexlab}[1]{#1}
\providecommand{\url}[1]{\texttt{#1}}
\expandafter\ifx\csname urlstyle\endcsname\relax
  \providecommand{\doi}[1]{doi: #1}\else
  \providecommand{\doi}{doi: \begingroup \urlstyle{rm}\Url}\fi

\bibitem[Aberle et~al.(2011)]{aberle2011reduced}
Denise~R. Aberle et~al.
\newblock Reduced lung-cancer mortality with low-dose computed tomographic
  screening.
\newblock \emph{New England Journal of Medicine}, 365\penalty0 (5):\penalty0
  395--409, 2011.
\newblock \doi{10.1056/NEJMoa1102873}.
\newblock URL \url{https://doi.org/10.1056/NEJMoa1102873}.
\newblock National Lung Screening Trial Research Team.

\bibitem[Assran et~al.(2023)Assran, Duval, Misra, Bojanowski, Vincent, Rabbat,
  LeCun, and Ballas]{assran2023ijepa}
Mahmoud Assran, Quentin Duval, Ishan Misra, Piotr Bojanowski, Pascal Vincent,
  Michael Rabbat, Yann LeCun, and Nicolas Ballas.
\newblock Self-supervised learning from images with a joint-embedding
  predictive architecture.
\newblock In \emph{Proceedings of the IEEE/CVF Conference on Computer Vision
  and Pattern Recognition (CVPR)}, pp.\  15619--15629, 2023.
\newblock URL
  \url{https://openaccess.thecvf.com/content/CVPR2023/html/Assran_Self-Supervised_Learning_From_Images_With_a_Joint-Embedding_Predictive_Architecture_CVPR_2023_paper.html}.

\bibitem[Assran et~al.(2025)Assran, Bardes, Fan, Garrido, Howes, Komeili,
  Muckley, Rizvi, Roberts, Sinha, Zholus, Arnaud, Gejji, Martin, Hogan, Dugas,
  Bojanowski, Khalidov, Labatut, Massa, Szafraniec, Krishnakumar, Li, Ma,
  Chandar, Meier, LeCun, Rabbat, and Ballas]{assran2025vjepa2}
Mido Assran, Adrien Bardes, David Fan, Quentin Garrido, Russell Howes, Mojtaba
  Komeili, Matthew Muckley, Ammar Rizvi, Claire Roberts, Koustuv Sinha, Artem
  Zholus, Sergio Arnaud, Abha Gejji, Ada Martin, Francois~Robert Hogan, Daniel
  Dugas, Piotr Bojanowski, Vasil Khalidov, Patrick Labatut, Francisco Massa,
  Marc Szafraniec, Kapil Krishnakumar, Yong Li, Xiaodong Ma, Sarath Chandar,
  Franziska Meier, Yann LeCun, Michael Rabbat, and Nicolas Ballas.
\newblock {V-JEPA 2}: Self-supervised video models enable understanding,
  prediction and planning.
\newblock \emph{arXiv preprint arXiv:2506.09985}, 2025.
\newblock \doi{10.48550/arXiv.2506.09985}.
\newblock URL \url{https://arxiv.org/abs/2506.09985}.

\bibitem[Baevski et~al.(2022)Baevski, Hsu, Xu, Babu, Gu, and
  Auli]{baevski2022data2vec}
Alexei Baevski, Wei-Ning Hsu, Qiantong Xu, Arun Babu, Jiatao Gu, and Michael
  Auli.
\newblock {data2vec}: A general framework for self-supervised learning in
  speech, vision and language.
\newblock In \emph{Proceedings of the 39th International Conference on Machine
  Learning}, volume 162 of \emph{Proceedings of Machine Learning Research},
  pp.\  1298--1312. PMLR, 2022.
\newblock URL \url{https://proceedings.mlr.press/v162/baevski22a.html}.

\bibitem[Baharoon et~al.(2026)Baharoon, Luo, Moritz, Kumar, Kim, Zhang, Zhu,
  Alabbad, Alhazmi, Mistry, Bijnens, Kleinschmidt, Chrisler, Suryadevara,
  Jaliparthi, Prudlo, Marino, Palacio, Akula, Zhou, Zhou, Hamamci, Adams,
  Alomaish, and Rajpurkar]{baharoon2025rexgroundingct}
Mohammed Baharoon, Luyang Luo, Michael Moritz, Abhinav Kumar, Sung~Eun Kim,
  Xiaoman Zhang, Miao Zhu, Mahmoud~H. Alabbad, Maha~S. Alhazmi, Neel~P. Mistry,
  Lucas Bijnens, Kent~R. Kleinschmidt, Brady Chrisler, Sathvik Suryadevara, Sri
  Sai~Dinesh Jaliparthi, Noah~M. Prudlo, Mark~D. Marino, Jeremy Palacio,
  Rithvik Akula, Di~Zhou, Hong-Yu Zhou, Ibrahim~E. Hamamci, Scott~J. Adams,
  Hassan~R. Alomaish, and Pranav Rajpurkar.
\newblock {ReXGroundingCT}: A {3D} chest {CT} dataset for segmentation of
  findings from free-text reports.
\newblock \emph{NEJM AI}, 3\penalty0 (7):\penalty0 AIdbp2501220, 2026.
\newblock \doi{10.1056/aidbp2501220}.
\newblock URL \url{https://doi.org/10.1056/aidbp2501220}.

\bibitem[Bardes et~al.(2024)Bardes, Garrido, Ponce, Chen, Rabbat, LeCun,
  Assran, and Ballas]{bardes2024vjepa}
Adrien Bardes, Quentin Garrido, Jean Ponce, Xinlei Chen, Michael Rabbat, Yann
  LeCun, Mido Assran, and Nicolas Ballas.
\newblock Revisiting feature prediction for learning visual representations
  from video.
\newblock \emph{Transactions on Machine Learning Research}, 2024.
\newblock URL \url{https://openreview.net/forum?id=QaCCuDfBk2}.

\bibitem[Bassi et~al.(2025)Bassi, Yavuz, Hamamci, Er, Chen, Li, Menze,
  Decherchi, Cavalli, Wang, Yang, Yuille, and Zhou]{bassi2025radgpt}
Pedro R. A.~S. Bassi, Mehmet~Can Yavuz, Ibrahim~Ethem Hamamci, Sezgin Er,
  Xiaoxi Chen, Wenxuan Li, Bjoern Menze, Sergio Decherchi, Andrea Cavalli, Kang
  Wang, Yang Yang, Alan Yuille, and Zongwei Zhou.
\newblock {RadGPT}: Constructing {3D} image-text tumor datasets.
\newblock In \emph{Proceedings of the IEEE/CVF International Conference on
  Computer Vision (ICCV)}, pp.\  23720--23730, 2025.
\newblock URL
  \url{https://openaccess.thecvf.com/content/ICCV2025/html/Bassi_RadGPT_Constructing_3D_Image-Text_Tumor_Datasets_ICCV_2025_paper.html}.

\bibitem[Benjamini \& Hochberg(1995)Benjamini and Hochberg]{benjamini1995fdr}
Yoav Benjamini and Yosef Hochberg.
\newblock Controlling the false discovery rate: A practical and powerful
  approach to multiple testing.
\newblock \emph{Journal of the Royal Statistical Society: Series B
  (Methodological)}, 57\penalty0 (1):\penalty0 289--300, 1995.
\newblock \doi{10.1111/j.2517-6161.1995.tb02031.x}.
\newblock URL \url{https://doi.org/10.1111/j.2517-6161.1995.tb02031.x}.

\bibitem[Blankemeier et~al.(2026)Blankemeier, Kumar, Cohen, Liu, Liu, Van~Veen,
  Gardezi, Yu, Paschali, Chen, Delbrouck, Reis, Holland, Truyts, Bluethgen, Wu,
  Lian, Jensen, Ostmeier, Varma, Valanarasu, Fang, Huo, Nabulsi, Ardila, Weng,
  Amaro~Junior, Ahuja, Fries, Shah, Zaharchuk, Willis, Yala, Johnston, Boutin,
  Wentland, Langlotz, Hom, Gatidis, and Chaudhari]{blankemeier2026merlin}
Louis Blankemeier, Ashwin Kumar, Joseph~Paul Cohen, Jiaming Liu, Longchao Liu,
  Dave Van~Veen, Syed Jamal~Safdar Gardezi, Hongkun Yu, Magdalini Paschali,
  Zhihong Chen, Jean-Benoit Delbrouck, Eduardo Reis, Robbie Holland, Cesar
  Truyts, Christian Bluethgen, Yufu Wu, Long Lian, Malte Engmann~Kjeldskov
  Jensen, Sophie Ostmeier, Maya Varma, Jeya Maria~Jose Valanarasu, Zhongnan
  Fang, Zepeng Huo, Zaid Nabulsi, Diego Ardila, Wei-Hung Weng, Edson
  Amaro~Junior, Neera Ahuja, Jason Fries, Nigam~H. Shah, Greg Zaharchuk, Marc
  Willis, Adam Yala, Andrew Johnston, Robert~D. Boutin, Andrew Wentland,
  Curtis~P. Langlotz, Jason Hom, Sergios Gatidis, and Akshay~S. Chaudhari.
\newblock Merlin: a computed tomography vision--language foundation model and
  dataset.
\newblock \emph{Nature}, 652:\penalty0 1318--1328, 2026.
\newblock \doi{10.1038/s41586-026-10181-8}.
\newblock URL \url{https://doi.org/10.1038/s41586-026-10181-8}.

\bibitem[Buess et~al.(2026)Buess, Kurin, Bhandary~Panambur, Arias-Vergara, and
  Maier]{buess2026alo}
Lukas Buess, Robert Kurin, Adarsh Bhandary~Panambur, Tomas Arias-Vergara, and
  Andreas Maier.
\newblock {ALO}: Addressing class imbalance in radiology report generation
  through anatomy-level oversampling.
\newblock In \emph{Proceedings of The 9th International Conference on Medical
  Imaging with Deep Learning}, volume 315 of \emph{Proceedings of Machine
  Learning Research}, pp.\  2998--3017. PMLR, 2026.
\newblock URL \url{https://proceedings.mlr.press/v315/buess26a.html}.

\bibitem[Cameron et~al.(2008)Cameron, Gelbach, and
  Miller]{cameron2008bootstrap}
A.~Colin Cameron, Jonah~B. Gelbach, and Douglas~L. Miller.
\newblock Bootstrap-based improvements for inference with clustered errors.
\newblock \emph{The Review of Economics and Statistics}, 90\penalty0
  (3):\penalty0 414--427, 2008.
\newblock \doi{10.1162/rest.90.3.414}.
\newblock URL \url{https://doi.org/10.1162/rest.90.3.414}.

\bibitem[Chevli et~al.(2026)Chevli, Brandt, Braren, Rueckert, and
  M{\"u}ller]{chevli2026big}
Maulik Chevli, Johannes Brandt, Rickmer Braren, Daniel Rueckert, and Philip
  M{\"u}ller.
\newblock Big, bright, or invisible: A frozen-feature benchmark of {3D CT}
  foundation models.
\newblock \emph{arXiv preprint arXiv:2608.05960}, 2026.
\newblock \doi{10.48550/arXiv.2608.05960}.
\newblock URL \url{https://arxiv.org/abs/2608.05960}.

\bibitem[Clark et~al.(2013)Clark, Vendt, Smith, Freymann, Kirby, Koppel, Moore,
  Phillips, Maffitt, Pringle, Tarbox, and Prior]{clark2013tcia}
Kenneth Clark, Bruce Vendt, Kirk Smith, John Freymann, Justin Kirby, Paul
  Koppel, Stephen Moore, Stanley Phillips, David Maffitt, Michael Pringle,
  Lawrence Tarbox, and Fred Prior.
\newblock The {Cancer Imaging Archive (TCIA)}: Maintaining and operating a
  public information repository.
\newblock \emph{Journal of Digital Imaging}, 26\penalty0 (6):\penalty0
  1045--1057, 2013.
\newblock \doi{10.1007/s10278-013-9622-7}.
\newblock URL \url{https://doi.org/10.1007/s10278-013-9622-7}.

\bibitem[Dinh \& Vo(2026)Dinh and Vo]{dinh2026paradox}
Anh Trac~Duc Dinh and Khang Nhat~Hoang Vo.
\newblock The {JEPA} paradox in language: The geometry of linguistic
  alternatives.
\newblock \emph{arXiv preprint arXiv:2607.23531}, 2026.
\newblock \doi{10.48550/arXiv.2607.23531}.
\newblock URL \url{https://arxiv.org/abs/2607.23531}.

\bibitem[Dosovitskiy et~al.(2021)Dosovitskiy, Beyer, Kolesnikov, Weissenborn,
  Zhai, Unterthiner, Dehghani, Minderer, Heigold, Gelly, Uszkoreit, and
  Houlsby]{dosovitskiy2021vit}
Alexey Dosovitskiy, Lucas Beyer, Alexander Kolesnikov, Dirk Weissenborn,
  Xiaohua Zhai, Thomas Unterthiner, Mostafa Dehghani, Matthias Minderer, Georg
  Heigold, Sylvain Gelly, Jakob Uszkoreit, and Neil Houlsby.
\newblock An image is worth 16x16 words: Transformers for image recognition at
  scale.
\newblock In \emph{International Conference on Learning Representations}, 2021.
\newblock URL \url{https://openreview.net/forum?id=YicbFdNTTy}.

\bibitem[Draelos et~al.(2020)Draelos, Dov, Mazurowski, Lo, Henao, Rubin, and
  Carin]{draelos2020radchestctdata}
Rachel~Lea Draelos, David Dov, Maciej~A. Mazurowski, Joseph~Y. Lo, Ricardo
  Henao, Geoffrey~D. Rubin, and Lawrence Carin.
\newblock {RAD-ChestCT Dataset (Version 1.0)}.
\newblock Dataset, Zenodo, 2020.
\newblock URL \url{https://doi.org/10.5281/zenodo.6406114}.

\bibitem[Draelos et~al.(2021)Draelos, Dov, Mazurowski, Lo, Henao, Rubin, and
  Carin]{draelos2021radchestct}
Rachel~Lea Draelos, David Dov, Maciej~A. Mazurowski, Joseph~Y. Lo, Ricardo
  Henao, Geoffrey~D. Rubin, and Lawrence Carin.
\newblock Machine-learning-based multiple abnormality prediction with
  large-scale chest computed tomography volumes.
\newblock \emph{Medical Image Analysis}, 67:\penalty0 101857, 2021.
\newblock \doi{10.1016/j.media.2020.101857}.
\newblock URL \url{https://doi.org/10.1016/j.media.2020.101857}.

\bibitem[Gneiting(2011)]{gneiting2011point}
Tilmann Gneiting.
\newblock Making and evaluating point forecasts.
\newblock \emph{Journal of the American Statistical Association}, 106\penalty0
  (494):\penalty0 746--762, 2011.
\newblock \doi{10.1198/jasa.2011.r10138}.
\newblock URL \url{https://doi.org/10.1198/jasa.2011.r10138}.

\bibitem[Hamamci et~al.(2026)Hamamci, Er, Wang, Almas, Simsek, Esirgun, Dogan,
  Durugol, Hou, Shit, Dai, Xu, Reynaud, Dasdelen, Wittmann, Amiranashvili,
  Simsar, Simsar, Erdemir, Alanbay, Sekuboyina, Lafci, Kaplan, Lu, Polacin,
  Kainz, Bluethgen, Batmanghelich, Ozdemir, and Menze]{hamamci2024ctclip}
Ibrahim~Ethem Hamamci, Sezgin Er, Chenyu Wang, Furkan Almas, Ayse~Gulnihan
  Simsek, Sevval~Nil Esirgun, Irem Dogan, Omer~Faruk Durugol, Benjamin Hou,
  Suprosanna Shit, Weicheng Dai, Murong Xu, Hadrien Reynaud, Muhammed~Furkan
  Dasdelen, Bastian Wittmann, Tamaz Amiranashvili, Enis Simsar, Mehmet Simsar,
  Emine~Bensu Erdemir, Abdullah Alanbay, Anjany Sekuboyina, Berkan Lafci, Ahmet
  Kaplan, Zhiyong Lu, Malgorzata Polacin, Bernhard Kainz, Christian Bluethgen,
  Kayhan Batmanghelich, Mehmet~Kemal Ozdemir, and Bjoern Menze.
\newblock Generalist foundation models from a multimodal dataset for {3D}
  computed tomography.
\newblock \emph{Nature Biomedical Engineering}, 10\penalty0 (8):\penalty0
  1610--1628, 2026.
\newblock \doi{10.1038/s41551-025-01599-y}.
\newblock URL \url{https://doi.org/10.1038/s41551-025-01599-y}.

\bibitem[Holm(1979)]{holm1979simple}
Sture Holm.
\newblock A simple sequentially rejective multiple test procedure.
\newblock \emph{Scandinavian Journal of Statistics}, 6\penalty0 (2):\penalty0
  65--70, 1979.
\newblock URL \url{https://www.jstor.org/stable/4615733}.

\bibitem[Huang et~al.(2026)Huang, Chen, Chen, Hyun, Loftus, Melmed, Orringer,
  Frontera, Dehkharghani, Masurkar, and Razavian]{huang2026neurojepa}
Haoxu Huang, Long Chen, Jingyun Chen, Jinu Hyun, James~Ryan Loftus, Kara
  Melmed, Daniel Orringer, Jennifer Frontera, Seena Dehkharghani, Arjun
  Masurkar, and Narges Razavian.
\newblock Learning sparse latent predictive foundation model for multimodal
  neuroimaging.
\newblock \emph{arXiv preprint arXiv:2606.14957}, 2026.
\newblock \doi{10.48550/arXiv.2606.14957}.
\newblock URL \url{https://arxiv.org/abs/2606.14957}.

\bibitem[Ilse et~al.(2018)Ilse, Tomczak, and Welling]{ilse2018attention}
Maximilian Ilse, Jakub Tomczak, and Max Welling.
\newblock Attention-based deep multiple instance learning.
\newblock In \emph{Proceedings of the 35th International Conference on Machine
  Learning}, volume~80 of \emph{Proceedings of Machine Learning Research}, pp.\
   2127--2136. PMLR, 2018.
\newblock URL \url{https://proceedings.mlr.press/v80/ilse18a.html}.

\bibitem[Ivezi{\'c} et~al.(2026)Ivezi{\'c}, Pleasure, Radhachandran,
  Panchavati, Athreya, Sant, Emert, Fishbein, Arnold, and
  Speier]{ivezic2026pretext}
Vedrana Ivezi{\'c}, Mara Pleasure, Ashwath Radhachandran, Saarang Panchavati,
  Shreeram Athreya, Vivek Sant, Benjamin Emert, Gregory Fishbein, Corey Arnold,
  and William Speier.
\newblock Pretext matters: An empirical study of {SSL} methods in medical
  imaging.
\newblock \emph{arXiv preprint arXiv:2603.22649}, 2026.
\newblock \doi{10.48550/arXiv.2603.22649}.
\newblock URL \url{https://arxiv.org/abs/2603.22649}.

\bibitem[Khan et~al.(2026)Khan, Husain, Jalan, and Jadhav]{khan2026radjepa}
Anas Anwarul~Haq Khan, Mariam Husain, Pratik Jalan, and Kshitij Jadhav.
\newblock {RadJEPA}: Radiology encoder for chest {X}-rays via joint embedding
  predictive architecture.
\newblock \emph{arXiv preprint arXiv:2601.15891}, 2026.
\newblock \doi{10.48550/arXiv.2601.15891}.
\newblock URL \url{https://arxiv.org/abs/2601.15891v4}.
\newblock Accepted at EMNLP 2026.

\bibitem[Kondepudi et~al.(2026)Kondepudi, Rao, Zhao, Lyu, Harake, Banerjee,
  Ogle, Joshi, Meissner, Hou, Jiang, Chowdury, Srinivasan, Athey, Gulani,
  Pandey, Lee, and Hollon]{kondepudi2026neurovfm}
Akhil Kondepudi, Akshay Rao, Chenhui Zhao, Yiwei Lyu, Samir Harake, Soumyanil
  Banerjee, Jacob Ogle, Rushikesh Joshi, Anna-Katharina Meissner, Xinhai Hou,
  Cheng Jiang, Asadur Chowdury, Ashok Srinivasan, Brian Athey, Vikas Gulani,
  Aditya Pandey, Honglak Lee, and Todd Hollon.
\newblock Health system learning enables generalist neuroimaging models.
\newblock \emph{Nature Medicine}, 32:\penalty0 2831--2837, 2026.
\newblock \doi{10.1038/s41591-026-04497-1}.
\newblock URL \url{https://www.nature.com/articles/s41591-026-04497-1}.

\bibitem[Littwin et~al.(2024)Littwin, Saremi, Advani, Thilak, Nakkiran, Huang,
  and Susskind]{littwin2024noisy}
Etai Littwin, Omid Saremi, Madhu Advani, Vimal Thilak, Preetum Nakkiran, Chen
  Huang, and Joshua Susskind.
\newblock How {JEPA} avoids noisy features: The implicit bias of deep linear
  self distillation networks.
\newblock In \emph{Advances in Neural Information Processing Systems},
  volume~37, pp.\  91300--91336, 2024.
\newblock \doi{10.52202/079017-2897}.
\newblock URL
  \url{https://proceedings.neurips.cc/paper_files/paper/2024/hash/a600f0f740605205133553cb74a1c107-Abstract-Conference.html}.

\bibitem[Loshchilov \& Hutter(2019)Loshchilov and Hutter]{loshchilov2019adamw}
Ilya Loshchilov and Frank Hutter.
\newblock Decoupled weight decay regularization.
\newblock In \emph{International Conference on Learning Representations}, 2019.
\newblock URL \url{https://openreview.net/forum?id=Bkg6RiCqY7}.

\bibitem[McInnes et~al.(2018)McInnes, Healy, and Melville]{mcinnes2018umap}
Leland McInnes, John Healy, and James Melville.
\newblock {UMAP}: Uniform manifold approximation and projection for dimension
  reduction.
\newblock \emph{arXiv preprint arXiv:1802.03426}, 2018.
\newblock \doi{10.48550/arXiv.1802.03426}.
\newblock URL \url{https://arxiv.org/abs/1802.03426}.

\bibitem[Moreno-Aguado et~al.(2026)Moreno-Aguado, Magall{\'o}n, Moreno, Fang,
  and Yang]{morenoaguado2026voxelfm}
Rub{\'e}n Moreno-Aguado, Alba Magall{\'o}n, Victor Moreno, Yingying Fang, and
  Guang Yang.
\newblock Learning robust visual features in computed tomography enables
  efficient transfer learning for clinical tasks.
\newblock \emph{arXiv preprint arXiv:2604.04133}, 2026.
\newblock \doi{10.48550/arXiv.2604.04133}.
\newblock URL \url{https://arxiv.org/abs/2604.04133}.

\bibitem[{National Lung Screening Trial Research Team}(2011)]{nlst2011}
{National Lung Screening Trial Research Team}.
\newblock The {National Lung Screening Trial}: overview and study design.
\newblock \emph{Radiology}, 258\penalty0 (1):\penalty0 243--253, 2011.
\newblock \doi{10.1148/radiol.10091808}.
\newblock URL \url{https://doi.org/10.1148/radiol.10091808}.

\bibitem[{National Lung Screening Trial Research Team}(2013)]{nlst2013tciadata}
{National Lung Screening Trial Research Team}.
\newblock Data from the {National Lung Screening Trial (NLST)}.
\newblock The Cancer Imaging Archive, 2013.
\newblock URL \url{https://doi.org/10.7937/TCIA.HMQ8-J677}.

\bibitem[Pendharkar(2026)]{pendharkar2026predictive}
Ayan Pendharkar.
\newblock Predictive objectives discard exogenous control-relevant features: A
  controlled mechanistic study.
\newblock \emph{arXiv preprint arXiv:2606.30068}, 2026.
\newblock \doi{10.48550/arXiv.2606.30068}.
\newblock URL \url{https://arxiv.org/abs/2606.30068}.

\bibitem[Radhachandran et~al.(2026)Radhachandran, Ivezi{\'c}, Athreya, Arnold,
  and Speier]{radhachandran2026usjepa}
Ashwath Radhachandran, Vedrana Ivezi{\'c}, Shreeram Athreya, Corey~W. Arnold,
  and William Speier.
\newblock {US-JEPA}: A joint embedding predictive architecture for ultrasound.
\newblock \emph{arXiv preprint arXiv:2602.19322}, 2026.
\newblock \doi{10.48550/arXiv.2602.19322}.
\newblock URL \url{https://arxiv.org/abs/2602.19322v2}.

\bibitem[Rahimi \& Recht(2007)Rahimi and Recht]{rahimi2007random}
Ali Rahimi and Benjamin Recht.
\newblock Random features for large-scale kernel machines.
\newblock In \emph{Advances in Neural Information Processing Systems},
  volume~20, 2007.
\newblock URL
  \url{https://proceedings.neurips.cc/paper/2007/hash/013a006f03dbc5392effeb8f18fda755-Abstract.html}.

\bibitem[Sobal et~al.(2022)Sobal, {Jyothir S V}, Jalagam, Carion, Cho, and
  LeCun]{sobal2022slow}
Vlad Sobal, {Jyothir S V}, Siddhartha Jalagam, Nicolas Carion, Kyunghyun Cho,
  and Yann LeCun.
\newblock Joint embedding predictive architectures focus on slow features.
\newblock \emph{arXiv preprint arXiv:2211.10831}, 2022.
\newblock URL \url{https://arxiv.org/abs/2211.10831}.

\bibitem[Su et~al.(2024)Su, Ahmed, Lu, Pan, Bo, and Liu]{su2024roformer}
Jianlin Su, Murtadha Ahmed, Yu~Lu, Shengfeng Pan, Wen Bo, and Yunfeng Liu.
\newblock {RoFormer}: Enhanced transformer with rotary position embedding.
\newblock \emph{Neurocomputing}, 568:\penalty0 127063, 2024.
\newblock \doi{10.1016/j.neucom.2023.127063}.
\newblock URL \url{https://doi.org/10.1016/j.neucom.2023.127063}.

\bibitem[Touvron et~al.(2021)Touvron, Cord, Sablayrolles, Synnaeve, and
  J{\'e}gou]{touvron2021cait}
Hugo Touvron, Matthieu Cord, Alexandre Sablayrolles, Gabriel Synnaeve, and
  Herv{\'e} J{\'e}gou.
\newblock Going deeper with image transformers.
\newblock In \emph{Proceedings of the IEEE/CVF International Conference on
  Computer Vision (ICCV)}, pp.\  32--42, 2021.
\newblock URL
  \url{https://openaccess.thecvf.com/content/ICCV2021/html/Touvron_Going_Deeper_With_Image_Transformers_ICCV_2021_paper.html}.

\bibitem[Trinh et~al.(2026)Trinh, Nguyen, and Bagci]{trinh2026radjepa3d}
Quoc-Huy Trinh, Minh-Van Nguyen, and Ulas Bagci.
\newblock {Rad-JEPA 3D}: Radiology joint-embedding predictive model for {3D}
  computed tomography.
\newblock \emph{arXiv preprint arXiv:2607.26196}, 2026.
\newblock \doi{10.48550/arXiv.2607.26196}.
\newblock URL \url{https://arxiv.org/abs/2607.26196}.

\bibitem[Van~Assel et~al.(2025)Van~Assel, Ibrahim, Biancalani, Regev, and
  Balestriero]{vanassel2025joint}
Hugues Van~Assel, Mark Ibrahim, Tommaso Biancalani, Aviv Regev, and Randall
  Balestriero.
\newblock Joint-embedding vs reconstruction: Provable benefits of latent space
  prediction for self-supervised learning.
\newblock In \emph{Advances in Neural Information Processing Systems},
  volume~38, pp.\  21897--21937, 2025.
\newblock \doi{10.52202/085713-0740}.
\newblock URL
  \url{https://proceedings.neurips.cc/paper_files/paper/2025/hash/1fa81061d6d4d7fea88f803d89ae9d6e-Abstract-Conference.html}.

\bibitem[Wald et~al.(2025)Wald, Hamamci, Gao, Bond-Taylor, Sharma, Ilse, Lo,
  Melnichenko, Schwaighofer, Codella, Wetscherek, Maier-Hein, Korfiatis,
  Salvatelli, Alvarez-Valle, and P{\'e}rez-Garc{\'i}a]{wald2026colipri}
Tassilo Wald, Ibrahim~Ethem Hamamci, Yuan Gao, Sam Bond-Taylor, Harshita
  Sharma, Maximilian Ilse, Cynthia Lo, Olesya Melnichenko, Anton Schwaighofer,
  Noel C.~F. Codella, Maria~Teodora Wetscherek, Klaus~H. Maier-Hein, Panagiotis
  Korfiatis, Valentina Salvatelli, Javier Alvarez-Valle, and Fernando
  P{\'e}rez-Garc{\'i}a.
\newblock Comprehensive language-image pre-training for {3D} medical image
  understanding.
\newblock \emph{arXiv preprint arXiv:2510.15042}, 2025.
\newblock \doi{10.48550/arXiv.2510.15042}.
\newblock URL \url{https://arxiv.org/abs/2510.15042v3}.
\newblock Version 3, revised August 2026.

\bibitem[Wasserthal et~al.(2023)Wasserthal, Breit, Meyer, Pradella, Hinck,
  Sauter, Heye, Boll, Cyriac, Yang, Bach, and
  Segeroth]{wasserthal2023totalsegmentator}
Jakob Wasserthal, Hanns-Christian Breit, Manfred~T. Meyer, Maurice Pradella,
  Daniel Hinck, Alexander~W. Sauter, Tobias Heye, Daniel~T. Boll, Joshy Cyriac,
  Shan Yang, Michael Bach, and Martin Segeroth.
\newblock {TotalSegmentator}: Robust segmentation of 104 anatomic structures in
  {CT} images.
\newblock \emph{Radiology: Artificial Intelligence}, 5\penalty0 (5):\penalty0
  e230024, 2023.
\newblock \doi{10.1148/ryai.230024}.
\newblock URL \url{https://doi.org/10.1148/ryai.230024}.

\bibitem[Yue et~al.(2025)Yue, Wang, Tao, Liu, Song, and
  Huang]{yue2025chexworld}
Yang Yue, Yulin Wang, Chenxin Tao, Pan Liu, Shiji Song, and Gao Huang.
\newblock {CheXWorld}: Exploring image world modeling for radiograph
  representation learning.
\newblock In \emph{Proceedings of the IEEE/CVF Conference on Computer Vision
  and Pattern Recognition (CVPR)}, pp.\  20778--20788, 2025.
\newblock URL
  \url{https://openaccess.thecvf.com/content/CVPR2025/html/Yue_CheXWorld_Exploring_Image_World_Modeling_for_Radiograph_Representation_Learning_CVPR_2025_paper.html}.

\end{thebibliography}
